\documentclass[twocolumn,aps,prb,10pt]{revtex4-2}

\usepackage{amsmath}
\usepackage{amssymb}
\usepackage{wasysym}
\usepackage{graphicx}
\usepackage{hyperref}
\usepackage{dsfont}
\usepackage{newtxtext}
\usepackage[varvw]{newtxmath}

\hypersetup{
    colorlinks,
    linkcolor={blue},
    citecolor={blue},
    urlcolor={blue}
}

\graphicspath{{./}{./figs/}}

\newcommand{\rme}{\mathrm{e}}
\newcommand{\rmi}{\mathrm{i}}

\newcommand{\NCTO}{Na$_2$Co$_2$TeO$_6$}
\newcommand{\eg}{{e_\mathrm{g}}}
\newcommand{\tg}{{t_\mathrm{2g}}}
\newcommand{\Oh}{\mathrm{O}_\mathrm{h}}

\usepackage{xcolor}
\usepackage[normalem]{ulem}

\begin{document}

\title{%
Spin vestigial orders in extended Heisenberg-Kitaev models near hidden SU(2) points: Application to Na$_2$Co$_2$TeO$_6$
}

\author{Niccol\`o Francini}
\author{Lukas Janssen}

\affiliation{Institut f\"ur Theoretische Physik and W\"urzburg-Dresden Cluster of Excellence ct.qmat, TU Dresden, 01062 Dresden, Germany}

\begin{abstract}
The honeycomb magnet Na$_2$Co$_2$TeO$_6$ has recently been argued to realize an approximate hidden SU(2) symmetry that can be understood by means of a duality transformation. Using large-scale classical Monte Carlo simulations, we study the finite-temperature phase diagram of the pertinent Heisenberg-Kitaev-$\Gamma$-$\Gamma'$ model near the hidden-SU(2)-symmetric point, in the presence of a six-spin ring exchange perturbation. At low temperatures, the model features collinear single-$\mathbf{q}$ zigzag and noncollinear triple-$\mathbf{q}$ ground states, depending on the sign of the ring exchange coupling. We show that in the vicinity of the hidden-SU(2)-symmetric point, the magnetic long-range orders melt in two stages. The corresponding finite-temperature transitions are continuous and fall into 2D Ising and 2D Potts universality classes, respectively. The two fluctuation-induced phases at intermediate temperatures spontaneously break spin rotational and lattice translational symmetries, respectively, but both leave time reversal symmetry intact. They are characterized by finite expectation values of a real, symmetric, traceless, second-rank tensor, and are naturally understood as vestigial orders of the underlying magnetic states. We identify these vestigial orders as $\mathds{Z}_3$ spin nematic and $\mathds{Z}_4$ spin current density wave phases, respectively. For increasing ring exchange perturbations, the width of the vestigial phases decreases, eventually giving rise to a direct first-order transition from the magnetically-ordered phase to the disordered paramagnet. We propose the $\mathds{Z}_4$ spin current density wave phase, which is the vestigial phase of the primary triple-$\mathbf q$ magnetic order, as a natural candidate for the paramagnetic 2D long-range-ordered state observed in Na$_2$Co$_2$TeO$_6$ in a small window above the antiferromagnetic ordering temperature.
\end{abstract}

\date{February 9, 2024}

\maketitle

\section{Introduction}
\label{sec:intro}

As one of the rare instances of an exactly solvable frustrated spin-$1/2$ model on a two-dimensional lattice, the Kitaev honeycomb model \cite{kitaev06} plays an essential role in the field of quantum magnetism. Its bond-dependent exchange interactions can be realized in spin-orbit-coupled magnetic Mott insulators with edge-sharing geometries of the magnetic ions in either a low-spin $d^5$ electron configuration~\cite{jackeli09, winter16}, or a high-spin $d^7$ electron configuration~\cite{liu18, sano18, liu20, winter22}.
Materials featuring the low-spin mechanism are $A_2$IrO$_3$ ($A = \text{Na}, \text{Li}$) and $\alpha$-RuCl$_3$~\cite{choi12, williams16, johnson15, sears15, cao16, banerjee16, banerjee17, sears17, wolter17, janssen17, winter17b, sears20, krueger20, janssen20, hentrich20, balz21, 
winter17, janssen19, takagi19, trebst22}, while candidates for the high-spin mechanism are the cobaltates \NCTO, Na$_3$Co$_2$SbO$_6$, and BaCo$_2$(AsO$_4$)$_2$~\cite{yan19, yao20, songvilay20, lin21, chen21, lee21, hong21, samarakoon21, kim22, mukherjee22, sanders22, yang22, yao22, li22, gu23, zhong20, shi21, zhang22, maksimov22b, krueger22, yao23, xiang23, zhang23, hong23}.
In any of the above examples, however, additional exchange interactions beyond the nearest-neighbor Kitaev interaction are present, and stabilize magnetic long-range order at low temperatures and in the absence of an external magnetic field.
While for $A_2$IrO$_3$ and $\alpha$-RuCl$_3$, a consensus on the natures of the ground states has been reached~\cite{choi12, williams16, johnson15, sears15, cao16, balz21}, the corresponding debate for the cobaltates is still ongoing.
For \NCTO, for instance, powder neutron diffraction measurements have been interpreted in terms of a collinear single-$\mathbf q$ zigzag ground state~\cite{yao20, songvilay20, sanders22}. Recent inelastic neutron scattering data on high-quality single crystals, however, have revealed a symmetry in the magnetic excitation spectrum that is inconsistent with a generic single-$\mathbf q$ ground state and points to noncollinear triple-$\mathbf q$ order~\cite{krueger22}.
A similar ambiguity between single- and multi-$\mathbf q$ states occurs in Na$_3$Co$_2$SbO$_6$~\cite{li22, gu23}.
While BaCo$_2$(AsO$_4$)$_2$ has initially been thought to realize a noncollinear spiral order~\cite{regnault77, zhong20, zhang23}, recent works suggest a collinear double-zigzag ground state with a \mbox{$+$\,$+$\,$-$\,$-$} pattern of zigzag chains~\cite{regnault18, maksimov22b}.
At finite temperatures and/or in magnetic fields, the cobaltates exhibit a variety of phase transitions and intermediate magnetic phases, the precise natures of which are currently under intense debate~\cite{lin21, chen21, lee21, hong21, zhong20, zhang23, yao23, xiang23, hong23}.
One of the major difficulties in this context is the lack of thorough understanding of the finite-temperature physics of pertinent effective minimal spin models, describing the cobaltates.

\begin{figure}[tb]
\includegraphics[width=\linewidth]{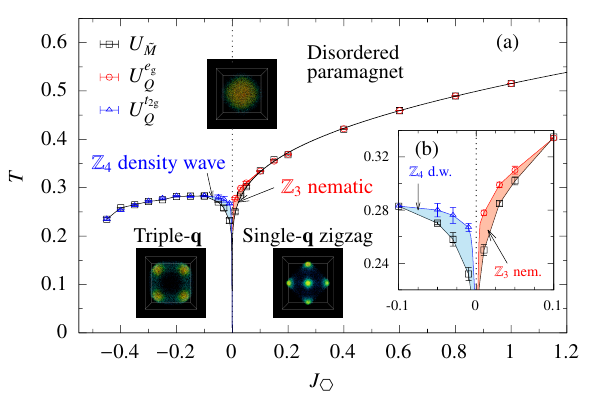}
\caption{%
(a)~Finite-temperature phase diagram of Heisenberg-Kitaev-$\Gamma$-$\Gamma'$ model as function of ring-exchange perturbation $J_{\hexagon}$ from classical Monte Carlo simulations.
Here, the bilinear couplings have been fixed as $(J, K, \Gamma, \Gamma') = (-1/9,-2/3,8/9,-4/9)$, such that $J_{\hexagon} = 0$ corresponds to the hidden-SU(2)-symmetric point.
At high temperatures, the model is in the disordered paramagnetic phase, characterized by an isotropic distribution of the dual magnetization $\tilde{\mathbf{M}}$, see upper inset [measured at $(J_{\hexagon}, T) = (0.2, 0.422)$].
At low temperatures, the model features magnetically-ordered collinear single-$\mathbf q$ zigzag and noncollinear triple-$\mathbf q$ phases for $J_{\hexagon} > 0$ and $J_{\hexagon} < 0$, respectively.
These are characterized by dual magnetizations $\tilde{\mathbf{M}}$ along the three cubic basis vectors and four cubic diagonals, respectively, as illustrated by the distributions shown in the two lower insets  [measured at $(J_{\hexagon}, T) = (0.4, 0.405)$ and $(-0.05, 0.233)$].
For $0 < |J_{\hexagon}| \lesssim 0.1$ near the hidden-SU(2)-symmetric point, the magnetic orders melt in two stages, with vestigial intermediate $\mathds{Z}_3$ spin nematic and $\mathds{Z}_4$ spin current density wave phases emerging at finite temperatures.
Black line for $J_{\hexagon} > 0$ represents fit according to Eq.~\eqref{eq:fit-shift-temperature}.
Other lines are guides to the eye.
(b)~Close-up view of vicinity of hidden-SU(2)-symmetric point, showing the spin vestigial phases.}
\label{fig:pd}
\end{figure}

In this work, we aim at filling this gap. By means of large-scale classical Monte Carlo simulations, we investigate the finite-temperature phase diagram of a minimal spin model relevant for \NCTO.
The model features all nearest-neighbor bilinear spin exchange interactions that are compatible with the symmetries of the material. Besides the standard Heisenberg exchange parametrized by $J$, this includes the Kitaev $K$ and off-diagonal $\Gamma$ and $\Gamma'$ interactions~\cite{chaloupka10, chaloupka13, rau14a, rau14b}. The $K$, $\Gamma$, and $\Gamma'$ interactions arise from spin-orbit coupling and reduce the standard SU(2) spin rotational symmetry to a discrete $C_3^*$ symmetry involving $2\pi/3$ rotations of pseudospins around the out-of-plane axis combined with $2\pi/3$ lattice rotations \cite{janssen16, janssen19}.
Within this large parameter space, we focus on the vicinity of an isolated point that features a \emph{hidden} SU(2) symmetry, which has recently been argued to realize a good starting point to understand the physics of \NCTO~\cite{krueger22}.
The hidden-SU(2)-symmetric point can be mapped to a standard nearest-neighbor Heisenberg model by means of a duality transformation~\cite{chaloupka15}.
Thus, at this point, long-range order is forbidden at any finite temperature as a consequence of the Mermin-Wagner theorem~\cite{mermin66}.
Perturbations away from the hidden-SU(2)-symmetric point will induce magnetic long-range order at low temperatures.
The nature of the induced order, however, crucially depends on the type and sign of the perturbation. This is due to the local nature of the duality transformation, which maps different members of the ground-state manifold at the hidden-SU(2)-symmetric point to different types of magnetic orders. For instance, the N\'eel state with staggered magnetization along the cubic [001] axis is mapped via the duality transformation to a collinear single-$\mathbf q$ zigzag state, while the N\'eel state with staggered magnetization along the cubic [111] direction is mapped to a noncollinear triple-$\mathbf q$ state~\cite{krueger22}.
While bilinear perturbations away from the hidden-SU(2)-symmetric point have been shown to lift the SU(2) ground-state degeneracy in favor of the collinear single-$\mathbf q$ zigzag states, nonbilinear perturbations can, depending on their sign, also induce noncollinear multi-$\mathbf q$ orderings.
Here, we focus on the six-spin ring exchange, which arises as leading correction to the nearest-neighbor Heisenberg exchange in the strong-coupling expansion of the single-band Hubbard model on the honeycomb lattice~\cite{yang12}.
For positive ring exchange, a collinear single-$\mathbf q$ zigzag state is realized at low temperatures.
Negative ring exchange, on the other hand, induce the noncollinear triple-$\mathbf q$ ordering that is believed to be realized in \NCTO~\cite{krueger22}.
We show that in the vicinity of the hidden-SU(2)-symmetric point, both magnetic orders melt in two stages, making room for novel long-range-ordered paramagnetic phases at intermediate temperatures, which separate the magnetically-ordered phases at low temperatures from the disordered paramagnet at high temperatures, see Fig.~\ref{fig:pd}.
The intermediate phases are characterized by finite expectation values of a real, symmetric, traceless, second-rank tensor and spontaneously break $C_3^*$ rotational symmetry and lattice translational symmetry, respectively, but leave time reversal symmetry intact. They can be understood as vestigial orders of the zigzag and triple-$\mathbf q$ states, respectively, in the sense that they feature a preferred axis in the dual spin space, while the two possible orientations along this axis remain still equivalent, and one of it will be eventually selected only in the low-temperature magnetically-ordered phase. 
Our large-scale classical Monte Carlo simulations indicate that both the high-temperature transition between the disordered paramagnet and the intermediate vestigial phase, as well as the low-temperature transition between the vestigial and the magnetic orders, are continuous. We characterize the universal critical behaviors and identify the corresponding universality classes as 2D Potts and 2D Ising, respectively.
We argue that the $\mathds{Z}_4$ spin current density wave phase that emerges at finite temperatures as vestigial order of the triple-$\mathbf q$ magnetically-ordered state is a natural candidate for the paramagnetic 2D long-range order observed in \NCTO\ in a small window above the antiferromagnetic ordering temperature.
Upon increasing the ring exchange perturbation beyond a certain finite threshold, the spin vestigial phases vanish and give way to a direct first-order transition between the magnetically-ordered low-temperature phase and the disordered high-temperature paramagnet, the strength of which increases upon increasing the perturbation.
All these results fully fit into the general picture of vestigial orders in systems with multicomponent order parameters~\cite{fernandes19}.

The remainder of the paper is organized as follows: In Sec.~\ref{sec:model}, we introduce the Heisenberg-Kitaev-$\Gamma$-$\Gamma'$ model with ring exchange perturbation and discuss the duality transformation that maps the hidden-SU(2)-symmetric point in this model to a Heisenberg model in terms of dual spins. Algorithmic details of our Monte Carlo simulations and an overview of the observables considered are given in Sec.~\ref{sec:monte-carlo-simulations}. Section~\ref{sec:phase-diagram} contains the discussion of the finite-temperature phase diagram. The nature of the various finite-temperature transitions is analyzed in Sec.~\ref{sec:transitions}. In Sec.~\ref{sec:ncto}, we discuss our findings in light of the experiments in \NCTO. Our conclusions are given in Sec.~\ref{sec:conclusions}. In the appendix, we present additional data elucidating the crossover behavior in the vicinity of the hidden-SU(2)-symmetric point.

\section{Model}
\label{sec:model}

\subsection{Heisenberg-Kitaev-\(\boldsymbol{\Gamma}\)-\(\boldsymbol{\Gamma'}\) model}

\begin{figure*}
\includegraphics[width=\linewidth]{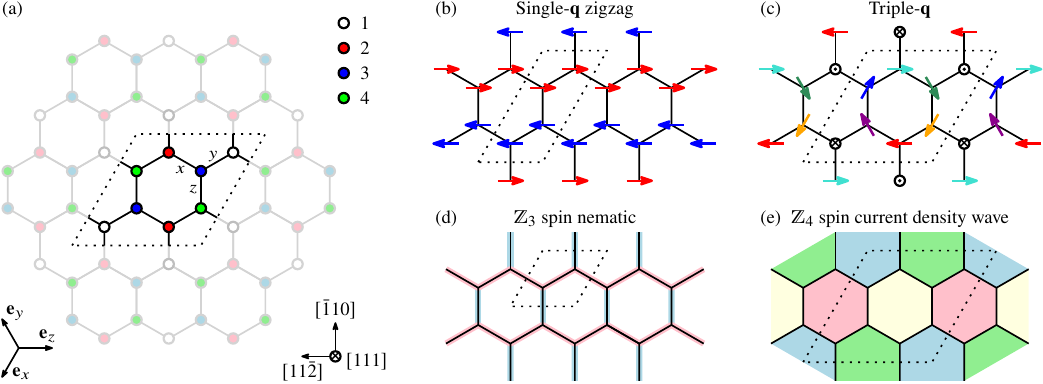}
\caption{%
(a)~Four-sublattice structure used to define the $\mathcal T_1 \mathcal T_4$ duality transformation. Dashed parallelogram indicates a corresponding unit cell, consisting of two sites per sublattice. The different colors indicate the different sublattices. The lower left inset indicates the projection of the cubic axes $\mathbf{e}_x$, $\mathbf{e}_y$, and $\mathbf{e}_z$ onto the honeycomb plane. The latter is spanned by the $[11\bar 2]$ and $[\bar 110]$ axes, as indicated in the lower right inset.
(b)~Representative spin configuration of collinear single-$\mathbf q$ zigzag state, arising from the duality transformation of a N\'eel state with staggered magnetization along the $[001]$ axis.
Arrows correspond to spin directions projected onto the honeycomb plane.
%
(c)~Same as (b), but for the noncollinear triple-$\mathbf q$ state, arising from the duality transformation of a N\'eel state with staggered magnetization along the $[111]$ axis.
(d)~Illustration of the $\mathds{Z}_3$ spin nematic order, which is the vestigial order of the single-$\mathbf q$ zigzag state shown in (b). Different colors indicate inequivalent bonds. The $\mathds{Z}_3$ spin nematic order respects time reversal symmetry, but breaks $C_3^*$ rotational symmetry. The corresponding unit cell (dashed parallelogram) coincides with the crystallographic unit cell.
(e)~Same as (d), but for the $\mathds{Z}_4$ spin current density wave order, which is the vestigial order of the triple-$\mathbf q$ state shown in (c). Different colors indicate inequivalent plaquettes. The $\mathds{Z}_4$ spin current density wave order respects time reversal symmetry, but breaks translational symmetry by doubling the length of both lattice vectors, corresponding to an eight-site unit cell (dashed parallelogram).}
\label{fig:duality}
\end{figure*}

On the level of nearest-neighbor interactions, the most general bilinear spin Hamiltonian, compatible with the $C_3^*$ symmetry of combined spin and lattice rotations~\cite{janssen16, janssen19}, is given by the Heisenberg-Kitaev-$\Gamma$-$\Gamma'$ model,
\begin{equation}
\label{eq:extended-KH-term}
    \begin{split}
    \mathcal H_{\mathrm{HK}\Gamma\Gamma'} = &\sum_{\gamma=x,y,z} \sum_{\langle ij \rangle_{\gamma}} 
    \Bigl[ J \mathbf{S}_i \cdot \mathbf{S}_j + K S_{i}^{\gamma}S_{j}^{\gamma}\\
    & + \Gamma(S_{i}^{\alpha} S_{j}^{\beta} + S_{i}^{\beta}S_{j}^{\alpha}) \\
    & + \Gamma'(S_{i}^{\gamma}S_{j}^{\alpha} + S_{i}^{\alpha}S_{j}^{\gamma} + S_{i}^{\gamma}S_{j}^{\beta} + S_{i}^{\beta}S_{j}^{\gamma}) \Bigr].
    \end{split}
\end{equation}
Here, $\langle ij \rangle_{\gamma}$ labels nearest neighbors along a $\gamma$ bond on the honeycomb lattice and $(\alpha, \beta, \gamma) = (x,y,z)$ or cyclic permutations thereof.
$(S^x_i, S^y_i, S^z_i)$ are the spin components in the cubic coordinate system,
and $\mathbf{S}_i = S^x_i \mathbf{e}_x + S^y_i \mathbf{e}_y + S^z_i \mathbf{e}_z$.
The coupling $J$ parametrizes the isotropic Heisenberg exchange, $K$ is the Kitaev coupling, and $\Gamma$ and $\Gamma'$ correspond to symmetric off-diagonal exchanges.

\subsection{\texorpdfstring{\(\boldsymbol{\mathcal{T}_1 \mathcal{T}_4}\)}{T1T4} hidden-SU(2)-symmetric point}

The Heisenberg-Kitaev-$\Gamma$-$\Gamma'$ model features five hidden-SU(2)-symmetric points that can be mapped to antiferromagnetic Heisenberg models in terms of dual spins~\cite{chaloupka15}. Recently, it has been shown that the magnetic excitation spectrum in the low-temperature ordered phase of \NCTO\ is modeled remarkably well using a parameter set proximate to one of these points, namely, the $\mathcal{T}_1\mathcal{T}_4$ point~\cite{krueger22}.
Most importantly, a model proximate to this hidden-SU(2)-symmetric point stabilizes the triple-$\mathbf q$ order, required to reproduce the experimentally-observed symmetry in the excitation spectrum. Moreover, the low-energy excitations with small gap $\sim 1\text{ meV}$ at both the $\boldsymbol{\Gamma}$ and the $\mathbf M$ points in the crystallographic Brillouin zone can be understood as pseudo-Goldstone modes arising from the  breaking of the approximate hidden SU(2) symmetry, with the size of the gap representing a measure of proximity to the hidden-SU(2)-symmetric point.

The $\mathcal T_1 \mathcal T_4$ point of hidden SU(2) symmetry can be found by combining a four-sublattice transformation $\mathcal T_4$~\cite{chaloupka10, chaloupka13} with a global rotation $\mathcal T_1$~\cite{chaloupka15}.
Here, $\mathcal{T}_4$ leaves spins on sublattice $1$ invariant, while inducing $\pi$ rotations of the spins on sublattices 2, 3, and 4 around the $[001] \parallel \mathbf{e}_z$, $[010] \parallel \mathbf{e}_y$, and $[100] \parallel \mathbf{e}_x$ axes in cubic spin space, see Fig.~\ref{fig:duality}(a).
The subsequent $\mathcal T_{1}$ transformation corresponds to a global $\pi$ rotation around the $[111]$ axis.
In total, the spins on the four sublattices transform as
\begin{equation}
\label{eq:duality-transformation}
    \mathcal T_1 \mathcal T_4: \quad 
    \mathbf{S}_i \mapsto T_{14} \mathbf S_i \coloneqq 
    \begin{cases}
        R_1 {\mathbf{S}}_i         &  \text{for $i \in $ sublattice 1,} \\
        R_1 R_{4}^z {\mathbf{S}}_i &  \text{for $i \in $ sublattice 2,} \\
        R_1 R_{4}^y {\mathbf{S}}_i &  \text{for $i \in $ sublattice 3,} \\
        R_1 R_{4}^x {\mathbf{S}}_i &  \text{for $i \in $ sublattice 4,} \\
    \end{cases}
\end{equation}
with rotation matrices
\begin{align}
\begin{split}
    R_{4}^x & = 
    \begin{pmatrix}
        1 & & \\
        & -1 & \\
        & & -1
    \end{pmatrix}, \qquad
    R_{4}^y = 
    \begin{pmatrix}
        -1 & & \\
        & 1 & \\
        & & -1
    \end{pmatrix}, \\ 
    R_{4}^z & = 
    \begin{pmatrix}
        -1 & & \\
        & -1 & \\
        & & 1
    \end{pmatrix}, \qquad
    R_{1} = 
    \frac{1}{3}
    \begin{pmatrix}
        -1 & 2 & 2 \\
        2 & -1 & 2 \\
        2 & 2 & -1
    \end{pmatrix}.
\end{split}
\end{align}
Under this transformation, an antiferromagnetic Heisenberg model $\tilde{\mathcal H}$ in terms of dual spins $\tilde{\mathbf S}_i = T_{14}^\top \mathbf{S}_i$,
\begin{equation}
\label{eq:dual-hamiltonian}
    \tilde{\mathcal H} = A \sum_{\langle ij \rangle} \tilde{\mathbf S}_i \cdot \tilde{\mathbf S}_j, \qquad A>0,
\end{equation}
maps to a Heisenberg-Kitaev-$\Gamma$-$\Gamma'$ model $\mathcal H_{\text{HK}\Gamma\Gamma'}$ with parameters~\cite{chaloupka15}
\begin{equation}
    \label{eq:SU(2)-values}
    (J, K, \Gamma, \Gamma') = (-1/9, -2/3, 8/9,-4/9)A.
\end{equation}
This defines the $\mathcal T_1 \mathcal T_4$ point of hidden SU(2) symmetry.
Importantly, the duality transformation respects the SU(2) spin algebra, such that the physics of the Heisenberg-Kitaev-$\Gamma$-$\Gamma'$ model at the hidden-SU(2)-symmetric point and the dual Heisenberg model can be exactly mapped onto each other not only for the static spin configurations in the classical limit, but also upon the inclusion of quantum and/or thermal fluctuations.

\subsection{Ring exchange perturbations}

In the vicinity of the hidden-SU(2)-symmetric point, the influence of further nonbilinear exchange interactions may become relevant.
Nonbilinear exchange interactions are important in a number of 3$d$ materials, including various \mbox{chromium-,} manganese-, and copper-based magnets~\cite{kvashnin20, fedorova15, dallapiazza12, larsen19}.
Here, we consider the six-spin ring exchange interaction, which arises as leading correction to the nearest-neighbor Heisenberg exchange in the strong-coupling expansion of the single-band Hubbard model on the honeycomb lattice \cite{yang12},
\begin{equation}
\label{eq:ring-term}
    \begin{split}
       \mathcal H_{\hexagon} & = \frac{J_{\hexagon}}{6} \sum_{\langle ijklmn \rangle} 
        \bigl[ 2(\mathbf{S}_i \cdot \mathbf{S}_{j})(\mathbf{S}_k \cdot \mathbf{S}_{l})(\mathbf{S}_m \cdot \mathbf{S}_{n}) \\
        &\quad - 6(\mathbf{S}_i \cdot \mathbf{S}_{k})(\mathbf{S}_j \cdot \mathbf{S}_{l})(\mathbf{S}_m \cdot \mathbf{S}_{n}) \\
        &\quad + 3(\mathbf{S}_i \cdot \mathbf{S}_{l})(\mathbf{S}_j \cdot \mathbf{S}_{k})(\mathbf{S}_m \cdot \mathbf{S}_{n}) \\
        &\quad + 3(\mathbf{S}_i \cdot \mathbf{S}_{k})(\mathbf{S}_j \cdot \mathbf{S}_{m})(\mathbf{S}_l \cdot \mathbf{S}_{n}) \\
        &\quad - (\mathbf{S}_i \cdot \mathbf{S}_{l})(\mathbf{S}_j \cdot \mathbf{S}_{m})(\mathbf{S}_k \cdot \mathbf{S}_{n}) \\
        &\quad + \text{cyclic permutation of }(ijklmn) \bigr]\,,
    \end{split}
\end{equation}
where the sum runs over the elementary hexagonal plaquettes following the sites $(ijklmn)$ in counterclockwise order.
The full Hamiltonian investigated in this work reads
\begin{equation}
\label{eq:full-model}
    \mathcal H= \mathcal H_{\mathrm{HK}\Gamma\Gamma'} + \mathcal H_{\hexagon}.
\end{equation}
We note that many other nonbilinear perturbations, which are compatible with the symmetries of the Heisenberg-Kitaev-$\Gamma$-$\Gamma'$ model, are in principle conceivable; however, these are expected to lead to qualitatively similar physics, as long as these perturbations remain small~\cite{krueger22}.

The ring exchange perturbation, just as bilinear perturbations within the Heisenberg-Kitaev-$\Gamma$-$\Gamma'$ theory space, explicitly breaks the hidden SU(2) symmetry present at the $\mathcal T_1 \mathcal T_4$ point given by the parameters in Eq.~\eqref{eq:SU(2)-values}. 
The perturbation, however, is naturally understood differently in the two different formulations of the model:
In the native formulation in terms of the original spins $\mathbf S_i$ in Eq.~\eqref{eq:extended-KH-term}, $\mathcal H_{\hexagon}$ reduces SU(2) to the usual $C_{3}^*$ symmetry involving $2\pi/3$ rotations of pseudospins around the out-of-plane axis combined with $2\pi/3$ lattice rotations \cite{janssen16, janssen19}.
In the dual framework in terms of the dual spins $\tilde{\mathbf S}_i$ in Eq.~\eqref{eq:dual-hamiltonian}, $\mathcal{H}_{\hexagon}$ is naturally understood as a cubic perturbation, corresponding to the breaking of $\mathrm{SU}(2)$ to the octahedral group $\Oh$.

\subsection{Magnetic orders}

At the hidden-SU(2)-symmetric point, magnetic long-range order is forbidden at any finite temperature as a consequence of the Mermin-Wagner theorem~\cite{mermin66}.
At zero temperature, there is an SU(2) degeneracy of ground states that can be constructed from the duality transformation of the N\'eel states in the dual Heisenberg formulation.
As the duality transformation acts differently on the four different sublattices, N\'eel states with different directions of the staggered magnetization in the dual formulation are mapped onto quite distinct ground states of the Heisenberg-Kitaev-$\Gamma$-$\Gamma'$ model at the hidden-SU(2)-symmetric point.
On the one hand, for instance, the N\'eel state with staggered magnetization along the $[001]$ direction, $\tilde{\mathbf S}_i = \pm S \mathbf{e}_z$, maps to the collinear single-$\mathbf q$ $z$-zigzag state with spin directions
\begin{equation}
\label{eq:zigzag}
    \mathbf{S}_i/S =
    \begin{cases}
        \pm (2\mathbf{e}_x + 2\mathbf{e}_y - \mathbf{e}_z)/3 & \text{for $i \in$ sublattices 1, 2},\\
        \mp (2\mathbf{e}_x + 2\mathbf{e}_y - \mathbf{e}_z)/3 & \text{for $i \in$ sublattices 3, 4},
    \end{cases}
\end{equation}
where the upper (lower) signs apply to sites on the crystallographic A (B) sublattice, and $\mathbf{e}_x$, $\mathbf{e}_y$, $\mathbf{e}_z$ denote the cubic basis vectors in spin space.
The $z$-zigzag state is illustrated in Fig.~\ref{fig:duality}(b).
On the other hand, the N\'eel state with staggered magnetization along the $[111]$ direction, $\tilde{\mathbf S}_i = \pm S (\mathbf{e}_x + \mathbf{e}_y + \mathbf{e}_z)/\sqrt{3}$, maps to the noncollinear triple-$\mathbf q$ state with spin directions
\begin{equation}
\label{eq:triple-q}
    \mathbf{S}_i/S =
    \begin{cases}
        \pm (\mathbf{e}_x + \mathbf{e}_y + \mathbf{e}_z)/\sqrt{3} & \text{for $i \in$ sublattice 1},\\
        \pm (\mathbf{e}_x + \mathbf{e}_y - 5 \mathbf{e}_z)/(3\sqrt{3}) & \text{for $i \in$ sublattice 2},\\
        \pm (\mathbf{e}_x - 5 \mathbf{e}_y + \mathbf{e}_z)/(3\sqrt{3}) & \text{for $i \in$ sublattice 3},\\
        \pm (-5\mathbf{e}_x + \mathbf{e}_y + \mathbf{e}_z)/(3\sqrt{3}) & \text{for $i \in$ sublattice 4},
    \end{cases}
\end{equation}
where the upper (lower) signs again apply to sites on the crystallographic A (B) sublattice.
The triple-$\mathbf q$ state is illustrated in Fig.~\ref{fig:duality}(c).

While the collinear single-$\mathbf q$ zigzag and noncollinear triple-$\mathbf q$ states are degenerate at the hidden-SU(2)-symmetric point, perturbations away from this point will lift the degeneracy.
Previously, it has been shown that bilinear perturbations within the space spanned by the parameters $J$, $K$, $\Gamma$, and $\Gamma'$ favor the single-$\mathbf q$ zigzag state for either sign of the perturbation. In contrast, nonbilinear perturbations will generically favor the single-$\mathbf q$ zigzag state for one sign and the triple-$\mathbf q$ state for the opposite sign of the nonbilinear perturbation~\cite{krueger22}.
By comparing the classical energies of the static spin configurations in Eqs.~\eqref{eq:zigzag} and \eqref{eq:triple-q}, it is easy to show that, in the vicinity of the hidden-SU(2)-symmetric point and at low temperatures, the single-$\mathbf q$ zigzag state is stabilized for $J_{\hexagon} > 0$, while the triple-$\mathbf q$ state is stabilized for $J_{\hexagon} < 0$.

We emphasize that the two different low-temperature magnetically-ordered states break different symmetries. 
The collinear single-$\mathbf q$ zigzag state is characterized by a preferred spin axis, thereby breaking the $C_{3}^*$ symmetry of combined lattice and spin rotation symmetry. It also doubles the length of one of the lattice vectors, i.e., the corresponding magnetic unit cell features four sites. 
The noncollinear triple-$\mathbf q$ state does not break the $C_{3}^*$ symmetry, but doubles the length of both lattice vectors, i.e., the corresponding magnetic unit cell now features eight sites.
It is this difference in symmetry that allows one to identify multi-$\mathbf q$ ground states in magnetic excitation spectra, and distinguish them from domain averages of single-$\mathbf q$ states~\cite{krueger22}.

\subsection{Spin vestigial orders}
\label{subsec:nematic}

Below, we present arguments for the emergence of vestigial orders at intermediate temperatures in the vicinity of the hidden-SU(2)-symmetric point.
Spin vestigial orders can be understood as partial meltings of primary magnetic orders before the system goes to a completely disordered paramagnetic state~\cite{fernandes12, fernandes19}.
Vestigial orders are characterized by composite order parameters, constructed from bilinears or multilinears of the individual order parameters of the corresponding primary states.
In our case, it is useful to define the composite order parameter in terms of the dual spins $\tilde{\mathbf S}_i = T_{14}^\top \mathbf S_{i}$.
From these, we can construct the three-dimensional, real, symmetric, and traceless tensorial order parameter on sites $i$ and $j$ as~\cite{shannon10}
\begin{equation}
    Q^{\alpha\beta}_{ij} = \tilde{S}_i^\alpha \tilde{S}_j^\beta + \tilde{S}_i^\beta \tilde{S}_j^\alpha - \tfrac{2}{3} (\mathbf{\tilde{S}}_i \cdot \mathbf{\tilde{S}}_j) \delta^{\alpha\beta},
\end{equation}
where $\alpha, \beta \in \{x,y,z\}$ indicate the different components of the tensor.
The above composite order parameter can be decomposed into its independent components by projection onto the five real Gell-Mann matrices $\Lambda^{(a)}$ as~\cite{janssen15}
\begin{equation}
       Q^{(a)}_{ij} = \frac{1}{2} \sum_{\alpha\beta} Q^{\alpha\beta}_{ij} \Lambda^{(a)}_{\beta\alpha}, \qquad a = 1, \dots, 5.
\end{equation}
In the case of cubic symmetry, it is useful to arrange the five components into groups of two and three, respectively, as
\begin{multline}
\label{eq:eg}
    \eg: \quad 
    \bigl(Q^{(1)}_{ij}, Q^{(2)}_{ij} \bigr) = 
    \bigl(
        \tfrac{1}{\sqrt{3}}[2 \tilde{S}^z_i \tilde{S}^z_j - \tilde{S}^x_i \tilde{S}^x_j - \tilde{S}^y_i \tilde{S}^y_j],
\\
        \tilde{S}^x_i \tilde{S}^x_j - \tilde{S}^y_i \tilde{S}^y_j
    \bigr),
\end{multline}
which transforms as a doublet under the octahedral group $\Oh$ in the dual spin space, and
\begin{align}
\label{eq:t2g}
    \tg: \quad 
    \bigl(Q^{(3)}_{ij}, Q^{(4)}_{ij}, Q^{(5)}_{ij}\bigr) = \bigl(
        \tilde{S}^y_i \tilde{S}^z_j,
        \tilde{S}^z_i \tilde{S}^x_j,
        \tilde{S}^x_i \tilde{S}^y_j
    \bigr) + (i \leftrightarrow j),
\end{align}
which transforms as a triplet.

The spin vestigial states spontaneously break the $\Oh$ symmetry in the dual formulation by selecting a preferred axis in the dual spin space, while leaving time reversal symmetry intact.
As a consequence, in the present case with cubic symmetry, there are two different spin vestigial orders, characterized by the doublet $\eg$ [Eq.~\eqref{eq:eg}] and triplet $\tg$ [Eq.~\eqref{eq:t2g}], respectively, of the composite order parameter.
A finite expectation value of a doublet $\eg$ component, together with a vanishing local magnetization, corresponds to a vestigial paramagnetic order, in which fluctuations single out one of the three cubic axes in the dual spin space.
This breaking of $\Oh$ in the dual formulation corresponds to a breaking of the $C_3^*$ rotational symmetry in the original formulation of spins.
This vestigial order is therefore a $\mathds{Z}_3$ spin nematic. It is illustrated in Fig.~\ref{fig:duality}(d).
The corresponding primary magnetic order is the collinear single-$\mathbf q$ zigzag state shown in Fig.~\ref{fig:duality}(b).
By contrast, a finite expectation value of a triplet $\tg$ component, together with a vanishing local magnetization, corresponds to a vestigial paramagnetic order, in which fluctuations single out one of the four cubic diagonals in the dual spin space.
This selection leaves the $C_3^*$ symmetry of the original spin formulation intact, but enlarges the unit cell by doubling the two lattice vectors.
Such vestigial order has previously been discussed in the context of doped graphene and can be understood as a $\mathds{Z}_4$ spin current density wave~\cite{chern12, fernandes19}.
It is illustrated in Fig.~\ref{fig:duality}(e).
The corresponding primary magnetic order is the noncollinear triple-$\mathbf q$ state shown in Fig.~\ref{fig:duality}(c).

\section{Monte Carlo simulations}
\label{sec:monte-carlo-simulations}

\subsection{Algorithmic details}

The model in Eq.~\eqref{eq:full-model} is simulated on a two-dimensional honeycomb lattice spanned by the lattice vectors $\mathbf a_1 = (3/2, \sqrt{3}/2)$ and $\mathbf a_2 = (3/2, -\sqrt{3}/2)$ with $N=2L^2$ sites and periodic boundary conditions at finite temperatures $T>0$. The spins are treated as classical three-dimensional vectors $\mathbf{S}_i=(S_i^x, S_i^y, S_i^z)$ of fixed length, and we choose units in which $\mathbf{S}_i^2 = 1$ and $k_\mathrm{B} = 1$. 
As a consequence, temperature $T$ and ring exchange coupling $J_{\hexagon}$ are measured in units of $A S^2/k_\mathrm{B}$ and $A/S^4$, respectively.
We employ large-scale classical Monte-Carlo simulations based on the Metropolis algorithm combined with an overrelaxation step with a ratio of 1:5. The typical order of magnitude of the accumulated statistics is $\mathcal O(10^5)$ configurations per point in parameter space, with each of these configurations taken after ten complete Metropolis and overrelaxation updates on the full lattice.
Note that the ring exchange perturbation includes 26 independent terms per plaquette, each one of it involving products of six spins. This significantly slows down the simulations in comparison with previous works on related bilinear models~\cite{price12, price13, chern17, janssen16, janssen17, andrade20}. 
Nevertheless, with our efficient code and high-performance computing resources available~\cite{nhr-alliance}, we are able to simulate systems with up to $N = 2 \times 128^2$ spins.

\subsection{Observables}

In this subsection, we describe the observables measured in the simulations.

\paragraph{Energy density.}

The energy density is given by the expectation value of the Hamiltonian
\begin{equation}
    \label{eq:energy-density}
    \varepsilon \coloneqq \langle \mathcal H \rangle /N,
\end{equation}
where $N$ corresponds to the total number of sites on the lattice.

\paragraph{Dual magnetization.}

Furthermore, we measure the staggered dual magnetization $\tilde M = \langle |\tilde{\mathbf M}| \rangle$ with
\begin{equation}
    \label{eq:dual-magnetization}
    \tilde{\mathbf{M}} \coloneqq \frac{1}{N} \sum_{i} (-1)^i \tilde{\mathbf{S}}_i,
\end{equation}
where $\tilde{\mathbf{S}}_i = T_{14}^\top \mathbf{S}_i$ is the dual spin at site $i$, with the transformation matrix $T_{14}$ as defined in Eq.~\eqref{eq:duality-transformation}. The factor $(-1)^i$ is one (minus one) for $i \in \mathrm{A}$ ($i \in \mathrm{B}$), where A and B correspond to the two crystallographic sublattices.

\paragraph{Composite order parameters.}

In order to distinguish single-$\mathbf q$ zigzag and triple-$\mathbf q$ orders in the simulations, and to identify the spin vestigial orders, we measure the ${\eg}$ composite order parameter $Q_\eg = 
\langle | \mathbf{Q}_\eg |\rangle$ with
\begin{align}
\label{eq:eg-op}
    \mathbf{Q}_{\eg} \coloneqq \tfrac{1}{2} \bigl(
        2\tilde{M}^2_z - \tilde{M}^2_x- \tilde{M}^2_y,
        \sqrt{3} [\tilde{M}^2_x-\tilde{M}^2_y]
    \bigr),
\end{align}
as well as the $\tg$ composite order parameter $Q_\tg = 
\langle | \mathbf{Q}_\tg |\rangle$ with
\begin{align}
\label{eq:t2g-op}
    \mathbf{Q}_{\tg} \coloneqq \sqrt{3} \bigl(
        \tilde{M}_y\tilde{M}_z, 
        \tilde{M}_z\tilde{M}_y,
        \tilde{M}_x\tilde{M}_y
    \bigr).
\end{align}
Here, the composite order parameters are normalized such that $(Q_\eg, Q_\tg) = (1, 0)$ for the single-$\mathbf q$ zigzag magnetic order in the low-temperature limit, and $(Q_\eg, Q_\tg) = (0, 1)$ for the triple-$\mathbf q$ magnetic order in the low-temperature limit. 

\paragraph{Susceptibilities.}

We also measure the susceptibilities of the different order parameters
\begin{equation}
\label{eq:susceptibility}
    \chi_{\mu} = \frac{N}{T} \bigl( \langle\mu^2 \rangle - \langle \mu \rangle^2\bigr) 
    \quad\mathrm{with}\quad 
    \mu = \tilde{\mathbf{M}},\mathbf{Q}_{\eg}, \mathbf{Q}_{\tg}.
\end{equation}

\paragraph{Binder cumulants.}

From the primary and composite order parameters, we can construct corresponding renormalization-group-invariant Binder cumulants
\begin{equation}
    \label{eq:binder-cumulant}
    U_{\mu}=\frac{\langle \mu^4 \rangle}{\langle \mu^2 \rangle^2}
    \quad\mathrm{with}\quad
    \mu = \tilde{\mathbf{M}},\mathbf{Q}_{\eg}, \mathbf{Q}_{\tg},
\end{equation}
which are useful to extract the location of the transition points from the finite-size simulations and to analyze the natures of the transitions.

\paragraph{Correlation length.}

Another useful renormalization-group-invariant quantity is given by the dimensionless correlation-length ratio $R_\xi=\xi/L$, where $\xi$ corresponds to the second-moment correlation length defined as
\begin{equation}
    \label{eq:corr-length}
    \xi=\frac{1}{2\sin{k_\text{min}/2}}\sqrt{\frac{\langle \tilde{G}(\mathbf 0)\rangle}{\langle \tilde{G}(\mathbf k_\text{min}) \rangle} - 1}\,.
\end{equation}
Here, $\mathbf{k}_\text{min}=\bigl(2\pi/(3L), 2\sqrt{3}\pi/(3L)\bigr)$ denotes the minimal momentum on the finite-size lattice, and $\tilde G(\mathbf p)$ corresponds to the static dual spin structure factor,
\begin{equation}
\label{eq:structure-factor}
    \tilde{G}(\mathbf p) = \frac{1}{N} \sum_{i,j} (-1)^{i+j} \tilde{\mathbf{S}}_{i} \cdot \tilde{\mathbf{S}}_{j} \rme^{-\rmi \mathbf p \cdot (\mathbf r_i - \mathbf r_j)},
\end{equation}
where $\mathbf r_i$ and $\mathbf r_j$ are the position vectors at sites $i$ and $j$, respectively.

\paragraph{Histograms.}

We furthermore study histograms of the Markov chain of the energy density $\varepsilon$ and the dual magnetization $\tilde M$.
Their behaviors at a phase transition indicate the nature of the transition:
While a continuous transition is reflected in a single maximum of the histograms of all observables, a first-order transition leads to a double-peak structure in (at least one of) the histograms.

\subsection{Finite-size scaling analysis}

\paragraph{Continuous transition.}

In the vicinity of a critical point, renormalization-group-invariant quantities are expected to scale as
\begin{equation}
    \label{eq:RG-invariant-critical-scaling}
    R=f_{R}((T-T_\mathrm{c})L^{1/\nu}) + \mathcal O(L^{-\omega}),
\end{equation}
where $R$ is a generic renormalization-group-invariant observable, such as the correlation-length ratio $R_\xi$ and the Binder cumulants $U_\mu$, 
$f_R$ is an (up to a rescaling of its argument) universal scaling function,
$T_\mathrm{c}$ is the critical temperature,
$\nu$ is the correlation-length exponent,
and
$\mathcal O(L^{-\omega})$ corresponds to corrections to scaling.

If the latter can be neglected, the curves for $R$ as function of temperature $T$ for different fixed lattices sizes will cross at $T_\mathrm{c}$.
%
%
We observe slight drifts in the crossing points as function of lattice size, which can be attributed to the presence of moderate scaling corrections.
In order to estimate the critical temperature in the thermodynamic limit, we plot the crossings of consecutive-size couples in the different renormalization-group-invariant observables as function of $1/L$ and extrapolate towards $1/L \to 0$.

Moreover, the graphs of $R$, plotted as function of $(T - T_\mathrm{c}) L^{1/\nu}$ for different fixed lattice sizes $L$, should collapse onto a single curve given by $f_R$, if scaling corrections are small.
In our analysis, we attempt such scaling of sets of increasing system sizes, neglecting the terms $\mathcal O(L^{-\omega})$ in Eq.~\eqref{eq:RG-invariant-critical-scaling}. This allows us to determine the effective correlation-length exponent $\nu$, which we again plot as function of $1/L$ and extrapolate towards the thermodynamic limit.

\paragraph{First-order transition.}

We emphasize that a first-order transition may also show scaling behavior, if the transition is governed by a discontinuity fixed point~\cite{nienhuis75, fisher82, binder87}.
The discontinuity fixed point features an infrared relevant direction with associated eigenvalue $\theta = d$ of the renormalization group stability matrix, where $d$ corresponds to the spatial dimension, leading to a finite-size scaling of the dimensionless correlation-length ratio
\begin{align}
\label{eq:first-order-scaling}
    R_\xi=f_{\xi}((T-T_\mathrm{c})L^{\theta}) + \mathcal O(L^{-\omega}),
\end{align}
which is formally equivalent to Eq.~\eqref{eq:RG-invariant-critical-scaling} for $R = R_\xi$ upon identifying $\theta = d$ with $1/\nu$.
A first-order phase transition arising from the presence of a discontinuity fixed point can therefore be understood as the limiting case of a continuous transition, in which the thermodynamic exponents approach the limits $\alpha \to 1$, $\beta \to 0$, $\gamma \to 1$, $\delta \to \infty$, $\nu \to 1/d$, and $\eta \to 2-d$.
We note that the values for $\alpha$, $\beta$, and $\delta$ are precisely those that should be expected in the presence of a finite latent heat at the first-order transition, and a jump in the magnetization as function of temperature and magnetization, respectively~\cite{fisher82}.

\section{Phase diagram}
\label{sec:phase-diagram}

\begin{figure*}[tb]
\includegraphics[width=\linewidth]{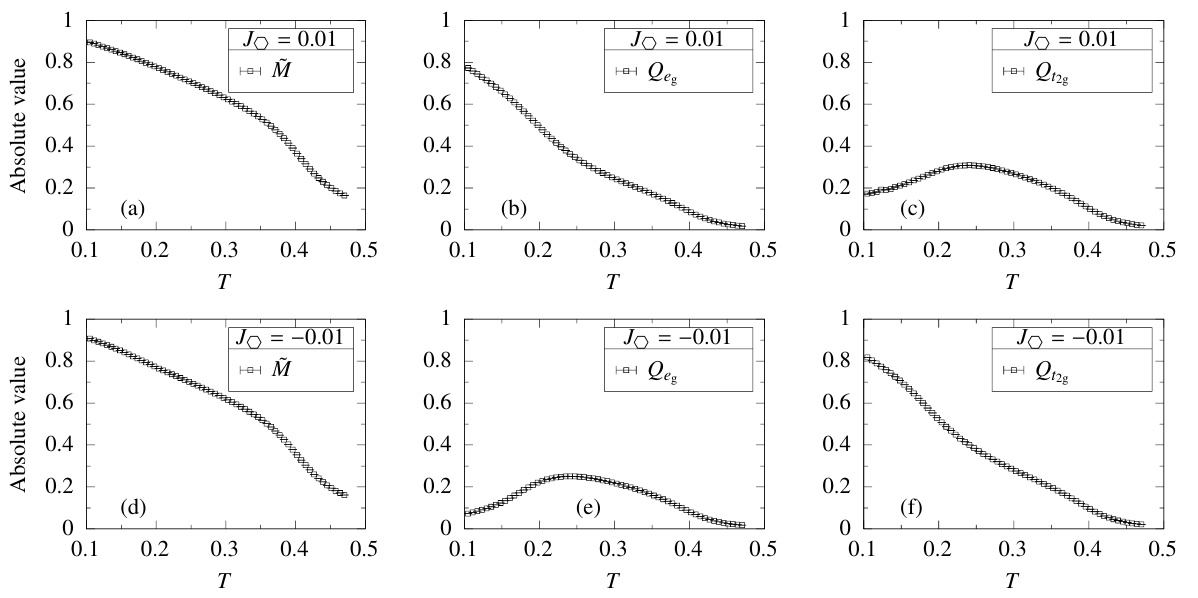}
\caption{%
(a)~Dual magnetization $\tilde{M}$ as function of temperature $T$ for $J_{\hexagon}=0.01 > 0$ and fixed lattice size $L=32$.
(b)~Same as (a), but for $\eg$ composite order parameter $Q_{\eg}$ indicating single-$\mathbf q$ zigzag order at low temperature, corresponding to a dual magnetization along the cubic axes directions.
(c)~Same as (a), but for $\tg$ composite order parameter $Q_{\tg}$, indicating the absence of triple-$\mathbf q$ order at low temperature.
(d)--(f)~Same as (a)--(c), but for $J_{\hexagon} = -0.01 < 0$, indicating triple-$\mathbf q$ order at low temperature, corresponding to a dual magnetization along the space diagonals in the cubic basis, and the absence of single-$\mathbf q$ zigzag order.}
\label{fig:order-parameters}
\end{figure*}

\subsection{Long-range-ordered phases}

For $J_{\hexagon} =0$, the model features a hidden SU(2) symmetry, which forbids long-range order at any finite temperature as a consequence of the Mermin-Wagner theorem~\cite{mermin66}.
The nature of the induced finite-temperature orders for $J_{\hexagon} \neq 0$ crucially depends on the sign of the ring exchange perturbation.
This is illustrated in Fig.~\ref{fig:order-parameters}, which shows the dual magnetization $\tilde M$ and the spin composite order parameters $Q_\eg$ and $Q_\tg$ as function of temperature $T$ for a fixed lattice size and representative values of $J_{\hexagon} > 0$ (top panels) and $J_{\hexagon} < 0$ (bottom panels), respectively.
In both cases, the finite dual magnetization indicates long-range magnetic order at low temperatures.

\begin{figure*}[tb]
\includegraphics[width=\linewidth]{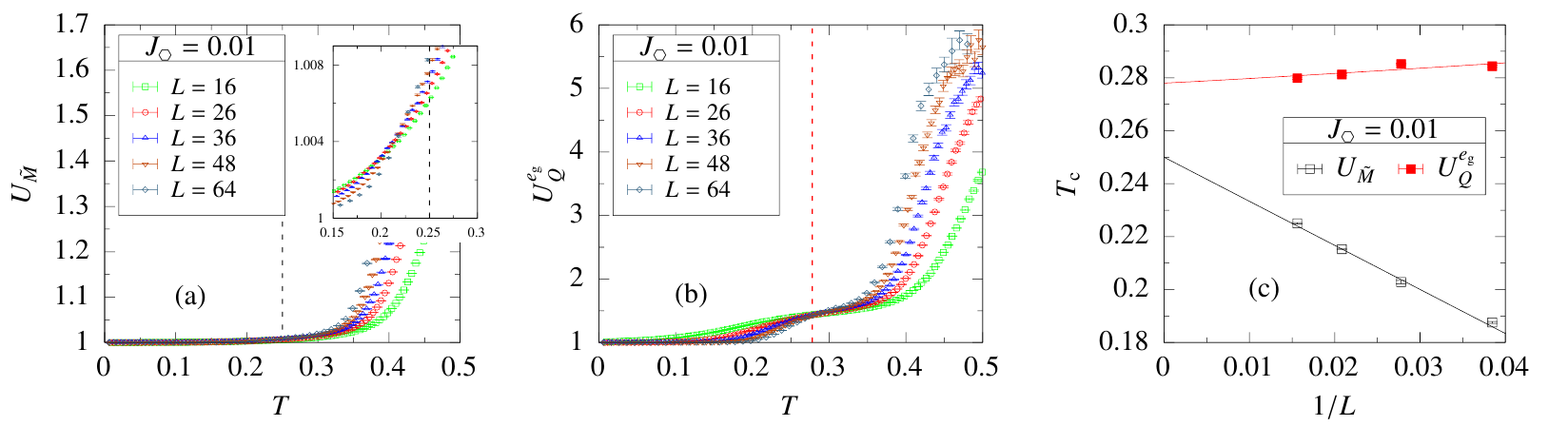}
\caption{%
(a)~Binder cumulant $U_{\tilde{M}}$ of dual magnetization as function of temperature $T$ for fixed $J_{\hexagon} = 0.01$ and different lattice sizes.
Inset shows vicinity of crossings. Vertical dashed line indicates the critical temperature in the thermodynamic limit.
(b)~Same as (a), but for Binder cumulant $U_{Q}^{\eg}$ of $\eg$ composite order parameter.
(c)~Critical temperatures $T_\mathrm{c}$ as a function of inverse system size $1/L$, obtained from the crossings of the Binder cumulant, indicating two different transitions in thermodynamic limit and the presence of an intermediate phase.}
\label{fig:critical-T}
\end{figure*}

For $J_{\hexagon} > 0$, the fact that $Q_\eg$ monotonically increases upon lowering the temperature, while $Q_\tg$ decreases after a broad hump, indicates that the favored direction of the dual magnetization is along the cubic axes $[100]$, $[010]$, and $[001]$.
This is also consistent with the Monte Carlo distributions of the dual magnetization shown in the lower right inset of Fig.~\ref{fig:pd}.
It demonstrates that the low-temperature state at  $J_{\hexagon} > 0$ features collinear single-$\mathbf q$ zigzag order with spin directions given as in Eq.~\eqref{eq:zigzag} or symmetry-related.

For $J_{\hexagon} < 0$, by contrast, $Q_\tg$ monotonically increases upon lowering the temperature, while $Q_\eg$ decreases after a broad hump. This shows that the favored direction of the dual magnetization in this case is along the cubic diagonals $[111]$, $[\bar 1 11]$, $[1\bar 1 1]$, and $[1 1 \bar 1]$, in agreement with the distribution shown in the lower left inset of Fig.~\ref{fig:pd}.
Consequently, the low-temperature state at $J_{\hexagon} < 0$ features noncollinear triple-$\mathbf q$ order with spin directions given as in Eq.~\eqref{eq:triple-q}.

The critical temperatures for $J_{\hexagon} > 0$ ($J_{\hexagon} < 0$) separating the different phases are extracted from the crossings of the Binder cumulants $U_{\tilde M}$ and $U_{Q}^\eg$ ($U_{Q}^\tg$) for the dual magnetization and the $\eg$ ($\tg$) composite order parameter, respectively, from consecutive system sizes.
An example for $J_{\hexagon} = 0.01$ is shown in Figs.~\ref{fig:critical-T}(a) and \ref{fig:critical-T}(b).
The resulting critical temperatures extracted from $U_{\tilde M}$ and $U_{Q}^\eg$, corresponding to the breaking of time reversal and $C_3^*$ symmetries, respectively, are shown as function of inverse system size $1/L$ in Fig.~\ref{fig:critical-T}(c).
Importantly, the critical temperatures $T_\mathrm{c1}$ and $T_\mathrm{c2}$ significantly deviate from each other, indicating the presence of an intermediate phase sandwiched between the low-temperature magnetically-ordered zigzag phase and the high-temperature disordered paramagnetic phase.
The intermediate phase at $T_\mathrm{c1} < T < T_\mathrm{c2}$ is characterized by a finite expectation value of the $\eg$ composite order parameter and a vanishing dual magnetization in the thermodynamic limit.
It realizes $\mathds{Z}_3$ spin nematic long-range order, as discussed in Sec.~\ref{subsec:nematic}.
An analogous behavior is found for $J_{\hexagon} < 0$ (not shown): There are two different critical temperatures $T_\mathrm{c1} < T_\mathrm{c2}$ extracted from the crossings of $U_{\tilde M}$ and $U_Q^\tg$, respectively. The intermediate phase at $T_\mathrm{c1} < T < T_\mathrm{c2}$ realizes $\mathds{Z}_4$ spin current density wave order, characterized by a finite expectation value of the $\tg$ composite order parameter and a vanishing dual magnetization in the thermodynamic limit.

For increasing $|J_{\hexagon}|$, the two critical temperatures approach each other and eventually merge at triple points located at $(J_{\hexagon}, T) \approx (-0.1, 0.28)$ and $(J_{\hexagon}, T) \approx (0.1, 0.33)$, respectively.
For $|J_{\hexagon}| > 0.1$, there is a direct transition between the low-temperature magnetically-ordered phase and the high-temperature disordered paramagnetic phase.
The resulting phase diagram is shown in Fig.~\ref{fig:pd}.

The finding of spin vestigial phases stabilized by thermal fluctuations in our numerical simulations of the Heisenberg-Kitaev-$\Gamma$-$\Gamma'$ model in the vicinity of the hidden-SU(2)-symmetric can be understood along the lines of previous renormalization group results obtained for 2D Heisenberg models with cubic anisotropy, which suggest a similar two-transition scenario~\cite{domany79, nagai85}.
Corresponding results have also been obtained numerically in a 2D Heisenberg model with cubic anisotropy~\cite{price13}.
This resonates with the fact that $\mathcal H_{\hexagon}$ can be understood as a cubic perturbation in the dual formulation, corresponding to the breaking of the hidden SU(2) to the hidden octahedral symmetry group $\Oh$.

\subsection{Hidden Mermin-Wagner physics}

\begin{figure*}[tb]
\includegraphics[width=\linewidth]{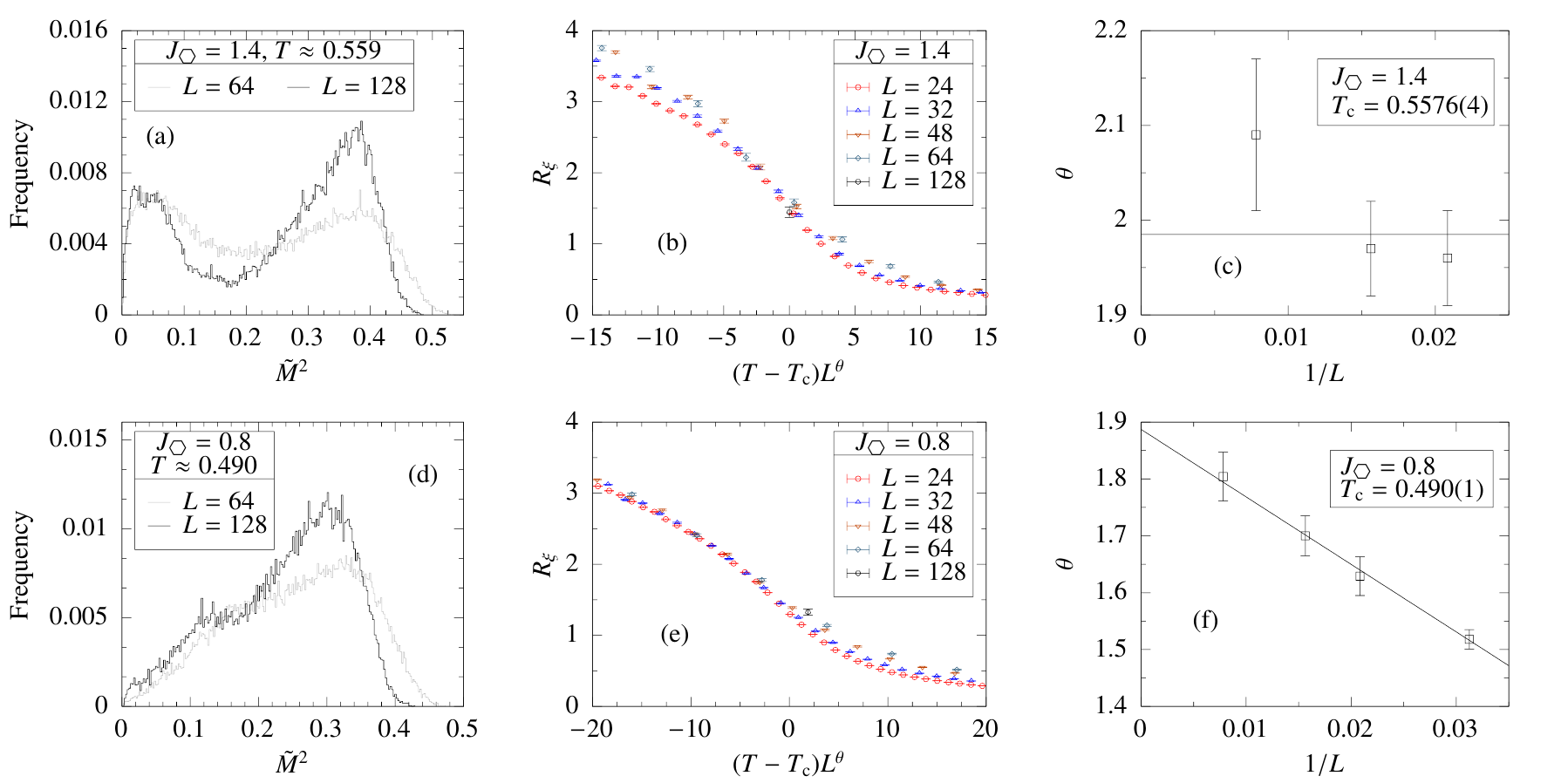}
\caption{%
(a)~Histogram of dual magnetization $\tilde{M}^2$ near the critical temperature $T_\mathrm{c} = 0.5576(4)$ for $J_{\hexagon} = 1.4$, indicating first-order behavior.
(b)~Finite-size scaling of correlation-length ratio $R_\xi = \xi/L$ as function of $(T-T_\mathrm{c1})L^{\theta}$ [Eq.~\eqref{eq:first-order-scaling}], using $\theta = 1.99$.
(c)~Exponent $\theta$ from finite-size scaling as function of system size $1/L$, indicating $\theta = 1.99(3)$, consistent with a first-order transition.
(d)--(f)~Same as (a)--(c), but near the critical temperature $T_\mathrm{c} = 0.490(1)$ for $J_{\hexagon} = 0.8$, showing that the first-order behavior weakens for decreasing $J_{\hexagon}$.
Nevertheless, the extrapolation to the thermodynamic limit shown in (f) yields $\theta=1.89(4)$ [this is also the value used in (e)], still consistent with first-order behavior.}
\label{fig:first-order-transition}
\end{figure*}

The presence of the hidden-SU(2)-symmetric point has a great influence on the shape of the phase diagram. From the absence of long-range-order at $J_{\hexagon} = 0$, we expect both $T_\mathrm{c1}$ and $T_\mathrm{c2}$ to vanish for decreasing $J_{\hexagon}$. However, the correlation length becomes exponentially large in the low-temperature limit~\cite{brezin76, kim94, alles99},
\begin{align}
\label{eq:J=0-correlation-length}
    \xi \propto \mathrm{e}^{-1/T},
\end{align}
for $J_{\hexagon} = 0$.
As a consequence, we expect the critical temperatures to vanish logarithmically slow near the hidden-SU(2)-symmetric point.
In fact, the crossover scaling theory, reviewed in the appendix, suggests the scaling~\cite{binder76}
\begin{align}
\label{eq:fit-shift-temperature}
    1/T_\mathrm{c}(J_{\hexagon}) = a + b \log{|1/J_{\hexagon}|},
\end{align}
for small $|J_{\hexagon}|$, with nonuniversal coefficients $a$ and $b$.
Fitting the above expectation to the data points for $T_{\mathrm{c1}}$ in the range of $0 < J_{\hexagon} < 1$ yields a remarkably good agreement for the fit parameters $a = 1.941(3)$ and $b = 0.463(4)$, see Fig.~\ref{fig:pd}.
We note that similar logarithmic behaviors of the critical temperatures are expected for $J_{\hexagon} < 0$. However, in this case, we refrain from fitting the data, as the phase boundary exhibits a local maximum at already a comparatively small value of $|J_{\hexagon}|$. This maximum in the critical temperature may be due to an instability of the triple-$\mathbf q$ state that arises at $J_{\hexagon} \approx -0.52$: Below this value of $J_{\hexagon}$, the ground state features a different magnetic long-range order, as can be verified by simple energy minimization of the classical Hamiltonian at $T=0$ (not shown). Elucidating the finite-temperature behavior in this regime is beyond the scope of the current work.

\section{Critical behaviors}
\label{sec:transitions}

In this section, we discuss the natures of the different finite-temperature transitions, lending further support on our result that the spin vestigial phases emerge in a finite intermediate-temperature window in the vicinity of the hidden-SU(2)-symmetric point.
We first focus on positive $J_{\hexagon} > 0$, which features a sizable regime in parameter space without any instability of the magnetic order at low temperature.
The behavior for $J_{\hexagon} < 0$, which turns out to be numerically more involved, will be discussed in light of the results for $J_{\hexagon} > 0$ below.

\subsection{Zigzag-to-paramagnet transition: First order}

We start by demonstrating that the direct transition at large $J_{\hexagon} > 0.1$ between the single-$\mathbf q$ zigzag order at low temperature and the paramagnetic phase at high temperature is of first-order nature, and governed by a discontinuity fixed point~\cite{nienhuis75, fisher82, binder87}.
Figure~\ref{fig:first-order-transition}(a) shows the histogram of the dual magnetization $\tilde M^2$ for a representative large value of $J_{\hexagon} = 1.4$ close to the corresponding transition temperature. The double-peak distribution clearly observed in the histogram demonstrates phase coexistence, implying a first-order transition.
A similar, although slightly less pronounced, double-peak distribution can be witnessed in the histogram of the energy density $\varepsilon$ (not shown).
Moreover, as depicted in Fig.~\ref{fig:first-order-transition}(b), the correlation-length ratio $R_{\xi}$ exhibits a finite-size scaling consistent with Eq.~\eqref{eq:first-order-scaling}. The corresponding exponent $\theta$, determined from the scaling collapse of consecutive system sizes, is shown as function of $1/L$ in Fig.~\ref{fig:first-order-transition}(c). Within numerical errors, the extracted value agrees with $\theta = d = 2$, which is the value expected for a first-order transition governed by a discontinuity fixed point~\cite{nienhuis75, fisher82, binder87}. 

For smaller value of $J_{\hexagon}$, but still above the triple point at $J_{\hexagon} \approx 0.1$, the first-order nature of the transition weakens significantly. 
Already for $J_{\hexagon} = 0.8$, the double-peak distribution in the histogram of the dual magnetization at the corresponding transition temperature is much less pronounced, see Fig.~\ref{fig:first-order-transition}(d).
In the energy histogram, a double-peak distribution is no longer visible for the current system sizes (not shown).
Furthermore, the exponent $\theta$ extracted from the finite-size collapse, shown in Fig.~\ref{fig:first-order-transition}(e), reveals significantly larger finite-size effects, see Fig.~\ref{fig:first-order-transition}(f).
Nevertheless, the value linearly extrapolated to the thermodynamic limit remains consistent with $\theta = d = 2$, reflecting the discontinuous nature of the transition.

The fact that the strength of the first-order transition significantly weakens upon decreasing $J_{\hexagon}$ indicates that the triple point at $J_{\hexagon} \approx 0.1$ may in fact be a bicritical point below which the, then two, finite-temperature transitions at $T_{\mathrm{c1}}$ and $T_\mathrm{c2}$ become continuous.
We now show that this is indeed the case.

\subsection{Zigzag-to-nematic transition: 2D Ising universality}

\begin{figure*}[tb]
\includegraphics[width=\linewidth]{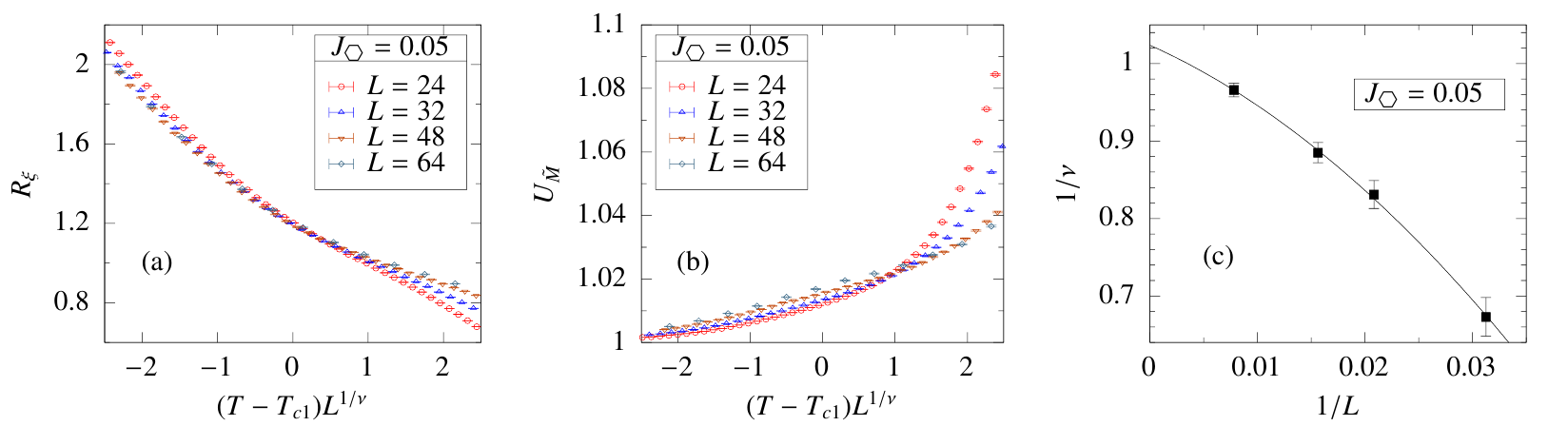}
\caption{%
(a)~Critical finite-size scaling of correlation-length ratio $R_\xi = \xi/L$ as function of $(T-T_\mathrm{c1})L^{1/\nu}$ for fixed $J_{\hexagon} = 0.05$ and different lattice sizes $L$, using $1/\nu = 1.02$ 
and $T_\mathrm{c1} = 0.302$, indicating a continuous finite-temperature transition.
(b)~Same as (a), but for the Binder cumulant $U_{\tilde{M}}$ of dual magnetization.
(c)~Finite-size scaling of correlation-length exponent $1/\nu$ as function of system size $1/L$, indicating $1/\nu = 1.02(5)$ in the thermodynamic limit, consistent with classical 2D Ising criticality.}
\label{fig:ising-transition}
\end{figure*}

We start with the lower transition at $T_\mathrm{c1}$, below which $\mathds Z_2$ time reversal symmetry is spontaneously broken.
As a consequence, we expect critical behavior in the 2D Ising universality class.
Figures~\ref{fig:ising-transition}(a) and \ref{fig:ising-transition}(b) show the scaling collapse for the correlation-length ratio $R_\xi$ and the Binder cumulant $U_{\tilde M}$, respectively, for a representative value of $J_{\hexagon} = 0.05$.
The quality of the scaling for small system sizes is mediocre, but improves for larger system sizes $L \geq 48$.
This indicates considerable corrections to scaling, in particular for $U_{\tilde M}$. These are likely due to the small value of $J_{\hexagon} \ll 1$, corresponding to approximate hidden SU(2) symmetry, as well as the presence of the second finite-temperature transition at $T_{\mathrm{c2}}$ near $T_{\mathrm{c1}}$.
The corrections to scaling are also visible in the drift of the corresponding critical exponent $1/\nu$ extracted from the scaling collapse of $R_\xi$ from consecutive system sizes, see Fig.~\ref{fig:ising-transition}(c).
Nevertheless, using a quadratic fitting ansatz to extrapolate to $1/L \to 0$, we obtain $1/\nu = 1.02(5)$ in the thermodynamic limit.
This agrees within errors with the expectation from 2D Ising universality, $1/\nu = 1$.

\subsection{Nematic-to-paramagnet transition: 2D Potts universality}

\begin{figure}[tb]
\includegraphics[width=\linewidth]{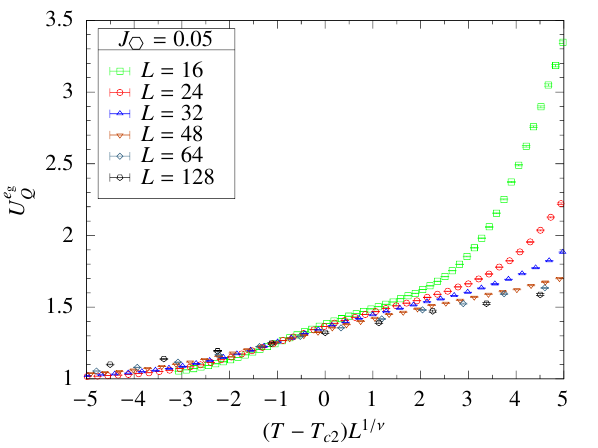}
\caption{%
Critical finite-size scaling of Binder cumulant $U_{Q}^{\eg}$ as function of $(T-T_\mathrm{c2})L^{1/\nu}$ for fixed $J_{\hexagon} = 0.05$ and different lattice sizes $L$, using $1/\nu = 6/5$ and $T_\mathrm{c2} = 0.310$, consistent with a continuous finite-temperature transition in the 2D three-state Potts universality class.
}
\label{fig:potts-transition}
\end{figure}

The transition at $T_\mathrm{c2}$ for $J_{\hexagon} > 0$ between the $\mathds{Z}_3$ spin nematic and the disordered paramagnet corresponds to a $C_3^*$-symmetry-breaking transition. If continuous, we therefore expect critical behavior in the three-state Potts universality class.
To confirm this expectation, we plot the Binder cumulant $U_Q^\eg$ corresponding to the $\eg$ composite order parameter as function of $(T-T_\mathrm{c2}) L^{1/\nu}$ for fixed representative value of $J_{\hexagon} = 0.05$ and different lattice sizes in Fig.~\ref{fig:potts-transition}.
The presence of large scaling corrections impedes an unbiased extraction of the correlation-length exponent in the present case, in contrast to the zigzag-to-nematic transition.
As a workaround, we assume the 2D three-state Potts exponent $1/\nu = 6/5$ from the outset and look for signs of convergence in the scaling collapse.
For the largest available system sizes $L=64$ and $L=128$, we indeed observe a moderate collapse of the curves at different system sizes.
We have verified that the scaling is worse if one assumes a first-order transition according to Eq.~\eqref{eq:first-order-scaling} with $\theta = d$.
This shows that the nematic-to-paramagnetic transition is continuous and hints that it falls into the three-state Potts universality class, as expected from universality arguments.
This result resonates with the real-space renormalization group analysis of the square-lattice Heisenberg model with face cubic anisotropy, which similarly suggests a continuous transition in the three-state Potts universality class at a temperature $T_{\mathrm{c2}}$, followed by an Ising transition at a lower temperature $T_{\mathrm{c1}} < T_{\mathrm{c2}}$~\cite{domany79}.

\subsection{Transitions for negative ring exchange}

As shown numerically above, the positive ring exchange term has, in terms of dual spins, the same effect on the Heisenberg-Kitaev-$\Gamma$-$\Gamma'$ model at the hidden SU(2) point as the face cubic anisotropy on the 2D Heisenberg model~\cite{domany79}. A similar analogy is therefore expected also for negative ring exchange $J_{\hexagon} < 0$.
The negative ring exchange corresponds, in terms of dual spins, to a corner cubic anisotropy, for which the general renormalization group analysis~\cite{domany79, nagai85} suggests a four-state Potts transition at a temperature $T_\mathrm{c2}$, followed by an Ising transition at a lower temperature $T_{\mathrm{c1}} < T_\mathrm{c2}$.
While this expectation is in principle consistent with our results for the phase diagram for $J_{\hexagon} < 0$, see Fig.~\ref{fig:pd}, a clear identification of the critical behavior at small negative $J_{\hexagon} \in (-0.1, 0)$ turns out to be difficult.
In particular, we do not observe any good finite-size-scaling collapse for any reasonable value of the exponent $1/\nu$. At the same time, there is no clear sign of a first-order transition, although very weak first-order behavior, visible only at lattice sizes beyond those currently reachable, is numerically impossible to exclude, as always.
The difficulty in the analysis may be attributed to three reasons:
First, the expected four-state Potts universality class is known to exhibit multiplicative logarithmic corrections~\cite{rebbi80, cardy80}, and the system sizes currently available for the model with ring exchange ($L \leq 128$) are likely insufficient to resolve these~\cite{salas97}.
Second, the proximity to the hidden SU(2) point leads to crossover effects, as shown in the appendix. 
Third, for the bilinear couplings $J$, $K$, $\Gamma$, and $\Gamma'$ at the hidden-SU(2)-symmetric point, Eq.~\eqref{eq:SU(2)-values}, the triple-$\mathbf{q}$ state becomes unstable for $J_{\hexagon} \lesssim -0.52$. This prevents the observation of a clear first-order transition between the triple-$\mathbf q$ state and the high-field paramagnet at large negative values of $J_{\hexagon}$.

\section{Application to \texorpdfstring{\NoCaseChange{\NCTO}}{Na2Co2TeO6}}
\label{sec:ncto}

The honeycomb magnet \NCTO\ has recently been argued to be well described by an effective spin Hamiltonian of the form of Eq.~\eqref{eq:full-model}, with bilinear exchange parameters proximate to the hidden-SU(2)-symmetric point [Eq.~\eqref{eq:SU(2)-values}], and small negative ring exchange perturbation $J_{\hexagon} < 0$~\cite{krueger22}.
In particular, it has been shown that the symmetry observed in the inelastic neutron scattering spectrum is inconsistent with a single-$\mathbf q$ zigzag magnetic ground state and can only be explained with a $C_3^*$-symmetric triple-$\mathbf q$ magnetic ordering characterized by eight spins in the hexagonal magnetic unit cell, as shown in Fig.~\ref{fig:duality}(c).
Careful magnetization and specific heat measurements exhibit a variety of anomalies as function of temperature within the long-range-ordered regime~\cite{lefrancois16, bera17, yao20, lin21, lee21}.
Of particular interest in light of our results is the fact that the single-crystal specific heat features a peak at a temperature $T_{\mathrm{c2}} = 30.97$~K \emph{above} the 3D antiferromagnetic ordering temperature at $T_\mathrm{c1} = 26.7$~K~\cite{chen21}.
The intermediate phase between $T_\mathrm{c1}$ and $T_\mathrm{c2}$ is characterized by a Bragg peak at the crystallographic $\mathbf M$ point in the 2D Brillouin zone and a rodlike shape of the neutron and X-ray diffraction pattern along the direction perpendicular to the 2D Brillouin zone~\cite{chen21}.
The Bragg peak structure reveals a 2D long-range order characterized by lattice translational symmetry breaking with an enlarged unit cell, wherein one or both of the 2D lattice vectors have doubled in magnitude.
Remarkably, the magnetic susceptibility curve as function of temperature does not exhibit any anomaly at $T_\mathrm{c2}$~\cite{yao20}. Furthermore, the spin-lattice and spin-spin relaxation rates measured in nuclear magnetic resonance (NMR) experiments show distinct peaks at $T_\mathrm{c1}$, but no pronounced anomalies at $T_\mathrm{c2}$~\cite{chen21, lee21}.
This might be indicative of a vestigial paramagnetic 2D long-range-ordered state between 26.7~K and 30.97~K.

A natural candidate to describe this finite-temperature intermediate phase in \NCTO\ is the $\mathds{Z}_4$ spin current density wave order discovered in this work in the vicinity of the hidden-SU(2)-symmetric point for small negative $J_{\hexagon} < 0$.
It is characterized by lattice translational symmetry breaking with an enlarged unit cell, wherein both of the 2D lattice vectors have doubled in magnitude, see Fig.~\ref{fig:duality}(e). This is consistent with the Bragg peak pattern observed in the diffraction experiments.
Moreover, as a paramagnetic state, it leaves time reversal intact, which would explain the absence of any pronounced anomaly in the susceptibility and NMR measurements.

To test this scenario, it would be desirable to elucidate the response of the $\mathds{Z}_4$ spin current density wave state to other perturbations that drive the system away from the hidden-SU(2)-symmetric point.
This includes nearest-neighbor interactions within the parameter space spanned by $J$, $K$, $\Gamma$, and $\Gamma'$, or further-neighbor interactions within the same honeycomb layer.
Of particular interest are interlayer interactions, which have previously been shown to be of significant relevance in other Kitaev magnets~\cite{janssen20, balz21}.
These might be expected to lead to a 2D-3D crossover in the critical behavior very close to the continuous finite-temperature transitions in and out of the spin vestigial phases.
A comprehensive theoretical characterization of this physics could help to understand the critical behavior seen in the X-ray diffraction of \NCTO\ for temperatures near and below $T_\mathrm{c2} = 30.97$~K.

\section{Conclusions}
\label{sec:conclusions}

In this work, we have studied the finite-temperature phase diagram of the honeycomb-lattice Heisenberg-Kitaev-$\Gamma$-$\Gamma'$ model near a hidden-SU(2)-symmetric point, including a six-spin ring exchange perturbation.
For positive (negative) ring exchange coupling $J_{\hexagon} > 0$ [$J_{\hexagon} < 0$], we find that the low-temperature ground state features collinear single-$\mathbf q$ zigzag (noncollinear triple-$\mathbf q$) magnetic long-range order.
These break time reversal and $C_3^*$ rotational (lattice translational) symmetries.
Upon increasing temperature, in the vicinity of the hidden-SU(2)-symmetric point, the magnetically-ordered phases melt in two stages.
In the intermediate vestigial phase, time reversal symmetry is restored, but the $C_3^*$ rotational (lattice translational) symmetry breaking of the corresponding primary single-$\mathbf q$ zigzag (triple-$\mathbf q$) phase remains.
We have explicitly confirmed, using an unbiased finite-size scaling analysis, that the continuous finite-temperature transition at $T_\mathrm{c1}$ between the magnetic order and its vestigial order is in the 2D Ising universality class.
The continuous transition between the $\mathds{Z}_3$ spin nematic vestigial order and the disordered paramagnet at the higher critical temperature $T_\mathrm{c2} > T_\mathrm{c1}$ falls into the three-state Potts universality class.
Similarly, for the transition between the $\mathds{Z}_4$ spin current density wave vestigial order and the disordered paramagnet, we expect a transition in the four-state Potts universality class.
For increasing ring exchange perturbation of either sign, the width of the intermediate phase, characterized by spin vestigial order, in the phase diagram shrinks, and eventually vanishes.
For large ring exchange, the transition becomes first order, signalled by a double-peak structure in the histograms of energy density and dual magnetization. The dimensionless correlation ratio exhibits finite-size scaling behavior with an exponent characteristic for a discontinuity fixed point~\cite{nienhuis75, fisher82, binder87}.

\begin{figure}[tb]
\includegraphics[width=\linewidth]{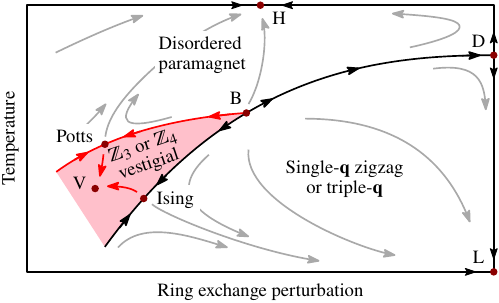}
\caption{%
Schematic renormalization group flow diagram emerging from our numerical results.
H and L denote the fully stable high-temperature and low-temperature fixed points, respectively.
The third fully stable fixed point V describes the corresponding spin vestigial phase, namely the $\mathds{Z}_3$ spin nematic ($\mathds{Z}_4$ spin current density wave) for positive (negative) ring exchange perturbation.
The Ising and three-state (four-state) Potts fixed points govern the continuous transitions in and out of the spin vestigial phase, respectively.
The direct first-order transition for sizable positive (negative) ring exchange perturbation between the magnetically-ordered single-$\mathbf q$ zigzag (triple-$\mathbf q$) phase and the disordered paramagnet is governed by the discontinuity fixed point D, which features an infrared relevant direction with associated eigenvalue of the stability matrix $\theta = d = 2$.
}
\label{fig:rgflow}
\end{figure}

A schematic renormalization group flow diagram that emerges from our numerical results is depicted in Fig.~\ref{fig:rgflow}.
Besides the usual high-temperature and low-temperature fixed points H and L, respectively, it features a third fully stable fixed point V, which describes the vestigial $\mathds{Z}_3$ spin nematic and $\mathds{Z}_4$ spin current density wave phases for positive and negative ring exchange, respectively.
The continuous finite-temperature transitions at $T_\mathrm{c1}$ and $T_\mathrm{c2}$ are governed by the critical Ising and Potts fixed points, respectively.
The first-order transition between the magnetically-ordered phase and the disordered paramagnet is governed by the discontinuity fixed point D. The discontinuity fixed point also features a single infrared relevant direction, however, with an associated eigenvalue of the stability matrix $\theta = d = 2$, consistent with a jump in the order parameter and a finite latent heat.
For the future, it might be interesting to map out this intricate flow diagram explicitly, e.g., by employing a Monte Carlo renormalization group approach~\cite{swendsen79, wellegehausen14}.

It is instructive to compare our results for the Heisenberg-Kitaev-$\Gamma$-$\Gamma'$ model near the hidden-SU(2)-symmetric point in the presence of a ring exchange perturbation with an earlier numerical study of the pure Heisenberg-Kitaev model with $\Gamma = \Gamma' = J_{\hexagon} = 0$~\cite{price12, price13}.
An intermediate phase at finite temperatures above the magnetic ordering temperature has been found in that case as well; however, the intermediate phase has been argued to feature algebraic order, while our spin vestigial phases are characterized by paramagnetic long-range order. The transition between the algebraically-ordered and the disordered paramagnet in the pure Heisenberg-Kitaev model is characterized by an essential singularity in the correlation length, characteristic of a Berezinskii–Kosterlitz–Thouless (BKT) transition~\cite{berezinskii71, kosterlitz73}.
As the hidden-SU(2)-symmetric point in the Heisenberg-Kitaev-$\Gamma$-$\Gamma'$ model is related to another hidden-SU(2)-symmetric point in the pure Heisenberg-Kitaev model~\cite{chaloupka15}, one might expect an analogous algebraically-ordered phase also in the Heisenberg-Kitaev-$\Gamma$-$\Gamma'$ model, once \emph{bilinear} perturbations that break the hidden SU(2) symmetry are taken into account.
This suggests an intricate interplay between the spin vestigial long-range orders and algebraically-ordered states in a model with both bilinear and nonbilinear exchange perturbations.

In this work, we have limited ourselves to the classical Heisenberg-Kitaev-$\Gamma$-$\Gamma'$ model with ring exchange perturbation.
In the quantum case, nonbilinear exchange perturbations can stabilize long-range ordered paramagnetic state, such as quantum spin nematics, even at zero temperature~\cite{shannon06, zhao12, pohle23}.
This indicates that the width of the spin vestigial phases in the finite-temperature phase diagram may increase upon the inclusion of quantum fluctuations. Studying the interplay of thermal and quantum fluctuations in the present model, within, for instance, a semiclassical expansion, clearly deserves future investigation.

Our numerical results may be of relevance for the Kitaev magnet \NCTO, which has recently been argued to realize a noncollinear triple-$\mathbf q$ ground state at low temperatures.
The $\mathds{Z}_4$ spin current density wave phase discovered in this work is a natural candidate to explain the 2D long-range-ordered state observed in \NCTO\ in a small temperature window \emph{above} the 3D antiferromagnetic ordering temperature $T_\mathrm{c1} = 26.7$~K.
However, in order to be able to confirm this scenario, more experimental and theoretical work is needed. On the theoretical side, in particular the effects of interlayer couplings should be included, which might give rise to a dimensional crossover in the critical behavior~\cite{svistov06, vojta18}. A realistic model of \NCTO\ should also explain the weak ferrimagnetism oberserved in high-quality single crystals, possibly arising from the two crystallographically inequivalent sublattices~\cite{yao20}.
These represent interesting directions for the future.

\paragraph*{Note added.} During the completion of this manuscript, we became aware of a parallel experimental work~\cite{sun23}, which reports the discovery of a $\mathds{Z}_3$ nematic phase as vestigial order of the zigzag antiferromagnet in few layers of the honeycomb magnet NiPS$_3$. Our theoretical results for this phase are consistent with the experimental observations in this material.
 
\begin{acknowledgments}

We thank Wilhelm Kr\"uger, Yuan Li, Rico Pohle, Urban Seifert, Matthias Vojta, and Manuel Weber for helpful discussions.
This work has been supported by the Deutsche Forschungsgemeinschaft (DFG) through SFB 1143 (A07, Project No.\ 247310070), the W\"urzburg-Dresden Cluster of Excellence \textit{ct.qmat} (EXC 2147, Project No.\ 390858490), and the Emmy Noether program (JA2306/4-1, Project No.\ 411750675). 
The authors are grateful to the Gemeinsame Wissenschaftskonferenz (GWK) for the generous provision of computing resources through the Center for Information Services and High Performance Computing (ZIH) at TU Dresden (Project No.\ 21599)~\cite{nhr-alliance}.

\end{acknowledgments}

\setcounter{equation}{0}
\renewcommand\theequation{A\arabic{equation}}

\section*{Appendix: Crossover behavior near hidden-SU(2)-symmetric point}

\begin{figure*}[tb]
\includegraphics[width=\linewidth]{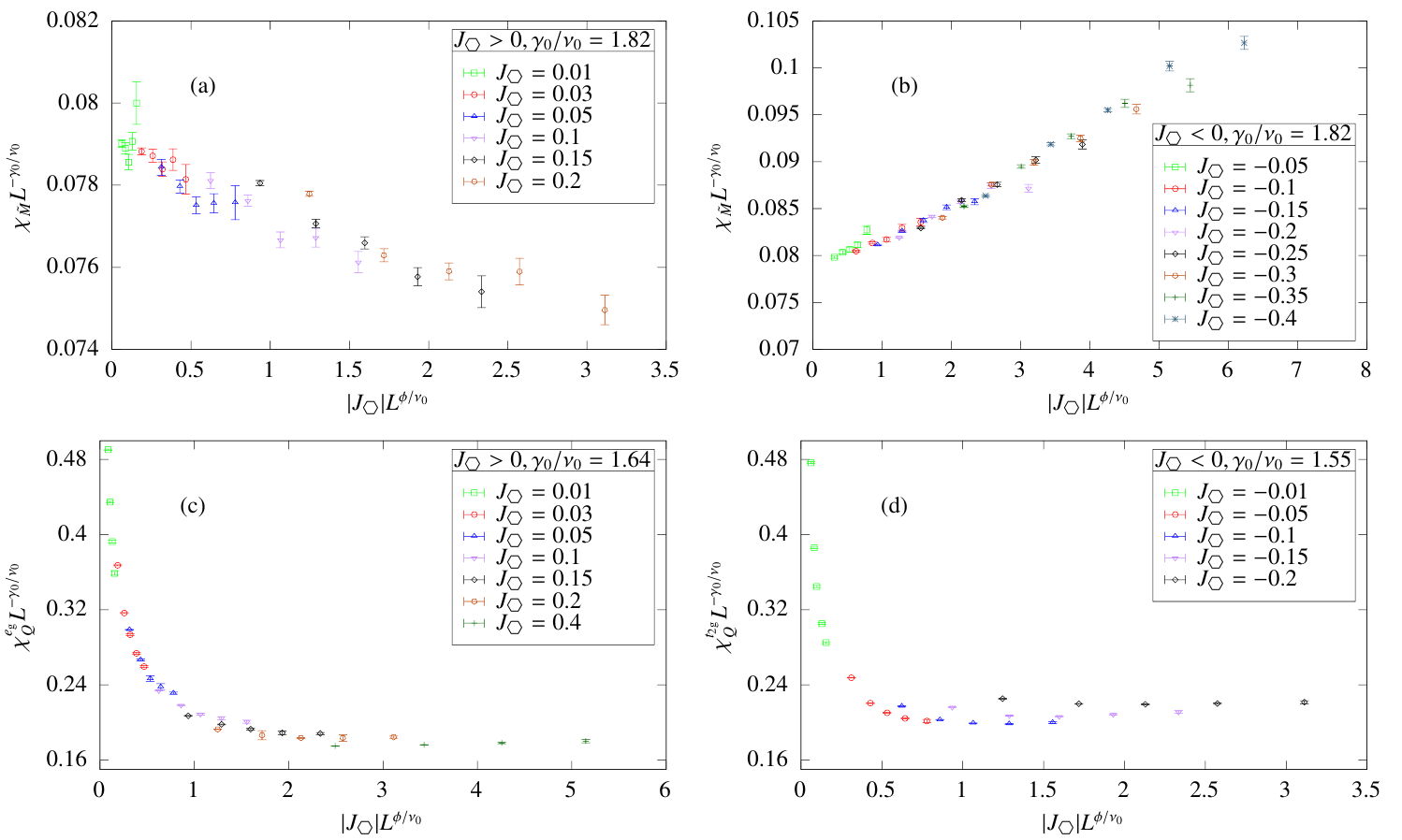}
\caption{%
(a)~Finite-size crossover scaling of maxima of magnetic susceptibility $\chi_{\tilde{M}} L^{-{\gamma}_0/\nu_0}$ of dual magnetization as function of $|J_{\hexagon}| L^{\phi/\nu_0}$, using ${\gamma}_0/\nu_0 = 1.82$, $\phi/\nu_0 = 2/3$, and $J_{\hexagon}>0$.
(b)~Same as (a), but for $J_{\hexagon}<0$. 
(c)~Same as (a), but for the susceptibility of the $\eg$ composite order parameter, using ${\gamma}_0/\nu_0 = 1.64$.
(d)~Same as (b), but for the susceptibility of the $\tg$ composite order parameter, using ${\gamma}_0/\nu_0 = 1.55$.}
\label{fig:crossover}
\end{figure*}

In the appendix, we review the crossover theory for critical points in the presence of small anisotropies. We will start with the standard scenario for critical points with power-law divergences, and then generalize to cases with exponential divergences, relevant for the hidden-SU(2)-symmetric point.

\paragraph{Crossover scaling theory for standard critical points in presence of small anisotropies.}
For a critical point with standard power-law critical behavior, the correlation length $\xi$ and the susceptibility $\chi$ scale with the reduced temperature $t=(T-T_\mathrm{c})/T_\mathrm{c}$ as
\begin{align}
    \label{eq:power-law-div}
    \xi & \propto t^{-\nu_0},
    & 
    \chi & \propto t^{-\gamma_0},
\end{align}
where $\nu_0$ and $\gamma_0$ are the corresponding universal critical exponents characterizing the isotropic fixed point.
Now, turning on an anisotropy term with coupling $J_{\hexagon}$, standard crossover theory~\cite{fisher74} suggests that the divergence changes as 
\begin{equation}
    \label{eq:standard-crossover}
    \chi \propto t^{-\gamma_0} \Phi(J_{\hexagon} t^{-\phi}),
\end{equation}
where $\Phi$ is a universal crossover scaling function and $\phi$ is a universal crossover exponent, with $\phi/\nu_0$ given by the scaling dimension of the anisotropy term at the isotropic fixed point.
Here, the reduced temperature $t$ is defined with respect to the critical temperature in the isotropic case, i.e., $t=[T-T_\mathrm{c}(J_{\hexagon}=0)]/T_\mathrm{c}(J_{\hexagon}=0)$. The crossover scaling function $\Phi(z)$ is normalized to $\Phi(0) = 1$ and diverges at a finite $z = z_\mathrm{c}$, which corresponds to the transition in the presence of the anisotropy.

\paragraph{Crossover scaling theory for cases with exponentially-divergent length scales.}

Let us now generalize the crossover theory for cases with exponentially-divergent length scales, as in the Heisenberg-Kitaev-$\Gamma$-$\Gamma'$ model at the hidden-SU(2)-symmetric point, see Eq.~\eqref{eq:J=0-correlation-length}.
For this, it is useful to rewrite Eqs.~\eqref{eq:power-law-div} and \eqref{eq:standard-crossover} as
\begin{equation} \label{eq:general-crossover}
\chi \propto \xi^{\gamma_0/\nu_0}\Phi(J_{\hexagon} \xi^{\phi/\nu_0}),
\end{equation}
which is expected to hold also in cases with exponentially-divergent length scales~\cite{binder76}. We note that while both $\nu_0$ and $\gamma_0$ are formally expected to diverge in the 2D Heisenberg case~\cite{kim94, alles99}, their ratio can be finite, as is the case in the BKT transition~\cite{kosterlitz74}. As the scaling dimension of the cubic anisotropy, which corresponds to the dual version of the ring exchange term in Eq.~\eqref{eq:ring-term}, is expected to be finite~\cite{calabrese02}, the crossover exponent $\phi$ will formally diverge at the hidden-SU(2)-point as well, in a way that $\phi/\nu_0$ remains finite.
From this, we obtain that the critical temperature $T_\mathrm{c}(J_{\hexagon})$ vanishes logarithmically with $J_{\hexagon}$, see Eq.~\eqref{eq:fit-shift-temperature}, which is also consistent with our numerical findings, see the black curve in Fig.~\ref{fig:pd}.

\paragraph{Finite-size crossover scaling theory.}

On finite lattices of linear size $L$, the correlation length is expected to scale at criticality as $\xi \propto L$.
From this, we obtain the finite-size crossover scaling law~\cite{binder92}
\begin{equation}
    \label{eq:crossover-FSS}
    \chi \propto L^{{\gamma_0}/\nu_0} \Phi(J_{\hexagon} L^{\phi/\nu_0}).
\end{equation}
Here, the scaling variable $J_{\hexagon} L^{{\phi/\nu_0}}$ can be understood to measure the strength of the crossover effects. For small $J_{\hexagon} L^{{\phi/\nu_0}} \ll 1$, the critical behavior will be strongly affected by crossover effects arising from the vicinity of the hidden-SU(2)-symmetric point, while for $J_{\hexagon} L^{{\phi/\nu_0}} \gg 1$ the critical behavior emerging from the anisotropic ring exchange term should be correctly represented on the finite-size lattices.

In our simulations at $J_{\hexagon} = 0$, we find that the susceptibilities of the composite order parameters $\chi_{Q}^{\eg}$ and $\chi_{Q}^{\tg}$ increase monotonically with decreasing $T$, as expected.
The low-temperature divergence for increasing system sizes allows us to obtain a rough estimate of the ratio $\gamma_0/\nu_0$ in Eq.~\eqref{eq:crossover-FSS}, upon setting $J_{\hexagon} = 0$ therein.
By contrast, for $\chi_{\tilde{M}}$, we observe a finite-temperature maximum arising from the pseudocritical behavior of the 2D Heisenberg model~\cite{tomita14, burgelman23}.
For the finite-size scaling analysis at $J_{\hexagon} = 0$, we therefore take this maximum as reference for the divergence of $\chi_{\tilde{M}}$.
This procedure gives a rough estimate for the ratio ${\gamma_{0}/\nu_0}$, but does not allow a precise evaluation of the exponent. Nevertheless, it is reassuring that the values, which we have obtained from Bayesian-inference-based~\cite{harada11} scaling collapses of the three different observables, turn out in the same ballpark, see Tab.~\ref{tab:gamma-0}.
If one assumes that the hidden-SU(2)-symmetric point is described at low temperature by the zero-temperature fixed point of the O(3) nonlinear sigma model, the theoretical expectation is $\gamma_0/\nu_0 = 2$~\cite{brezin76}, which is roughly consistent with our results.

\begin{table}[tb]
    \caption{Estimates of ratio ${\gamma}_0/\nu_0$ from simulations at $J_{\hexagon}=0$, obtained from finite-size scaling of susceptibilities of dual magnetization $\tilde M$ and composite order parameters $Q_\eg$ and $Q_\tg$.}
    \label{tab:gamma-0}
    \begin{tabular*}{\linewidth}{@{\extracolsep{\fill} } c c c c}
    \hline
    \hline
         & $\chi_{\tilde{M}}$ & $\chi_{Q}^{\eg}$ & $\chi_{Q}^{\tg}$ \\
         \hline
         ${\gamma}_0/\nu_0$ & 1.82(1) & $\sim 1.64$ & $\sim 1.55$ \\ 
    \hline
    \hline
    \end{tabular*}
\end{table}

Taking the maxima of the susceptibility curves for fixed $J_{\hexagon}\neq 0$ and fixed lattice size $L$ as function of temperature $T$, we test the finite-size crossover scaling hypothesis in Eq.~\eqref{eq:crossover-FSS} for several values of the ratio ${\phi}/\nu_0$, using the values for $\gamma_0/\nu_0$ as obtained from the analysis at $J_{\hexagon} = 0$. A decent scaling on multiple observables is obtained for ${\phi}/\nu_0\approx 2/3$, see Fig.~\ref{fig:crossover}. Note that the range of $|J_{\hexagon}|L^{{\phi/\nu_0}}$, for which the scaling is reasonable, depends on the observable. 
Overall, the results agree with the expectation from the crossover scaling hypothesis, indicating the strong influence of the hidden-SU(2)-symmetric point on the behavior of the system at small~$|J_{\hexagon}|$.

\bibliographystyle{longapsrev4-2}
\bibliography{HK-finite-T}

\begin{thebibliography}{106}%
\makeatletter
\providecommand \@ifxundefined [1]{%
 \@ifx{#1\undefined}
}%
\providecommand \@ifnum [1]{%
 \ifnum #1\expandafter \@firstoftwo
 \else \expandafter \@secondoftwo
 \fi
}%
\providecommand \@ifx [1]{%
 \ifx #1\expandafter \@firstoftwo
 \else \expandafter \@secondoftwo
 \fi
}%
\providecommand \natexlab [1]{#1}%
\providecommand \enquote  [1]{``#1''}%
\providecommand \bibnamefont  [1]{#1}%
\providecommand \bibfnamefont [1]{#1}%
\providecommand \citenamefont [1]{#1}%
\providecommand \href@noop [0]{\@secondoftwo}%
\providecommand \href [0]{\begingroup \@sanitize@url \@href}%
\providecommand \@href[1]{\@@startlink{#1}\@@href}%
\providecommand \@@href[1]{\endgroup#1\@@endlink}%
\providecommand \@sanitize@url [0]{\catcode `\\12\catcode `\$12\catcode
  `\&12\catcode `\#12\catcode `\^12\catcode `\_12\catcode `\%12\relax}%
\providecommand \@@startlink[1]{}%
\providecommand \@@endlink[0]{}%
\providecommand \url  [0]{\begingroup\@sanitize@url \@url }%
\providecommand \@url [1]{\endgroup\@href {#1}{\urlprefix }}%
\providecommand \urlprefix  [0]{URL }%
\providecommand \Eprint [0]{\href }%
\providecommand \doibase [0]{https://doi.org/}%
\providecommand \selectlanguage [0]{\@gobble}%
\providecommand \bibinfo  [0]{\@secondoftwo}%
\providecommand \bibfield  [0]{\@secondoftwo}%
\providecommand \translation [1]{[#1]}%
\providecommand \BibitemOpen [0]{}%
\providecommand \bibitemStop [0]{}%
\providecommand \bibitemNoStop [0]{.\EOS\space}%
\providecommand \EOS [0]{\spacefactor3000\relax}%
\providecommand \BibitemShut  [1]{\csname bibitem#1\endcsname}%
\let\auto@bib@innerbib\@empty
\bibitem [{\citenamefont {Kitaev}(2006)}]{kitaev06}%
  \BibitemOpen
  \bibfield  {author} {\bibinfo {author} {\bibfnamefont {A.}~\bibnamefont
  {Kitaev}},\ }\bibfield  {title} {\bibinfo {title} {Anyons in an exactly
  solved model and beyond},\ }\href
  {https://doi.org/https://doi.org/10.1016/j.aop.2005.10.005} {\bibfield
  {journal} {\bibinfo  {journal} {Ann. Phys. (N. Y.)}\ }\textbf {\bibinfo
  {volume} {321}},\ \bibinfo {pages} {2} (\bibinfo {year} {2006})}\BibitemShut
  {NoStop}%
\bibitem [{\citenamefont {Jackeli}\ and\ \citenamefont
  {Khaliullin}(2009)}]{jackeli09}%
  \BibitemOpen
  \bibfield  {author} {\bibinfo {author} {\bibfnamefont {G.}~\bibnamefont
  {Jackeli}}\ and\ \bibinfo {author} {\bibfnamefont {G.}~\bibnamefont
  {Khaliullin}},\ }\bibfield  {title} {\bibinfo {title} {Mott Insulators in the
  Strong Spin-Orbit Coupling Limit: From Heisenberg to a Quantum Compass and
  Kitaev Models},\ }\href {https://doi.org/10.1103/PhysRevLett.102.017205}
  {\bibfield  {journal} {\bibinfo  {journal} {Phys. Rev. Lett.}\ }\textbf
  {\bibinfo {volume} {102}},\ \bibinfo {pages} {017205} (\bibinfo {year}
  {2009})}\BibitemShut {NoStop}%
\bibitem [{\citenamefont {Winter}\ \emph {et~al.}(2016)\citenamefont {Winter},
  \citenamefont {Li}, \citenamefont {Jeschke},\ and\ \citenamefont
  {Valent\'{\i}}}]{winter16}%
  \BibitemOpen
  \bibfield  {author} {\bibinfo {author} {\bibfnamefont {S.~M.}\ \bibnamefont
  {Winter}}, \bibinfo {author} {\bibfnamefont {Y.}~\bibnamefont {Li}}, \bibinfo
  {author} {\bibfnamefont {H.~O.}\ \bibnamefont {Jeschke}},\ and\ \bibinfo
  {author} {\bibfnamefont {R.}~\bibnamefont {Valent\'{\i}}},\ }\bibfield
  {title} {\bibinfo {title} {Challenges in design of Kitaev materials: Magnetic
  interactions from competing energy scales},\ }\href
  {https://doi.org/10.1103/PhysRevB.93.214431} {\bibfield  {journal} {\bibinfo
  {journal} {Phys. Rev. B}\ }\textbf {\bibinfo {volume} {93}},\ \bibinfo
  {pages} {214431} (\bibinfo {year} {2016})}\BibitemShut {NoStop}%
\bibitem [{\citenamefont {Liu}\ and\ \citenamefont {Khaliullin}(2018)}]{liu18}%
  \BibitemOpen
  \bibfield  {author} {\bibinfo {author} {\bibfnamefont {H.}~\bibnamefont
  {Liu}}\ and\ \bibinfo {author} {\bibfnamefont {G.}~\bibnamefont
  {Khaliullin}},\ }\bibfield  {title} {\bibinfo {title} {Pseudospin exchange
  interactions in ${d}^{7}$ cobalt compounds: Possible realization of the
  Kitaev model},\ }\href {https://doi.org/10.1103/PhysRevB.97.014407}
  {\bibfield  {journal} {\bibinfo  {journal} {Phys. Rev. B}\ }\textbf {\bibinfo
  {volume} {97}},\ \bibinfo {pages} {014407} (\bibinfo {year}
  {2018})}\BibitemShut {NoStop}%
\bibitem [{\citenamefont {Sano}\ \emph {et~al.}(2018)\citenamefont {Sano},
  \citenamefont {Kato},\ and\ \citenamefont {Motome}}]{sano18}%
  \BibitemOpen
  \bibfield  {author} {\bibinfo {author} {\bibfnamefont {R.}~\bibnamefont
  {Sano}}, \bibinfo {author} {\bibfnamefont {Y.}~\bibnamefont {Kato}},\ and\
  \bibinfo {author} {\bibfnamefont {Y.}~\bibnamefont {Motome}},\ }\bibfield
  {title} {\bibinfo {title} {Kitaev-Heisenberg Hamiltonian for high-spin
  ${d}^{7}$ Mott insulators},\ }\href
  {https://doi.org/10.1103/PhysRevB.97.014408} {\bibfield  {journal} {\bibinfo
  {journal} {Phys. Rev. B}\ }\textbf {\bibinfo {volume} {97}},\ \bibinfo
  {pages} {014408} (\bibinfo {year} {2018})}\BibitemShut {NoStop}%
\bibitem [{\citenamefont {Liu}\ \emph {et~al.}(2020)\citenamefont {Liu},
  \citenamefont {Chaloupka},\ and\ \citenamefont {Khaliullin}}]{liu20}%
  \BibitemOpen
  \bibfield  {author} {\bibinfo {author} {\bibfnamefont {H.}~\bibnamefont
  {Liu}}, \bibinfo {author} {\bibfnamefont {J.}~\bibnamefont {Chaloupka}},\
  and\ \bibinfo {author} {\bibfnamefont {G.}~\bibnamefont {Khaliullin}},\
  }\bibfield  {title} {\bibinfo {title} {Kitaev Spin Liquid in $3d$ Transition
  Metal Compounds},\ }\href {https://doi.org/10.1103/PhysRevLett.125.047201}
  {\bibfield  {journal} {\bibinfo  {journal} {Phys. Rev. Lett.}\ }\textbf
  {\bibinfo {volume} {125}},\ \bibinfo {pages} {047201} (\bibinfo {year}
  {2020})}\BibitemShut {NoStop}%
\bibitem [{\citenamefont {Winter}(2022)}]{winter22}%
  \BibitemOpen
  \bibfield  {author} {\bibinfo {author} {\bibfnamefont {S.~M.}\ \bibnamefont
  {Winter}},\ }\bibfield  {title} {\bibinfo {title} {Magnetic couplings in
  edge-sharing high-spin $d^7$ compounds},\ }\href
  {https://doi.org/10.1088/2515-7639/ac94f8} {\bibfield  {journal} {\bibinfo
  {journal} {J. Phys. Mat.}\ }\textbf {\bibinfo {volume} {5}},\ \bibinfo
  {pages} {045003} (\bibinfo {year} {2022})}\BibitemShut {NoStop}%
\bibitem [{\citenamefont {Choi}\ \emph {et~al.}(2012)\citenamefont {Choi},
  \citenamefont {Coldea}, \citenamefont {Kolmogorov}, \citenamefont
  {Lancaster}, \citenamefont {Mazin}, \citenamefont {Blundell}, \citenamefont
  {Radaelli}, \citenamefont {Singh}, \citenamefont {Gegenwart}, \citenamefont
  {Choi}, \citenamefont {Cheong}, \citenamefont {Baker}, \citenamefont
  {Stock},\ and\ \citenamefont {Taylor}}]{choi12}%
  \BibitemOpen
  \bibfield  {author} {\bibinfo {author} {\bibfnamefont {S.~K.}\ \bibnamefont
  {Choi}}, \bibinfo {author} {\bibfnamefont {R.}~\bibnamefont {Coldea}},
  \bibinfo {author} {\bibfnamefont {A.~N.}\ \bibnamefont {Kolmogorov}},
  \bibinfo {author} {\bibfnamefont {T.}~\bibnamefont {Lancaster}}, \bibinfo
  {author} {\bibfnamefont {I.~I.}\ \bibnamefont {Mazin}}, \bibinfo {author}
  {\bibfnamefont {S.~J.}\ \bibnamefont {Blundell}}, \bibinfo {author}
  {\bibfnamefont {P.~G.}\ \bibnamefont {Radaelli}}, \bibinfo {author}
  {\bibfnamefont {Y.}~\bibnamefont {Singh}}, \bibinfo {author} {\bibfnamefont
  {P.}~\bibnamefont {Gegenwart}}, \bibinfo {author} {\bibfnamefont {K.~R.}\
  \bibnamefont {Choi}}, \bibinfo {author} {\bibfnamefont {S.-W.}\ \bibnamefont
  {Cheong}}, \bibinfo {author} {\bibfnamefont {P.~J.}\ \bibnamefont {Baker}},
  \bibinfo {author} {\bibfnamefont {C.}~\bibnamefont {Stock}},\ and\ \bibinfo
  {author} {\bibfnamefont {J.}~\bibnamefont {Taylor}},\ }\bibfield  {title}
  {\bibinfo {title} {Spin Waves and Revised Crystal Structure of Honeycomb
  Iridate ${\mathrm{Na}}_{2}{\mathrm{IrO}}_{3}$},\ }\href
  {https://doi.org/10.1103/PhysRevLett.108.127204} {\bibfield  {journal}
  {\bibinfo  {journal} {Phys. Rev. Lett.}\ }\textbf {\bibinfo {volume} {108}},\
  \bibinfo {pages} {127204} (\bibinfo {year} {2012})}\BibitemShut {NoStop}%
\bibitem [{\citenamefont {Williams}\ \emph {et~al.}(2016)\citenamefont
  {Williams}, \citenamefont {Johnson}, \citenamefont {Freund}, \citenamefont
  {Choi}, \citenamefont {Jesche}, \citenamefont {Kimchi}, \citenamefont
  {Manni}, \citenamefont {Bombardi}, \citenamefont {Manuel}, \citenamefont
  {Gegenwart},\ and\ \citenamefont {Coldea}}]{williams16}%
  \BibitemOpen
  \bibfield  {author} {\bibinfo {author} {\bibfnamefont {S.~C.}\ \bibnamefont
  {Williams}}, \bibinfo {author} {\bibfnamefont {R.~D.}\ \bibnamefont
  {Johnson}}, \bibinfo {author} {\bibfnamefont {F.}~\bibnamefont {Freund}},
  \bibinfo {author} {\bibfnamefont {S.}~\bibnamefont {Choi}}, \bibinfo {author}
  {\bibfnamefont {A.}~\bibnamefont {Jesche}}, \bibinfo {author} {\bibfnamefont
  {I.}~\bibnamefont {Kimchi}}, \bibinfo {author} {\bibfnamefont
  {S.}~\bibnamefont {Manni}}, \bibinfo {author} {\bibfnamefont
  {A.}~\bibnamefont {Bombardi}}, \bibinfo {author} {\bibfnamefont
  {P.}~\bibnamefont {Manuel}}, \bibinfo {author} {\bibfnamefont
  {P.}~\bibnamefont {Gegenwart}},\ and\ \bibinfo {author} {\bibfnamefont
  {R.}~\bibnamefont {Coldea}},\ }\bibfield  {title} {\bibinfo {title}
  {Incommensurate counterrotating magnetic order stabilized by Kitaev
  interactions in the layered honeycomb
  $\ensuremath{\alpha}\text{{-}}{\mathrm{Li}}_{2}{\mathrm{IrO}}_{3}$},\ }\href
  {https://doi.org/10.1103/PhysRevB.93.195158} {\bibfield  {journal} {\bibinfo
  {journal} {Phys. Rev. B}\ }\textbf {\bibinfo {volume} {93}},\ \bibinfo
  {pages} {195158} (\bibinfo {year} {2016})}\BibitemShut {NoStop}%
\bibitem [{\citenamefont {Johnson}\ \emph {et~al.}(2015)\citenamefont
  {Johnson}, \citenamefont {Williams}, \citenamefont {Haghighirad},
  \citenamefont {Singleton}, \citenamefont {Zapf}, \citenamefont {Manuel},
  \citenamefont {Mazin}, \citenamefont {Li}, \citenamefont {Jeschke},
  \citenamefont {Valent\'{\i}},\ and\ \citenamefont {Coldea}}]{johnson15}%
  \BibitemOpen
  \bibfield  {author} {\bibinfo {author} {\bibfnamefont {R.~D.}\ \bibnamefont
  {Johnson}}, \bibinfo {author} {\bibfnamefont {S.~C.}\ \bibnamefont
  {Williams}}, \bibinfo {author} {\bibfnamefont {A.~A.}\ \bibnamefont
  {Haghighirad}}, \bibinfo {author} {\bibfnamefont {J.}~\bibnamefont
  {Singleton}}, \bibinfo {author} {\bibfnamefont {V.}~\bibnamefont {Zapf}},
  \bibinfo {author} {\bibfnamefont {P.}~\bibnamefont {Manuel}}, \bibinfo
  {author} {\bibfnamefont {I.~I.}\ \bibnamefont {Mazin}}, \bibinfo {author}
  {\bibfnamefont {Y.}~\bibnamefont {Li}}, \bibinfo {author} {\bibfnamefont
  {H.~O.}\ \bibnamefont {Jeschke}}, \bibinfo {author} {\bibfnamefont
  {R.}~\bibnamefont {Valent\'{\i}}},\ and\ \bibinfo {author} {\bibfnamefont
  {R.}~\bibnamefont {Coldea}},\ }\bibfield  {title} {\bibinfo {title}
  {Monoclinic crystal structure of $\alpha$-RuCl$_3$ and the zigzag
  antiferromagnetic ground state},\ }\href
  {https://doi.org/10.1103/PhysRevB.92.235119} {\bibfield  {journal} {\bibinfo
  {journal} {Phys. Rev. B}\ }\textbf {\bibinfo {volume} {92}},\ \bibinfo
  {pages} {235119} (\bibinfo {year} {2015})}\BibitemShut {NoStop}%
\bibitem [{\citenamefont {Sears}\ \emph {et~al.}(2015)\citenamefont {Sears},
  \citenamefont {Songvilay}, \citenamefont {Plumb}, \citenamefont {Clancy},
  \citenamefont {Qiu}, \citenamefont {Zhao}, \citenamefont {Parshall},\ and\
  \citenamefont {Kim}}]{sears15}%
  \BibitemOpen
  \bibfield  {author} {\bibinfo {author} {\bibfnamefont {J.~A.}\ \bibnamefont
  {Sears}}, \bibinfo {author} {\bibfnamefont {M.}~\bibnamefont {Songvilay}},
  \bibinfo {author} {\bibfnamefont {K.~W.}\ \bibnamefont {Plumb}}, \bibinfo
  {author} {\bibfnamefont {J.~P.}\ \bibnamefont {Clancy}}, \bibinfo {author}
  {\bibfnamefont {Y.}~\bibnamefont {Qiu}}, \bibinfo {author} {\bibfnamefont
  {Y.}~\bibnamefont {Zhao}}, \bibinfo {author} {\bibfnamefont {D.}~\bibnamefont
  {Parshall}},\ and\ \bibinfo {author} {\bibfnamefont {Y.-J.}\ \bibnamefont
  {Kim}},\ }\bibfield  {title} {\bibinfo {title} {Magnetic order in
  $\alpha$-RuCl$_3$: A honeycomb-lattice quantum magnet with strong spin-orbit
  coupling},\ }\href {https://doi.org/10.1103/PhysRevB.91.144420} {\bibfield
  {journal} {\bibinfo  {journal} {Phys. Rev. B}\ }\textbf {\bibinfo {volume}
  {91}},\ \bibinfo {pages} {144420} (\bibinfo {year} {2015})}\BibitemShut
  {NoStop}%
\bibitem [{\citenamefont {Cao}\ \emph {et~al.}(2016)\citenamefont {Cao},
  \citenamefont {Banerjee}, \citenamefont {Yan}, \citenamefont {Bridges},
  \citenamefont {Lumsden}, \citenamefont {Mandrus}, \citenamefont {Tennant},
  \citenamefont {Chakoumakos},\ and\ \citenamefont {Nagler}}]{cao16}%
  \BibitemOpen
  \bibfield  {author} {\bibinfo {author} {\bibfnamefont {H.~B.}\ \bibnamefont
  {Cao}}, \bibinfo {author} {\bibfnamefont {A.}~\bibnamefont {Banerjee}},
  \bibinfo {author} {\bibfnamefont {J.-Q.}\ \bibnamefont {Yan}}, \bibinfo
  {author} {\bibfnamefont {C.~A.}\ \bibnamefont {Bridges}}, \bibinfo {author}
  {\bibfnamefont {M.~D.}\ \bibnamefont {Lumsden}}, \bibinfo {author}
  {\bibfnamefont {D.~G.}\ \bibnamefont {Mandrus}}, \bibinfo {author}
  {\bibfnamefont {D.~A.}\ \bibnamefont {Tennant}}, \bibinfo {author}
  {\bibfnamefont {B.~C.}\ \bibnamefont {Chakoumakos}},\ and\ \bibinfo {author}
  {\bibfnamefont {S.~E.}\ \bibnamefont {Nagler}},\ }\bibfield  {title}
  {\bibinfo {title} {Low-temperature crystal and magnetic structure of
  $\ensuremath{\alpha}${-}${\mathrm{RuCl}}_{3}$},\ }\href
  {https://doi.org/10.1103/PhysRevB.93.134423} {\bibfield  {journal} {\bibinfo
  {journal} {Phys. Rev. B}\ }\textbf {\bibinfo {volume} {93}},\ \bibinfo
  {pages} {134423} (\bibinfo {year} {2016})}\BibitemShut {NoStop}%
\bibitem [{\citenamefont {Banerjee}\ \emph {et~al.}(2016)\citenamefont
  {Banerjee}, \citenamefont {Bridges}, \citenamefont {Yan}, \citenamefont
  {Aczel}, \citenamefont {Li}, \citenamefont {Stone}, \citenamefont {Granroth},
  \citenamefont {Lumsden}, \citenamefont {Yiu}, \citenamefont {Knolle},
  \citenamefont {Bhattacharjee}, \citenamefont {Kovrizhin}, \citenamefont
  {Moessner}, \citenamefont {Tennant}, \citenamefont {Mandrus},\ and\
  \citenamefont {Nagler}}]{banerjee16}%
  \BibitemOpen
  \bibfield  {author} {\bibinfo {author} {\bibfnamefont {A.}~\bibnamefont
  {Banerjee}}, \bibinfo {author} {\bibfnamefont {C.~A.}\ \bibnamefont
  {Bridges}}, \bibinfo {author} {\bibfnamefont {J.~Q.}\ \bibnamefont {Yan}},
  \bibinfo {author} {\bibfnamefont {A.~A.}\ \bibnamefont {Aczel}}, \bibinfo
  {author} {\bibfnamefont {L.}~\bibnamefont {Li}}, \bibinfo {author}
  {\bibfnamefont {M.~B.}\ \bibnamefont {Stone}}, \bibinfo {author}
  {\bibfnamefont {G.~E.}\ \bibnamefont {Granroth}}, \bibinfo {author}
  {\bibfnamefont {M.~D.}\ \bibnamefont {Lumsden}}, \bibinfo {author}
  {\bibfnamefont {Y.}~\bibnamefont {Yiu}}, \bibinfo {author} {\bibfnamefont
  {J.}~\bibnamefont {Knolle}}, \bibinfo {author} {\bibfnamefont
  {S.}~\bibnamefont {Bhattacharjee}}, \bibinfo {author} {\bibfnamefont {D.~L.}\
  \bibnamefont {Kovrizhin}}, \bibinfo {author} {\bibfnamefont {R.}~\bibnamefont
  {Moessner}}, \bibinfo {author} {\bibfnamefont {D.~A.}\ \bibnamefont
  {Tennant}}, \bibinfo {author} {\bibfnamefont {D.~G.}\ \bibnamefont
  {Mandrus}},\ and\ \bibinfo {author} {\bibfnamefont {S.~E.}\ \bibnamefont
  {Nagler}},\ }\bibfield  {title} {\bibinfo {title} {Proximate Kitaev quantum
  spin liquid behaviour in a honeycomb magnet},\ }\href
  {https://doi.org/10.1038/nmat4604} {\bibfield  {journal} {\bibinfo  {journal}
  {Nat. Mat.}\ }\textbf {\bibinfo {volume} {15}},\ \bibinfo {pages} {733}
  (\bibinfo {year} {2016})}\BibitemShut {NoStop}%
\bibitem [{\citenamefont {Banerjee}\ \emph {et~al.}(2017)\citenamefont
  {Banerjee}, \citenamefont {Yan}, \citenamefont {Knolle}, \citenamefont
  {Bridges}, \citenamefont {Stone}, \citenamefont {Lumsden}, \citenamefont
  {Mandrus}, \citenamefont {Tennant}, \citenamefont {Moessner},\ and\
  \citenamefont {Nagler}}]{banerjee17}%
  \BibitemOpen
  \bibfield  {author} {\bibinfo {author} {\bibfnamefont {A.}~\bibnamefont
  {Banerjee}}, \bibinfo {author} {\bibfnamefont {J.}~\bibnamefont {Yan}},
  \bibinfo {author} {\bibfnamefont {J.}~\bibnamefont {Knolle}}, \bibinfo
  {author} {\bibfnamefont {C.~A.}\ \bibnamefont {Bridges}}, \bibinfo {author}
  {\bibfnamefont {M.~B.}\ \bibnamefont {Stone}}, \bibinfo {author}
  {\bibfnamefont {M.~D.}\ \bibnamefont {Lumsden}}, \bibinfo {author}
  {\bibfnamefont {D.~G.}\ \bibnamefont {Mandrus}}, \bibinfo {author}
  {\bibfnamefont {D.~A.}\ \bibnamefont {Tennant}}, \bibinfo {author}
  {\bibfnamefont {R.}~\bibnamefont {Moessner}},\ and\ \bibinfo {author}
  {\bibfnamefont {S.~E.}\ \bibnamefont {Nagler}},\ }\bibfield  {title}
  {\bibinfo {title} {Neutron scattering in the proximate quantum spin liquid
  $\alpha$-RuCl$_3$},\ }\href {https://doi.org/10.1126/science.aah6015}
  {\bibfield  {journal} {\bibinfo  {journal} {Science}\ }\textbf {\bibinfo
  {volume} {356}},\ \bibinfo {pages} {1055} (\bibinfo {year}
  {2017})}\BibitemShut {NoStop}%
\bibitem [{\citenamefont {Sears}\ \emph {et~al.}(2017)\citenamefont {Sears},
  \citenamefont {Zhao}, \citenamefont {Xu}, \citenamefont {Lynn},\ and\
  \citenamefont {Kim}}]{sears17}%
  \BibitemOpen
  \bibfield  {author} {\bibinfo {author} {\bibfnamefont {J.~A.}\ \bibnamefont
  {Sears}}, \bibinfo {author} {\bibfnamefont {Y.}~\bibnamefont {Zhao}},
  \bibinfo {author} {\bibfnamefont {Z.}~\bibnamefont {Xu}}, \bibinfo {author}
  {\bibfnamefont {J.~W.}\ \bibnamefont {Lynn}},\ and\ \bibinfo {author}
  {\bibfnamefont {Y.-J.}\ \bibnamefont {Kim}},\ }\bibfield  {title} {\bibinfo
  {title} {Phase diagram of $\alpha$-RuCl$_3$ in an in-plane magnetic field},\
  }\href {https://doi.org/10.1103/PhysRevB.95.180411} {\bibfield  {journal}
  {\bibinfo  {journal} {Phys. Rev. B}\ }\textbf {\bibinfo {volume} {95}},\
  \bibinfo {pages} {180411} (\bibinfo {year} {2017})}\BibitemShut {NoStop}%
\bibitem [{\citenamefont {Wolter}\ \emph {et~al.}(2017)\citenamefont {Wolter},
  \citenamefont {Corredor}, \citenamefont {Janssen}, \citenamefont {Nenkov},
  \citenamefont {Sch\"onecker}, \citenamefont {Do}, \citenamefont {Choi},
  \citenamefont {Albrecht}, \citenamefont {Hunger}, \citenamefont {Doert},
  \citenamefont {Vojta},\ and\ \citenamefont {B\"uchner}}]{wolter17}%
  \BibitemOpen
  \bibfield  {author} {\bibinfo {author} {\bibfnamefont {A.~U.~B.}\
  \bibnamefont {Wolter}}, \bibinfo {author} {\bibfnamefont {L.~T.}\
  \bibnamefont {Corredor}}, \bibinfo {author} {\bibfnamefont {L.}~\bibnamefont
  {Janssen}}, \bibinfo {author} {\bibfnamefont {K.}~\bibnamefont {Nenkov}},
  \bibinfo {author} {\bibfnamefont {S.}~\bibnamefont {Sch\"onecker}}, \bibinfo
  {author} {\bibfnamefont {S.-H.}\ \bibnamefont {Do}}, \bibinfo {author}
  {\bibfnamefont {K.-Y.}\ \bibnamefont {Choi}}, \bibinfo {author}
  {\bibfnamefont {R.}~\bibnamefont {Albrecht}}, \bibinfo {author}
  {\bibfnamefont {J.}~\bibnamefont {Hunger}}, \bibinfo {author} {\bibfnamefont
  {T.}~\bibnamefont {Doert}}, \bibinfo {author} {\bibfnamefont
  {M.}~\bibnamefont {Vojta}},\ and\ \bibinfo {author} {\bibfnamefont
  {B.}~\bibnamefont {B\"uchner}},\ }\bibfield  {title} {\bibinfo {title}
  {Field-induced quantum criticality in the Kitaev system $\alpha$-RuCl$_3$},\
  }\href {https://doi.org/10.1103/PhysRevB.96.041405} {\bibfield  {journal}
  {\bibinfo  {journal} {Phys. Rev. B}\ }\textbf {\bibinfo {volume} {96}},\
  \bibinfo {pages} {041405} (\bibinfo {year} {2017})}\BibitemShut {NoStop}%
\bibitem [{\citenamefont {Janssen}\ \emph {et~al.}(2017)\citenamefont
  {Janssen}, \citenamefont {Andrade},\ and\ \citenamefont {Vojta}}]{janssen17}%
  \BibitemOpen
  \bibfield  {author} {\bibinfo {author} {\bibfnamefont {L.}~\bibnamefont
  {Janssen}}, \bibinfo {author} {\bibfnamefont {E.~C.}\ \bibnamefont
  {Andrade}},\ and\ \bibinfo {author} {\bibfnamefont {M.}~\bibnamefont
  {Vojta}},\ }\bibfield  {title} {\bibinfo {title} {Magnetization processes of
  zigzag states on the honeycomb lattice: Identifying spin models for
  $\alpha$-RuCl$_3$ and ${\mathrm{Na}}_{2}{\mathrm{IrO}}_{3}$},\ }\href
  {https://doi.org/10.1103/PhysRevB.96.064430} {\bibfield  {journal} {\bibinfo
  {journal} {Phys. Rev. B}\ }\textbf {\bibinfo {volume} {96}},\ \bibinfo
  {pages} {064430} (\bibinfo {year} {2017})}\BibitemShut {NoStop}%
\bibitem [{\citenamefont {Winter}\ \emph
  {et~al.}(2017{\natexlab{a}})\citenamefont {Winter}, \citenamefont {Riedl},
  \citenamefont {Maksimov}, \citenamefont {Chernyshev}, \citenamefont
  {Honecker},\ and\ \citenamefont {Valent{\'\i}}}]{winter17b}%
  \BibitemOpen
  \bibfield  {author} {\bibinfo {author} {\bibfnamefont {S.~M.}\ \bibnamefont
  {Winter}}, \bibinfo {author} {\bibfnamefont {K.}~\bibnamefont {Riedl}},
  \bibinfo {author} {\bibfnamefont {P.~A.}\ \bibnamefont {Maksimov}}, \bibinfo
  {author} {\bibfnamefont {A.~L.}\ \bibnamefont {Chernyshev}}, \bibinfo
  {author} {\bibfnamefont {A.}~\bibnamefont {Honecker}},\ and\ \bibinfo
  {author} {\bibfnamefont {R.}~\bibnamefont {Valent{\'\i}}},\ }\bibfield
  {title} {\bibinfo {title} {Breakdown of magnons in a strongly spin-orbital
  coupled magnet},\ }\href {https://doi.org/10.1038/s41467-017-01177-0}
  {\bibfield  {journal} {\bibinfo  {journal} {Nat. Commun.}\ }\textbf {\bibinfo
  {volume} {8}},\ \bibinfo {pages} {1152} (\bibinfo {year}
  {2017}{\natexlab{a}})}\BibitemShut {NoStop}%
\bibitem [{\citenamefont {Sears}\ \emph {et~al.}(2020)\citenamefont {Sears},
  \citenamefont {Chern}, \citenamefont {Kim}, \citenamefont {Bereciartua},
  \citenamefont {Francoual}, \citenamefont {Kim},\ and\ \citenamefont
  {Kim}}]{sears20}%
  \BibitemOpen
  \bibfield  {author} {\bibinfo {author} {\bibfnamefont {J.~A.}\ \bibnamefont
  {Sears}}, \bibinfo {author} {\bibfnamefont {L.~E.}\ \bibnamefont {Chern}},
  \bibinfo {author} {\bibfnamefont {S.}~\bibnamefont {Kim}}, \bibinfo {author}
  {\bibfnamefont {P.~J.}\ \bibnamefont {Bereciartua}}, \bibinfo {author}
  {\bibfnamefont {S.}~\bibnamefont {Francoual}}, \bibinfo {author}
  {\bibfnamefont {Y.~B.}\ \bibnamefont {Kim}},\ and\ \bibinfo {author}
  {\bibfnamefont {Y.-J.}\ \bibnamefont {Kim}},\ }\bibfield  {title} {\bibinfo
  {title} {Ferromagnetic Kitaev interaction and the origin of large magnetic
  anisotropy in $\alpha$-RuCl$_3$},\ }\href
  {https://doi.org/10.1038/s41567-020-0874-0} {\bibfield  {journal} {\bibinfo
  {journal} {Nat. Phys.}\ }\textbf {\bibinfo {volume} {16}},\ \bibinfo {pages}
  {837} (\bibinfo {year} {2020})}\BibitemShut {NoStop}%
\bibitem [{\citenamefont {Kr\"uger}\ \emph {et~al.}(2020)\citenamefont
  {Kr\"uger}, \citenamefont {Vojta},\ and\ \citenamefont
  {Janssen}}]{krueger20}%
  \BibitemOpen
  \bibfield  {author} {\bibinfo {author} {\bibfnamefont {W.~G.~F.}\
  \bibnamefont {Kr\"uger}}, \bibinfo {author} {\bibfnamefont {M.}~\bibnamefont
  {Vojta}},\ and\ \bibinfo {author} {\bibfnamefont {L.}~\bibnamefont
  {Janssen}},\ }\bibfield  {title} {\bibinfo {title} {Heisenberg-Kitaev models
  on hyperhoneycomb and stripy-honeycomb lattices: 3D-2D equivalence of ordered
  states and phase diagrams},\ }\href
  {https://doi.org/10.1103/PhysRevResearch.2.012021} {\bibfield  {journal}
  {\bibinfo  {journal} {Phys. Rev. Res.}\ }\textbf {\bibinfo {volume} {2}},\
  \bibinfo {pages} {012021} (\bibinfo {year} {2020})}\BibitemShut {NoStop}%
\bibitem [{\citenamefont {Janssen}\ \emph {et~al.}(2020)\citenamefont
  {Janssen}, \citenamefont {Koch},\ and\ \citenamefont {Vojta}}]{janssen20}%
  \BibitemOpen
  \bibfield  {author} {\bibinfo {author} {\bibfnamefont {L.}~\bibnamefont
  {Janssen}}, \bibinfo {author} {\bibfnamefont {S.}~\bibnamefont {Koch}},\ and\
  \bibinfo {author} {\bibfnamefont {M.}~\bibnamefont {Vojta}},\ }\bibfield
  {title} {\bibinfo {title} {Magnon dispersion and dynamic spin response in
  three-dimensional spin models for $\alpha$-RuCl$_3$},\ }\href
  {https://doi.org/10.1103/PhysRevB.101.174444} {\bibfield  {journal} {\bibinfo
   {journal} {Phys. Rev. B}\ }\textbf {\bibinfo {volume} {101}},\ \bibinfo
  {pages} {174444} (\bibinfo {year} {2020})}\BibitemShut {NoStop}%
\bibitem [{\citenamefont {Hentrich}\ \emph {et~al.}(2020)\citenamefont
  {Hentrich}, \citenamefont {Hong}, \citenamefont {Gillig}, \citenamefont
  {Caglieris}, \citenamefont {\ifmmode~\check{C}\else \v{C}\fi{}ulo},
  \citenamefont {Shahrokhvand}, \citenamefont {Zeitler}, \citenamefont
  {Roslova}, \citenamefont {Isaeva}, \citenamefont {Doert}, \citenamefont
  {Janssen}, \citenamefont {Vojta}, \citenamefont {B\"uchner},\ and\
  \citenamefont {Hess}}]{hentrich20}%
  \BibitemOpen
  \bibfield  {author} {\bibinfo {author} {\bibfnamefont {R.}~\bibnamefont
  {Hentrich}}, \bibinfo {author} {\bibfnamefont {X.}~\bibnamefont {Hong}},
  \bibinfo {author} {\bibfnamefont {M.}~\bibnamefont {Gillig}}, \bibinfo
  {author} {\bibfnamefont {F.}~\bibnamefont {Caglieris}}, \bibinfo {author}
  {\bibfnamefont {M.}~\bibnamefont {\ifmmode~\check{C}\else \v{C}\fi{}ulo}},
  \bibinfo {author} {\bibfnamefont {M.}~\bibnamefont {Shahrokhvand}}, \bibinfo
  {author} {\bibfnamefont {U.}~\bibnamefont {Zeitler}}, \bibinfo {author}
  {\bibfnamefont {M.}~\bibnamefont {Roslova}}, \bibinfo {author} {\bibfnamefont
  {A.}~\bibnamefont {Isaeva}}, \bibinfo {author} {\bibfnamefont
  {T.}~\bibnamefont {Doert}}, \bibinfo {author} {\bibfnamefont
  {L.}~\bibnamefont {Janssen}}, \bibinfo {author} {\bibfnamefont
  {M.}~\bibnamefont {Vojta}}, \bibinfo {author} {\bibfnamefont
  {B.}~\bibnamefont {B\"uchner}},\ and\ \bibinfo {author} {\bibfnamefont
  {C.}~\bibnamefont {Hess}},\ }\bibfield  {title} {\bibinfo {title} {High-field
  thermal transport properties of the Kitaev quantum magnet
  $\ensuremath{\alpha}\text{\ensuremath{-}}\mathrm{Ru}{\mathrm{Cl}}_{3}$:
  Evidence for low-energy excitations beyond the critical field},\ }\href
  {https://doi.org/10.1103/PhysRevB.102.235155} {\bibfield  {journal} {\bibinfo
   {journal} {Phys. Rev. B}\ }\textbf {\bibinfo {volume} {102}},\ \bibinfo
  {pages} {235155} (\bibinfo {year} {2020})}\BibitemShut {NoStop}%
\bibitem [{\citenamefont {Balz}\ \emph {et~al.}(2021)\citenamefont {Balz},
  \citenamefont {Janssen}, \citenamefont {Lampen-Kelley}, \citenamefont
  {Banerjee}, \citenamefont {Liu}, \citenamefont {Yan}, \citenamefont
  {Mandrus}, \citenamefont {Vojta},\ and\ \citenamefont {Nagler}}]{balz21}%
  \BibitemOpen
  \bibfield  {author} {\bibinfo {author} {\bibfnamefont {C.}~\bibnamefont
  {Balz}}, \bibinfo {author} {\bibfnamefont {L.}~\bibnamefont {Janssen}},
  \bibinfo {author} {\bibfnamefont {P.}~\bibnamefont {Lampen-Kelley}}, \bibinfo
  {author} {\bibfnamefont {A.}~\bibnamefont {Banerjee}}, \bibinfo {author}
  {\bibfnamefont {Y.~H.}\ \bibnamefont {Liu}}, \bibinfo {author} {\bibfnamefont
  {J.-Q.}\ \bibnamefont {Yan}}, \bibinfo {author} {\bibfnamefont {D.~G.}\
  \bibnamefont {Mandrus}}, \bibinfo {author} {\bibfnamefont {M.}~\bibnamefont
  {Vojta}},\ and\ \bibinfo {author} {\bibfnamefont {S.~E.}\ \bibnamefont
  {Nagler}},\ }\bibfield  {title} {\bibinfo {title} {Field-induced intermediate
  ordered phase and anisotropic interlayer interactions in $\alpha$-RuCl$_3$},\
  }\href {https://doi.org/10.1103/PhysRevB.103.174417} {\bibfield  {journal}
  {\bibinfo  {journal} {Phys. Rev. B}\ }\textbf {\bibinfo {volume} {103}},\
  \bibinfo {pages} {174417} (\bibinfo {year} {2021})}\BibitemShut {NoStop}%
\bibitem [{\citenamefont {Winter}\ \emph
  {et~al.}(2017{\natexlab{b}})\citenamefont {Winter}, \citenamefont {Tsirlin},
  \citenamefont {Daghofer}, \citenamefont {van~den Brink}, \citenamefont
  {Singh}, \citenamefont {Gegenwart},\ and\ \citenamefont
  {Valent{\'{\i}}}}]{winter17}%
  \BibitemOpen
  \bibfield  {author} {\bibinfo {author} {\bibfnamefont {S.~M.}\ \bibnamefont
  {Winter}}, \bibinfo {author} {\bibfnamefont {A.~A.}\ \bibnamefont {Tsirlin}},
  \bibinfo {author} {\bibfnamefont {M.}~\bibnamefont {Daghofer}}, \bibinfo
  {author} {\bibfnamefont {J.}~\bibnamefont {van~den Brink}}, \bibinfo {author}
  {\bibfnamefont {Y.}~\bibnamefont {Singh}}, \bibinfo {author} {\bibfnamefont
  {P.}~\bibnamefont {Gegenwart}},\ and\ \bibinfo {author} {\bibfnamefont
  {R.}~\bibnamefont {Valent{\'{\i}}}},\ }\bibfield  {title} {\bibinfo {title}
  {Models and materials for generalized Kitaev magnetism},\ }\href
  {https://doi.org/10.1088/1361-648x/aa8cf5} {\bibfield  {journal} {\bibinfo
  {journal} {J. Phys. Condens. Matter}\ }\textbf {\bibinfo {volume} {29}},\
  \bibinfo {pages} {493002} (\bibinfo {year} {2017}{\natexlab{b}})}\BibitemShut
  {NoStop}%
\bibitem [{\citenamefont {Janssen}\ and\ \citenamefont
  {Vojta}(2019)}]{janssen19}%
  \BibitemOpen
  \bibfield  {author} {\bibinfo {author} {\bibfnamefont {L.}~\bibnamefont
  {Janssen}}\ and\ \bibinfo {author} {\bibfnamefont {M.}~\bibnamefont
  {Vojta}},\ }\bibfield  {title} {\bibinfo {title} {Heisenberg-Kitaev physics
  in magnetic fields},\ }\href {https://doi.org/10.1088/1361-648x/ab283e}
  {\bibfield  {journal} {\bibinfo  {journal} {J. Phys. Condens. Matter}\
  }\textbf {\bibinfo {volume} {31}},\ \bibinfo {pages} {423002} (\bibinfo
  {year} {2019})}\BibitemShut {NoStop}%
\bibitem [{\citenamefont {Takagi}\ \emph {et~al.}(2019)\citenamefont {Takagi},
  \citenamefont {Takayama}, \citenamefont {Jackeli}, \citenamefont
  {Khaliullin},\ and\ \citenamefont {Nagler}}]{takagi19}%
  \BibitemOpen
  \bibfield  {author} {\bibinfo {author} {\bibfnamefont {H.}~\bibnamefont
  {Takagi}}, \bibinfo {author} {\bibfnamefont {T.}~\bibnamefont {Takayama}},
  \bibinfo {author} {\bibfnamefont {G.}~\bibnamefont {Jackeli}}, \bibinfo
  {author} {\bibfnamefont {G.}~\bibnamefont {Khaliullin}},\ and\ \bibinfo
  {author} {\bibfnamefont {S.~E.}\ \bibnamefont {Nagler}},\ }\bibfield  {title}
  {\bibinfo {title} {Concept and realization of Kitaev quantum spin liquids},\
  }\href {https://doi.org/10.1038/s42254-019-0038-2} {\bibfield  {journal}
  {\bibinfo  {journal} {Nat. Rev. Phys.}\ }\textbf {\bibinfo {volume} {1}},\
  \bibinfo {pages} {264} (\bibinfo {year} {2019})}\BibitemShut {NoStop}%
\bibitem [{\citenamefont {Trebst}\ and\ \citenamefont
  {Hickey}(2022)}]{trebst22}%
  \BibitemOpen
  \bibfield  {author} {\bibinfo {author} {\bibfnamefont {S.}~\bibnamefont
  {Trebst}}\ and\ \bibinfo {author} {\bibfnamefont {C.}~\bibnamefont
  {Hickey}},\ }\bibfield  {title} {\bibinfo {title} {Kitaev materials},\ }\href
  {https://doi.org/https://doi.org/10.1016/j.physrep.2021.11.003} {\bibfield
  {journal} {\bibinfo  {journal} {Phys. Rep.}\ }\textbf {\bibinfo {volume}
  {950}},\ \bibinfo {pages} {1} (\bibinfo {year} {2022})}\BibitemShut {NoStop}%
\bibitem [{\citenamefont {Yan}\ \emph {et~al.}(2019)\citenamefont {Yan},
  \citenamefont {Okamoto}, \citenamefont {Wu}, \citenamefont {Zheng},
  \citenamefont {Zhou}, \citenamefont {Cao},\ and\ \citenamefont
  {McGuire}}]{yan19}%
  \BibitemOpen
  \bibfield  {author} {\bibinfo {author} {\bibfnamefont {J.-Q.}\ \bibnamefont
  {Yan}}, \bibinfo {author} {\bibfnamefont {S.}~\bibnamefont {Okamoto}},
  \bibinfo {author} {\bibfnamefont {Y.}~\bibnamefont {Wu}}, \bibinfo {author}
  {\bibfnamefont {Q.}~\bibnamefont {Zheng}}, \bibinfo {author} {\bibfnamefont
  {H.~D.}\ \bibnamefont {Zhou}}, \bibinfo {author} {\bibfnamefont {H.~B.}\
  \bibnamefont {Cao}},\ and\ \bibinfo {author} {\bibfnamefont {M.~A.}\
  \bibnamefont {McGuire}},\ }\bibfield  {title} {\bibinfo {title} {Magnetic
  order in single crystals of Na$_3$Co$_2$SbO$_6$ with a honeycomb arrangement
  of $3d^7$ Co$^{2+}$ ions},\ }\href
  {https://doi.org/10.1103/PhysRevMaterials.3.074405} {\bibfield  {journal}
  {\bibinfo  {journal} {Phys. Rev. Materials}\ }\textbf {\bibinfo {volume}
  {3}},\ \bibinfo {pages} {074405} (\bibinfo {year} {2019})}\BibitemShut
  {NoStop}%
\bibitem [{\citenamefont {Yao}\ and\ \citenamefont {Li}(2020)}]{yao20}%
  \BibitemOpen
  \bibfield  {author} {\bibinfo {author} {\bibfnamefont {W.}~\bibnamefont
  {Yao}}\ and\ \bibinfo {author} {\bibfnamefont {Y.}~\bibnamefont {Li}},\
  }\bibfield  {title} {\bibinfo {title} {Ferrimagnetism and anisotropic phase
  tunability by magnetic fields in
  ${\mathrm{Na}}_{2}{\mathrm{Co}}_{2}{\mathrm{TeO}}_{6}$},\ }\href
  {https://doi.org/10.1103/PhysRevB.101.085120} {\bibfield  {journal} {\bibinfo
   {journal} {Phys. Rev. B}\ }\textbf {\bibinfo {volume} {101}},\ \bibinfo
  {pages} {085120} (\bibinfo {year} {2020})}\BibitemShut {NoStop}%
\bibitem [{\citenamefont {Songvilay}\ \emph {et~al.}(2020)\citenamefont
  {Songvilay}, \citenamefont {Robert}, \citenamefont {Petit}, \citenamefont
  {Rodriguez-Rivera}, \citenamefont {Ratcliff}, \citenamefont {Damay},
  \citenamefont {Bal\'edent}, \citenamefont {Jim\'enez-Ruiz}, \citenamefont
  {Lejay}, \citenamefont {Pachoud}, \citenamefont {Hadj-Azzem}, \citenamefont
  {Simonet},\ and\ \citenamefont {Stock}}]{songvilay20}%
  \BibitemOpen
  \bibfield  {author} {\bibinfo {author} {\bibfnamefont {M.}~\bibnamefont
  {Songvilay}}, \bibinfo {author} {\bibfnamefont {J.}~\bibnamefont {Robert}},
  \bibinfo {author} {\bibfnamefont {S.}~\bibnamefont {Petit}}, \bibinfo
  {author} {\bibfnamefont {J.~A.}\ \bibnamefont {Rodriguez-Rivera}}, \bibinfo
  {author} {\bibfnamefont {W.~D.}\ \bibnamefont {Ratcliff}}, \bibinfo {author}
  {\bibfnamefont {F.}~\bibnamefont {Damay}}, \bibinfo {author} {\bibfnamefont
  {V.}~\bibnamefont {Bal\'edent}}, \bibinfo {author} {\bibfnamefont
  {M.}~\bibnamefont {Jim\'enez-Ruiz}}, \bibinfo {author} {\bibfnamefont
  {P.}~\bibnamefont {Lejay}}, \bibinfo {author} {\bibfnamefont
  {E.}~\bibnamefont {Pachoud}}, \bibinfo {author} {\bibfnamefont
  {A.}~\bibnamefont {Hadj-Azzem}}, \bibinfo {author} {\bibfnamefont
  {V.}~\bibnamefont {Simonet}},\ and\ \bibinfo {author} {\bibfnamefont
  {C.}~\bibnamefont {Stock}},\ }\bibfield  {title} {\bibinfo {title} {Kitaev
  interactions in the Co honeycomb antiferromagnets
  ${\mathrm{Na}}_{3}{\mathrm{Co}}_{2}{\mathrm{SbO}}_{6}$ and
  ${\mathrm{Na}}_{2}{\mathrm{Co}}_{2}{\mathrm{TeO}}_{6}$},\ }\href
  {https://doi.org/10.1103/PhysRevB.102.224429} {\bibfield  {journal} {\bibinfo
   {journal} {Phys. Rev. B}\ }\textbf {\bibinfo {volume} {102}},\ \bibinfo
  {pages} {224429} (\bibinfo {year} {2020})}\BibitemShut {NoStop}%
\bibitem [{\citenamefont {Lin}\ \emph {et~al.}(2021)\citenamefont {Lin},
  \citenamefont {Jeong}, \citenamefont {Kim}, \citenamefont {Wang},
  \citenamefont {Huang}, \citenamefont {Masuda}, \citenamefont {Asai},
  \citenamefont {Itoh}, \citenamefont {G{\"u}nther}, \citenamefont {Russina},
  \citenamefont {Lu}, \citenamefont {Sheng}, \citenamefont {Wang},
  \citenamefont {Wang}, \citenamefont {Wang}, \citenamefont {Ren},
  \citenamefont {Xi}, \citenamefont {Tong}, \citenamefont {Ling}, \citenamefont
  {Liu}, \citenamefont {Wu}, \citenamefont {Mei}, \citenamefont {Qu},
  \citenamefont {Zhou}, \citenamefont {Wang}, \citenamefont {Park},
  \citenamefont {Wan},\ and\ \citenamefont {Ma}}]{lin21}%
  \BibitemOpen
  \bibfield  {author} {\bibinfo {author} {\bibfnamefont {G.}~\bibnamefont
  {Lin}}, \bibinfo {author} {\bibfnamefont {J.}~\bibnamefont {Jeong}}, \bibinfo
  {author} {\bibfnamefont {C.}~\bibnamefont {Kim}}, \bibinfo {author}
  {\bibfnamefont {Y.}~\bibnamefont {Wang}}, \bibinfo {author} {\bibfnamefont
  {Q.}~\bibnamefont {Huang}}, \bibinfo {author} {\bibfnamefont
  {T.}~\bibnamefont {Masuda}}, \bibinfo {author} {\bibfnamefont
  {S.}~\bibnamefont {Asai}}, \bibinfo {author} {\bibfnamefont {S.}~\bibnamefont
  {Itoh}}, \bibinfo {author} {\bibfnamefont {G.}~\bibnamefont {G{\"u}nther}},
  \bibinfo {author} {\bibfnamefont {M.}~\bibnamefont {Russina}}, \bibinfo
  {author} {\bibfnamefont {Z.}~\bibnamefont {Lu}}, \bibinfo {author}
  {\bibfnamefont {J.}~\bibnamefont {Sheng}}, \bibinfo {author} {\bibfnamefont
  {L.}~\bibnamefont {Wang}}, \bibinfo {author} {\bibfnamefont {J.}~\bibnamefont
  {Wang}}, \bibinfo {author} {\bibfnamefont {G.}~\bibnamefont {Wang}}, \bibinfo
  {author} {\bibfnamefont {Q.}~\bibnamefont {Ren}}, \bibinfo {author}
  {\bibfnamefont {C.}~\bibnamefont {Xi}}, \bibinfo {author} {\bibfnamefont
  {W.}~\bibnamefont {Tong}}, \bibinfo {author} {\bibfnamefont {L.}~\bibnamefont
  {Ling}}, \bibinfo {author} {\bibfnamefont {Z.}~\bibnamefont {Liu}}, \bibinfo
  {author} {\bibfnamefont {L.}~\bibnamefont {Wu}}, \bibinfo {author}
  {\bibfnamefont {J.}~\bibnamefont {Mei}}, \bibinfo {author} {\bibfnamefont
  {Z.}~\bibnamefont {Qu}}, \bibinfo {author} {\bibfnamefont {H.}~\bibnamefont
  {Zhou}}, \bibinfo {author} {\bibfnamefont {X.}~\bibnamefont {Wang}}, \bibinfo
  {author} {\bibfnamefont {J.-G.}\ \bibnamefont {Park}}, \bibinfo {author}
  {\bibfnamefont {Y.}~\bibnamefont {Wan}},\ and\ \bibinfo {author}
  {\bibfnamefont {J.}~\bibnamefont {Ma}},\ }\bibfield  {title} {\bibinfo
  {title} {Field-induced quantum spin disordered state in spin-1/2 honeycomb
  magnet Na$_2$Co$_2$TeO$_6$},\ }\href
  {https://doi.org/10.1038/s41467-021-25567-7} {\bibfield  {journal} {\bibinfo
  {journal} {Nat. Commun.}\ }\textbf {\bibinfo {volume} {12}},\ \bibinfo
  {pages} {5559} (\bibinfo {year} {2021})}\BibitemShut {NoStop}%
\bibitem [{\citenamefont {Chen}\ \emph {et~al.}(2021)\citenamefont {Chen},
  \citenamefont {Li}, \citenamefont {Hu}, \citenamefont {Hu}, \citenamefont
  {Yue}, \citenamefont {Sutarto}, \citenamefont {He}, \citenamefont {Iida},
  \citenamefont {Kamazawa}, \citenamefont {Yu}, \citenamefont {Lin},\ and\
  \citenamefont {Li}}]{chen21}%
  \BibitemOpen
  \bibfield  {author} {\bibinfo {author} {\bibfnamefont {W.}~\bibnamefont
  {Chen}}, \bibinfo {author} {\bibfnamefont {X.}~\bibnamefont {Li}}, \bibinfo
  {author} {\bibfnamefont {Z.}~\bibnamefont {Hu}}, \bibinfo {author}
  {\bibfnamefont {Z.}~\bibnamefont {Hu}}, \bibinfo {author} {\bibfnamefont
  {L.}~\bibnamefont {Yue}}, \bibinfo {author} {\bibfnamefont {R.}~\bibnamefont
  {Sutarto}}, \bibinfo {author} {\bibfnamefont {F.}~\bibnamefont {He}},
  \bibinfo {author} {\bibfnamefont {K.}~\bibnamefont {Iida}}, \bibinfo {author}
  {\bibfnamefont {K.}~\bibnamefont {Kamazawa}}, \bibinfo {author}
  {\bibfnamefont {W.}~\bibnamefont {Yu}}, \bibinfo {author} {\bibfnamefont
  {X.}~\bibnamefont {Lin}},\ and\ \bibinfo {author} {\bibfnamefont
  {Y.}~\bibnamefont {Li}},\ }\bibfield  {title} {\bibinfo {title} {Spin-orbit
  phase behavior of ${\mathrm{Na}}_{2}{\mathrm{Co}}_{2}{\mathrm{TeO}}_{6}$ at
  low temperatures},\ }\href {https://doi.org/10.1103/PhysRevB.103.L180404}
  {\bibfield  {journal} {\bibinfo  {journal} {Phys. Rev. B}\ }\textbf {\bibinfo
  {volume} {103}},\ \bibinfo {pages} {L180404} (\bibinfo {year}
  {2021})}\BibitemShut {NoStop}%
\bibitem [{\citenamefont {Lee}\ \emph {et~al.}(2021)\citenamefont {Lee},
  \citenamefont {Lee}, \citenamefont {Choi}, \citenamefont {Jang},
  \citenamefont {Kalaivanan}, \citenamefont {Sankar},\ and\ \citenamefont
  {Choi}}]{lee21}%
  \BibitemOpen
  \bibfield  {author} {\bibinfo {author} {\bibfnamefont {C.~H.}\ \bibnamefont
  {Lee}}, \bibinfo {author} {\bibfnamefont {S.}~\bibnamefont {Lee}}, \bibinfo
  {author} {\bibfnamefont {Y.~S.}\ \bibnamefont {Choi}}, \bibinfo {author}
  {\bibfnamefont {Z.~H.}\ \bibnamefont {Jang}}, \bibinfo {author}
  {\bibfnamefont {R.}~\bibnamefont {Kalaivanan}}, \bibinfo {author}
  {\bibfnamefont {R.}~\bibnamefont {Sankar}},\ and\ \bibinfo {author}
  {\bibfnamefont {K.-Y.}\ \bibnamefont {Choi}},\ }\bibfield  {title} {\bibinfo
  {title} {Multistage development of anisotropic magnetic correlations in the
  Co-based honeycomb lattice
  ${\mathrm{Na}}_{2}{\mathrm{Co}}_{2}{\mathrm{TeO}}_{6}$},\ }\href
  {https://doi.org/10.1103/PhysRevB.103.214447} {\bibfield  {journal} {\bibinfo
   {journal} {Phys. Rev. B}\ }\textbf {\bibinfo {volume} {103}},\ \bibinfo
  {pages} {214447} (\bibinfo {year} {2021})}\BibitemShut {NoStop}%
\bibitem [{\citenamefont {Hong}\ \emph {et~al.}(2021)\citenamefont {Hong},
  \citenamefont {Gillig}, \citenamefont {Hentrich}, \citenamefont {Yao},
  \citenamefont {Kocsis}, \citenamefont {Witte}, \citenamefont {Schreiner},
  \citenamefont {Baumann}, \citenamefont {P\'erez}, \citenamefont {Wolter},
  \citenamefont {Li}, \citenamefont {B\"uchner},\ and\ \citenamefont
  {Hess}}]{hong21}%
  \BibitemOpen
  \bibfield  {author} {\bibinfo {author} {\bibfnamefont {X.}~\bibnamefont
  {Hong}}, \bibinfo {author} {\bibfnamefont {M.}~\bibnamefont {Gillig}},
  \bibinfo {author} {\bibfnamefont {R.}~\bibnamefont {Hentrich}}, \bibinfo
  {author} {\bibfnamefont {W.}~\bibnamefont {Yao}}, \bibinfo {author}
  {\bibfnamefont {V.}~\bibnamefont {Kocsis}}, \bibinfo {author} {\bibfnamefont
  {A.~R.}\ \bibnamefont {Witte}}, \bibinfo {author} {\bibfnamefont
  {T.}~\bibnamefont {Schreiner}}, \bibinfo {author} {\bibfnamefont
  {D.}~\bibnamefont {Baumann}}, \bibinfo {author} {\bibfnamefont
  {N.}~\bibnamefont {P\'erez}}, \bibinfo {author} {\bibfnamefont {A.~U.~B.}\
  \bibnamefont {Wolter}}, \bibinfo {author} {\bibfnamefont {Y.}~\bibnamefont
  {Li}}, \bibinfo {author} {\bibfnamefont {B.}~\bibnamefont {B\"uchner}},\ and\
  \bibinfo {author} {\bibfnamefont {C.}~\bibnamefont {Hess}},\ }\bibfield
  {title} {\bibinfo {title} {Strongly scattered phonon heat transport of the
  candidate Kitaev material
  ${\mathrm{Na}}_{2}{\mathrm{Co}}_{2}{\mathrm{TeO}}_{6}$},\ }\href
  {https://doi.org/10.1103/PhysRevB.104.144426} {\bibfield  {journal} {\bibinfo
   {journal} {Phys. Rev. B}\ }\textbf {\bibinfo {volume} {104}},\ \bibinfo
  {pages} {144426} (\bibinfo {year} {2021})}\BibitemShut {NoStop}%
\bibitem [{\citenamefont {Samarakoon}\ \emph {et~al.}(2021)\citenamefont
  {Samarakoon}, \citenamefont {Chen}, \citenamefont {Zhou},\ and\ \citenamefont
  {Garlea}}]{samarakoon21}%
  \BibitemOpen
  \bibfield  {author} {\bibinfo {author} {\bibfnamefont {A.~M.}\ \bibnamefont
  {Samarakoon}}, \bibinfo {author} {\bibfnamefont {Q.}~\bibnamefont {Chen}},
  \bibinfo {author} {\bibfnamefont {H.}~\bibnamefont {Zhou}},\ and\ \bibinfo
  {author} {\bibfnamefont {V.~O.}\ \bibnamefont {Garlea}},\ }\bibfield  {title}
  {\bibinfo {title} {Static and dynamic magnetic properties of honeycomb
  lattice antiferromagnets ${\mathrm{Na}}_{2}{M}_{2}{\mathrm{TeO}}_{6}$,
  $M=\mathrm{Co}$ and Ni},\ }\href
  {https://doi.org/10.1103/PhysRevB.104.184415} {\bibfield  {journal} {\bibinfo
   {journal} {Phys. Rev. B}\ }\textbf {\bibinfo {volume} {104}},\ \bibinfo
  {pages} {184415} (\bibinfo {year} {2021})}\BibitemShut {NoStop}%
\bibitem [{\citenamefont {Kim}\ \emph {et~al.}(2022)\citenamefont {Kim},
  \citenamefont {Jeong}, \citenamefont {Lin}, \citenamefont {Park},
  \citenamefont {Masuda}, \citenamefont {Asai}, \citenamefont {Itoh},
  \citenamefont {Kim}, \citenamefont {Zhou}, \citenamefont {Ma},\ and\
  \citenamefont {Park}}]{kim22}%
  \BibitemOpen
  \bibfield  {author} {\bibinfo {author} {\bibfnamefont {C.}~\bibnamefont
  {Kim}}, \bibinfo {author} {\bibfnamefont {J.}~\bibnamefont {Jeong}}, \bibinfo
  {author} {\bibfnamefont {G.}~\bibnamefont {Lin}}, \bibinfo {author}
  {\bibfnamefont {P.}~\bibnamefont {Park}}, \bibinfo {author} {\bibfnamefont
  {T.}~\bibnamefont {Masuda}}, \bibinfo {author} {\bibfnamefont
  {S.}~\bibnamefont {Asai}}, \bibinfo {author} {\bibfnamefont {S.}~\bibnamefont
  {Itoh}}, \bibinfo {author} {\bibfnamefont {H.-S.}\ \bibnamefont {Kim}},
  \bibinfo {author} {\bibfnamefont {H.}~\bibnamefont {Zhou}}, \bibinfo {author}
  {\bibfnamefont {J.}~\bibnamefont {Ma}},\ and\ \bibinfo {author}
  {\bibfnamefont {J.-G.}\ \bibnamefont {Park}},\ }\bibfield  {title} {\bibinfo
  {title} {Antiferromagnetic Kitaev interaction in $J_\text{eff} = 1/2$ cobalt
  honeycomb materials Na$_3$Co$_2$SbO$_6$ and Na$_2$Co$_2$TeO$_6$},\ }\href
  {https://doi.org/10.1088/1361-648x/ac2644} {\bibfield  {journal} {\bibinfo
  {journal} {J. Phys. Condens. Matter}\ }\textbf {\bibinfo {volume} {34}},\
  \bibinfo {pages} {045802} (\bibinfo {year} {2022})}\BibitemShut {NoStop}%
\bibitem [{\citenamefont {Mukherjee}\ \emph {et~al.}(2022)\citenamefont
  {Mukherjee}, \citenamefont {Manna}, \citenamefont {Saha}, \citenamefont
  {Majumdar},\ and\ \citenamefont {Giri}}]{mukherjee22}%
  \BibitemOpen
  \bibfield  {author} {\bibinfo {author} {\bibfnamefont {S.}~\bibnamefont
  {Mukherjee}}, \bibinfo {author} {\bibfnamefont {G.}~\bibnamefont {Manna}},
  \bibinfo {author} {\bibfnamefont {P.}~\bibnamefont {Saha}}, \bibinfo {author}
  {\bibfnamefont {S.}~\bibnamefont {Majumdar}},\ and\ \bibinfo {author}
  {\bibfnamefont {S.}~\bibnamefont {Giri}},\ }\bibfield  {title} {\bibinfo
  {title} {Ferroelectric order with a linear high-field magnetoelectric
  coupling in ${\mathrm{Na}}_{2}{\mathrm{Co}}_{2}{\mathrm{TeO}}_{6}$: A
  proposed Kitaev compound},\ }\href
  {https://doi.org/10.1103/PhysRevMaterials.6.054407} {\bibfield  {journal}
  {\bibinfo  {journal} {Phys. Rev. Materials}\ }\textbf {\bibinfo {volume}
  {6}},\ \bibinfo {pages} {054407} (\bibinfo {year} {2022})}\BibitemShut
  {NoStop}%
\bibitem [{\citenamefont {Sanders}\ \emph {et~al.}(2022)\citenamefont
  {Sanders}, \citenamefont {Mole}, \citenamefont {Liu}, \citenamefont {Brown},
  \citenamefont {Yu}, \citenamefont {Ling},\ and\ \citenamefont
  {Rachel}}]{sanders22}%
  \BibitemOpen
  \bibfield  {author} {\bibinfo {author} {\bibfnamefont {A.~L.}\ \bibnamefont
  {Sanders}}, \bibinfo {author} {\bibfnamefont {R.~A.}\ \bibnamefont {Mole}},
  \bibinfo {author} {\bibfnamefont {J.}~\bibnamefont {Liu}}, \bibinfo {author}
  {\bibfnamefont {A.~J.}\ \bibnamefont {Brown}}, \bibinfo {author}
  {\bibfnamefont {D.}~\bibnamefont {Yu}}, \bibinfo {author} {\bibfnamefont
  {C.~D.}\ \bibnamefont {Ling}},\ and\ \bibinfo {author} {\bibfnamefont
  {S.}~\bibnamefont {Rachel}},\ }\bibfield  {title} {\bibinfo {title} {Dominant
  Kitaev interactions in the honeycomb materials
  ${\mathrm{Na}}_{3}{\mathrm{Co}}_{2}{\mathrm{SbO}}_{6}$ and
  ${\mathrm{Na}}_{2}{\mathrm{Co}}_{2}{\mathrm{TeO}}_{6}$},\ }\href
  {https://doi.org/10.1103/PhysRevB.106.014413} {\bibfield  {journal} {\bibinfo
   {journal} {Phys. Rev. B}\ }\textbf {\bibinfo {volume} {106}},\ \bibinfo
  {pages} {014413} (\bibinfo {year} {2022})}\BibitemShut {NoStop}%
\bibitem [{\citenamefont {Yang}\ \emph {et~al.}(2022)\citenamefont {Yang},
  \citenamefont {Kim}, \citenamefont {Choi}, \citenamefont {Lee}, \citenamefont
  {Lin}, \citenamefont {Ma}, \citenamefont {Kratochv\'{\i}lov\'a},
  \citenamefont {Proschek}, \citenamefont {Moon}, \citenamefont {Lee},
  \citenamefont {Oh},\ and\ \citenamefont {Park}}]{yang22}%
  \BibitemOpen
  \bibfield  {author} {\bibinfo {author} {\bibfnamefont {H.}~\bibnamefont
  {Yang}}, \bibinfo {author} {\bibfnamefont {C.}~\bibnamefont {Kim}}, \bibinfo
  {author} {\bibfnamefont {Y.}~\bibnamefont {Choi}}, \bibinfo {author}
  {\bibfnamefont {J.~H.}\ \bibnamefont {Lee}}, \bibinfo {author} {\bibfnamefont
  {G.}~\bibnamefont {Lin}}, \bibinfo {author} {\bibfnamefont {J.}~\bibnamefont
  {Ma}}, \bibinfo {author} {\bibfnamefont {M.}~\bibnamefont
  {Kratochv\'{\i}lov\'a}}, \bibinfo {author} {\bibfnamefont {P.}~\bibnamefont
  {Proschek}}, \bibinfo {author} {\bibfnamefont {E.-G.}\ \bibnamefont {Moon}},
  \bibinfo {author} {\bibfnamefont {K.~H.}\ \bibnamefont {Lee}}, \bibinfo
  {author} {\bibfnamefont {Y.~S.}\ \bibnamefont {Oh}},\ and\ \bibinfo {author}
  {\bibfnamefont {J.-G.}\ \bibnamefont {Park}},\ }\bibfield  {title} {\bibinfo
  {title} {Significant thermal Hall effect in the $3d$ cobalt Kitaev system
  ${\mathrm{Na}}_{2}{\mathrm{Co}}_{2}\mathrm{Te}{\mathrm{O}}_{6}$},\ }\href
  {https://doi.org/10.1103/PhysRevB.106.L081116} {\bibfield  {journal}
  {\bibinfo  {journal} {Phys. Rev. B}\ }\textbf {\bibinfo {volume} {106}},\
  \bibinfo {pages} {L081116} (\bibinfo {year} {2022})}\BibitemShut {NoStop}%
\bibitem [{\citenamefont {Yao}\ \emph {et~al.}(2022)\citenamefont {Yao},
  \citenamefont {Iida}, \citenamefont {Kamazawa},\ and\ \citenamefont
  {Li}}]{yao22}%
  \BibitemOpen
  \bibfield  {author} {\bibinfo {author} {\bibfnamefont {W.}~\bibnamefont
  {Yao}}, \bibinfo {author} {\bibfnamefont {K.}~\bibnamefont {Iida}}, \bibinfo
  {author} {\bibfnamefont {K.}~\bibnamefont {Kamazawa}},\ and\ \bibinfo
  {author} {\bibfnamefont {Y.}~\bibnamefont {Li}},\ }\bibfield  {title}
  {\bibinfo {title} {Excitations in the Ordered and Paramagnetic States of
  Honeycomb Magnet ${\mathrm{Na}}_{2}{\mathrm{Co}}_{2}{\mathrm{TeO}}_{6}$},\
  }\href {https://doi.org/10.1103/PhysRevLett.129.147202} {\bibfield  {journal}
  {\bibinfo  {journal} {Phys. Rev. Lett.}\ }\textbf {\bibinfo {volume} {129}},\
  \bibinfo {pages} {147202} (\bibinfo {year} {2022})}\BibitemShut {NoStop}%
\bibitem [{\citenamefont {Li}\ \emph {et~al.}(2022)\citenamefont {Li},
  \citenamefont {Gu}, \citenamefont {Chen}, \citenamefont {Garlea},
  \citenamefont {Iida}, \citenamefont {Kamazawa}, \citenamefont {Li},
  \citenamefont {Deng}, \citenamefont {Xiao}, \citenamefont {Zheng},
  \citenamefont {Ye}, \citenamefont {Peng}, \citenamefont {Zaliznyak},
  \citenamefont {Tranquada},\ and\ \citenamefont {Li}}]{li22}%
  \BibitemOpen
  \bibfield  {author} {\bibinfo {author} {\bibfnamefont {X.}~\bibnamefont
  {Li}}, \bibinfo {author} {\bibfnamefont {Y.}~\bibnamefont {Gu}}, \bibinfo
  {author} {\bibfnamefont {Y.}~\bibnamefont {Chen}}, \bibinfo {author}
  {\bibfnamefont {V.~O.}\ \bibnamefont {Garlea}}, \bibinfo {author}
  {\bibfnamefont {K.}~\bibnamefont {Iida}}, \bibinfo {author} {\bibfnamefont
  {K.}~\bibnamefont {Kamazawa}}, \bibinfo {author} {\bibfnamefont
  {Y.}~\bibnamefont {Li}}, \bibinfo {author} {\bibfnamefont {G.}~\bibnamefont
  {Deng}}, \bibinfo {author} {\bibfnamefont {Q.}~\bibnamefont {Xiao}}, \bibinfo
  {author} {\bibfnamefont {X.}~\bibnamefont {Zheng}}, \bibinfo {author}
  {\bibfnamefont {Z.}~\bibnamefont {Ye}}, \bibinfo {author} {\bibfnamefont
  {Y.}~\bibnamefont {Peng}}, \bibinfo {author} {\bibfnamefont {I.~A.}\
  \bibnamefont {Zaliznyak}}, \bibinfo {author} {\bibfnamefont {J.~M.}\
  \bibnamefont {Tranquada}},\ and\ \bibinfo {author} {\bibfnamefont
  {Y.}~\bibnamefont {Li}},\ }\bibfield  {title} {\bibinfo {title} {Giant
  Magnetic In-Plane Anisotropy and Competing Instabilities in
  ${\mathrm{Na}}_{3}{\mathrm{Co}}_{2}{\mathrm{SbO}}_{6}$},\ }\href
  {https://doi.org/10.1103/PhysRevX.12.041024} {\bibfield  {journal} {\bibinfo
  {journal} {Phys. Rev. X}\ }\textbf {\bibinfo {volume} {12}},\ \bibinfo
  {pages} {041024} (\bibinfo {year} {2022})}\BibitemShut {NoStop}%
\bibitem [{\citenamefont {Gu}\ \emph {et~al.}(2023)\citenamefont {Gu},
  \citenamefont {Li}, \citenamefont {Chen}, \citenamefont {Iida}, \citenamefont
  {Nakao}, \citenamefont {Munakata}, \citenamefont {Garlea}, \citenamefont
  {Li}, \citenamefont {Deng}, \citenamefont {Zaliznyak}, \citenamefont
  {Tranquada},\ and\ \citenamefont {Li}}]{gu23}%
  \BibitemOpen
  \bibfield  {author} {\bibinfo {author} {\bibfnamefont {Y.}~\bibnamefont
  {Gu}}, \bibinfo {author} {\bibfnamefont {X.}~\bibnamefont {Li}}, \bibinfo
  {author} {\bibfnamefont {Y.}~\bibnamefont {Chen}}, \bibinfo {author}
  {\bibfnamefont {K.}~\bibnamefont {Iida}}, \bibinfo {author} {\bibfnamefont
  {A.}~\bibnamefont {Nakao}}, \bibinfo {author} {\bibfnamefont
  {K.}~\bibnamefont {Munakata}}, \bibinfo {author} {\bibfnamefont {V.~O.}\
  \bibnamefont {Garlea}}, \bibinfo {author} {\bibfnamefont {Y.}~\bibnamefont
  {Li}}, \bibinfo {author} {\bibfnamefont {G.}~\bibnamefont {Deng}}, \bibinfo
  {author} {\bibfnamefont {I.~A.}\ \bibnamefont {Zaliznyak}}, \bibinfo {author}
  {\bibfnamefont {J.~M.}\ \bibnamefont {Tranquada}},\ and\ \bibinfo {author}
  {\bibfnamefont {Y.}~\bibnamefont {Li}},\ }\bibinfo {title} {Easy-plane
  multi-$\mathbf{q}$ magnetic ground state of Na$_3$Co$_2$SbO$_6$},\ \Eprint
  {https://arxiv.org/abs/2306.07175} {arXiv:2306.07175} \BibitemShut {NoStop}%
\bibitem [{\citenamefont {Zhong}\ \emph {et~al.}(2020)\citenamefont {Zhong},
  \citenamefont {Gao}, \citenamefont {Ong},\ and\ \citenamefont
  {Cava}}]{zhong20}%
  \BibitemOpen
  \bibfield  {author} {\bibinfo {author} {\bibfnamefont {R.}~\bibnamefont
  {Zhong}}, \bibinfo {author} {\bibfnamefont {T.}~\bibnamefont {Gao}}, \bibinfo
  {author} {\bibfnamefont {N.~P.}\ \bibnamefont {Ong}},\ and\ \bibinfo {author}
  {\bibfnamefont {R.~J.}\ \bibnamefont {Cava}},\ }\bibfield  {title} {\bibinfo
  {title} {Weak-field induced nonmagnetic state in a Co-based honeycomb},\
  }\href {https://doi.org/10.1126/sciadv.aay6953} {\bibfield  {journal}
  {\bibinfo  {journal} {Sci. Adv.}\ }\textbf {\bibinfo {volume} {6}},\ \bibinfo
  {pages} {eaay6953} (\bibinfo {year} {2020})}\BibitemShut {NoStop}%
\bibitem [{\citenamefont {Shi}\ \emph {et~al.}(2021)\citenamefont {Shi},
  \citenamefont {Wang}, \citenamefont {Zhong}, \citenamefont {Wang},
  \citenamefont {Hu}, \citenamefont {Zhang}, \citenamefont {Liu}, \citenamefont
  {Dong}, \citenamefont {Wang},\ and\ \citenamefont {Wang}}]{shi21}%
  \BibitemOpen
  \bibfield  {author} {\bibinfo {author} {\bibfnamefont {L.~Y.}\ \bibnamefont
  {Shi}}, \bibinfo {author} {\bibfnamefont {X.~M.}\ \bibnamefont {Wang}},
  \bibinfo {author} {\bibfnamefont {R.~D.}\ \bibnamefont {Zhong}}, \bibinfo
  {author} {\bibfnamefont {Z.~X.}\ \bibnamefont {Wang}}, \bibinfo {author}
  {\bibfnamefont {T.~C.}\ \bibnamefont {Hu}}, \bibinfo {author} {\bibfnamefont
  {S.~J.}\ \bibnamefont {Zhang}}, \bibinfo {author} {\bibfnamefont {Q.~M.}\
  \bibnamefont {Liu}}, \bibinfo {author} {\bibfnamefont {T.}~\bibnamefont
  {Dong}}, \bibinfo {author} {\bibfnamefont {F.}~\bibnamefont {Wang}},\ and\
  \bibinfo {author} {\bibfnamefont {N.~L.}\ \bibnamefont {Wang}},\ }\bibfield
  {title} {\bibinfo {title} {Magnetic excitations of the field-induced states
  in ${\mathrm{BaCo}}_{2}{({\mathrm{AsO}}_{4})}_{2}$ probed by time-domain
  terahertz spectroscopy},\ }\href
  {https://doi.org/10.1103/PhysRevB.104.144408} {\bibfield  {journal} {\bibinfo
   {journal} {Phys. Rev. B}\ }\textbf {\bibinfo {volume} {104}},\ \bibinfo
  {pages} {144408} (\bibinfo {year} {2021})}\BibitemShut {NoStop}%
\bibitem [{\citenamefont {Zhang}\ \emph
  {et~al.}(2023{\natexlab{a}})\citenamefont {Zhang}, \citenamefont {Xu},
  \citenamefont {Halloran}, \citenamefont {Zhong}, \citenamefont {Broholm},
  \citenamefont {Cava}, \citenamefont {Drichko},\ and\ \citenamefont
  {Armitage}}]{zhang22}%
  \BibitemOpen
  \bibfield  {author} {\bibinfo {author} {\bibfnamefont {X.}~\bibnamefont
  {Zhang}}, \bibinfo {author} {\bibfnamefont {Y.}~\bibnamefont {Xu}}, \bibinfo
  {author} {\bibfnamefont {T.}~\bibnamefont {Halloran}}, \bibinfo {author}
  {\bibfnamefont {R.}~\bibnamefont {Zhong}}, \bibinfo {author} {\bibfnamefont
  {C.}~\bibnamefont {Broholm}}, \bibinfo {author} {\bibfnamefont {R.~J.}\
  \bibnamefont {Cava}}, \bibinfo {author} {\bibfnamefont {N.}~\bibnamefont
  {Drichko}},\ and\ \bibinfo {author} {\bibfnamefont {N.~P.}\ \bibnamefont
  {Armitage}},\ }\bibfield  {title} {\bibinfo {title} {A magnetic continuum in
  the cobalt-based honeycomb magnet BaCo$_2$(AsO$_4$)$_2$},\ }\href
  {https://doi.org/10.1038/s41563-022-01403-1} {\bibfield  {journal} {\bibinfo
  {journal} {Nat. Mat.}\ }\textbf {\bibinfo {volume} {22}},\ \bibinfo {pages}
  {58} (\bibinfo {year} {2023}{\natexlab{a}})}\BibitemShut {NoStop}%
\bibitem [{\citenamefont {Maksimov}\ \emph {et~al.}(2022)\citenamefont
  {Maksimov}, \citenamefont {Ushakov}, \citenamefont {Pchelkina}, \citenamefont
  {Li}, \citenamefont {Winter},\ and\ \citenamefont {Streltsov}}]{maksimov22b}%
  \BibitemOpen
  \bibfield  {author} {\bibinfo {author} {\bibfnamefont {P.~A.}\ \bibnamefont
  {Maksimov}}, \bibinfo {author} {\bibfnamefont {A.~V.}\ \bibnamefont
  {Ushakov}}, \bibinfo {author} {\bibfnamefont {Z.~V.}\ \bibnamefont
  {Pchelkina}}, \bibinfo {author} {\bibfnamefont {Y.}~\bibnamefont {Li}},
  \bibinfo {author} {\bibfnamefont {S.~M.}\ \bibnamefont {Winter}},\ and\
  \bibinfo {author} {\bibfnamefont {S.~V.}\ \bibnamefont {Streltsov}},\
  }\bibfield  {title} {\bibinfo {title} {Ab initio guided minimal model for the
  ``Kitaev'' material ${\mathrm{BaCo}}_{2}$(${\mathrm{AsO}}_{4}{)}_{2}$:
  Importance of direct hopping, third-neighbor exchange, and quantum
  fluctuations},\ }\href {https://doi.org/10.1103/PhysRevB.106.165131}
  {\bibfield  {journal} {\bibinfo  {journal} {Phys. Rev. B}\ }\textbf {\bibinfo
  {volume} {106}},\ \bibinfo {pages} {165131} (\bibinfo {year}
  {2022})}\BibitemShut {NoStop}%
\bibitem [{\citenamefont {Kr\"uger}\ \emph {et~al.}(2023)\citenamefont
  {Kr\"uger}, \citenamefont {Chen}, \citenamefont {Jin}, \citenamefont {Li},\
  and\ \citenamefont {Janssen}}]{krueger22}%
  \BibitemOpen
  \bibfield  {author} {\bibinfo {author} {\bibfnamefont {W.~G.~F.}\
  \bibnamefont {Kr\"uger}}, \bibinfo {author} {\bibfnamefont {W.}~\bibnamefont
  {Chen}}, \bibinfo {author} {\bibfnamefont {X.}~\bibnamefont {Jin}}, \bibinfo
  {author} {\bibfnamefont {Y.}~\bibnamefont {Li}},\ and\ \bibinfo {author}
  {\bibfnamefont {L.}~\bibnamefont {Janssen}},\ }\bibfield  {title} {\bibinfo
  {title} {Triple-q Order in
  ${\mathrm{Na}}_{2}{\mathrm{Co}}_{2}{\mathrm{TeO}}_{6}$ from Proximity to
  Hidden-SU(2)-Symmetric Point},\ }\href
  {https://doi.org/10.1103/PhysRevLett.131.146702} {\bibfield  {journal}
  {\bibinfo  {journal} {Phys. Rev. Lett.}\ }\textbf {\bibinfo {volume} {131}},\
  \bibinfo {pages} {146702} (\bibinfo {year} {2023})}\BibitemShut {NoStop}%
\bibitem [{\citenamefont {Yao}\ \emph {et~al.}(2023)\citenamefont {Yao},
  \citenamefont {Zhao}, \citenamefont {Qiu}, \citenamefont {Balz},
  \citenamefont {Stewart}, \citenamefont {Lynn},\ and\ \citenamefont
  {Li}}]{yao23}%
  \BibitemOpen
  \bibfield  {author} {\bibinfo {author} {\bibfnamefont {W.}~\bibnamefont
  {Yao}}, \bibinfo {author} {\bibfnamefont {Y.}~\bibnamefont {Zhao}}, \bibinfo
  {author} {\bibfnamefont {Y.}~\bibnamefont {Qiu}}, \bibinfo {author}
  {\bibfnamefont {C.}~\bibnamefont {Balz}}, \bibinfo {author} {\bibfnamefont
  {J.~R.}\ \bibnamefont {Stewart}}, \bibinfo {author} {\bibfnamefont {J.~W.}\
  \bibnamefont {Lynn}},\ and\ \bibinfo {author} {\bibfnamefont
  {Y.}~\bibnamefont {Li}},\ }\bibfield  {title} {\bibinfo {title} {Magnetic
  ground state of the Kitaev
  ${\mathrm{Na}}_{2}{\mathrm{Co}}_{2}{\mathrm{TeO}}_{6}$ spin liquid
  candidate},\ }\href {https://doi.org/10.1103/PhysRevResearch.5.L022045}
  {\bibfield  {journal} {\bibinfo  {journal} {Phys. Rev. Res.}\ }\textbf
  {\bibinfo {volume} {5}},\ \bibinfo {pages} {L022045} (\bibinfo {year}
  {2023})}\BibitemShut {NoStop}%
\bibitem [{\citenamefont {Xiang}\ \emph {et~al.}(2023)\citenamefont {Xiang},
  \citenamefont {Dhakal}, \citenamefont {Ozerov}, \citenamefont {Jiang},
  \citenamefont {Mou}, \citenamefont {Ozarowski}, \citenamefont {Huang},
  \citenamefont {Zhou}, \citenamefont {Fang}, \citenamefont {Winter},
  \citenamefont {Jiang},\ and\ \citenamefont {Smirnov}}]{xiang23}%
  \BibitemOpen
  \bibfield  {author} {\bibinfo {author} {\bibfnamefont {L.}~\bibnamefont
  {Xiang}}, \bibinfo {author} {\bibfnamefont {R.}~\bibnamefont {Dhakal}},
  \bibinfo {author} {\bibfnamefont {M.}~\bibnamefont {Ozerov}}, \bibinfo
  {author} {\bibfnamefont {Y.}~\bibnamefont {Jiang}}, \bibinfo {author}
  {\bibfnamefont {B.~S.}\ \bibnamefont {Mou}}, \bibinfo {author} {\bibfnamefont
  {A.}~\bibnamefont {Ozarowski}}, \bibinfo {author} {\bibfnamefont
  {Q.}~\bibnamefont {Huang}}, \bibinfo {author} {\bibfnamefont
  {H.}~\bibnamefont {Zhou}}, \bibinfo {author} {\bibfnamefont {J.}~\bibnamefont
  {Fang}}, \bibinfo {author} {\bibfnamefont {S.~M.}\ \bibnamefont {Winter}},
  \bibinfo {author} {\bibfnamefont {Z.}~\bibnamefont {Jiang}},\ and\ \bibinfo
  {author} {\bibfnamefont {D.}~\bibnamefont {Smirnov}},\ }\bibfield  {title}
  {\bibinfo {title} {Disorder-Enriched Magnetic Excitations in a
  Heisenberg-Kitaev Quantum Magnet
  ${\mathrm{Na}}_{2}{\mathrm{Co}}_{2}{\mathrm{TeO}}_{6}$},\ }\href
  {https://doi.org/10.1103/PhysRevLett.131.076701} {\bibfield  {journal}
  {\bibinfo  {journal} {Phys. Rev. Lett.}\ }\textbf {\bibinfo {volume} {131}},\
  \bibinfo {pages} {076701} (\bibinfo {year} {2023})}\BibitemShut {NoStop}%
\bibitem [{\citenamefont {Zhang}\ \emph
  {et~al.}(2023{\natexlab{b}})\citenamefont {Zhang}, \citenamefont {Lee},
  \citenamefont {Woods}, \citenamefont {Peria}, \citenamefont {Thomas},
  \citenamefont {Movshovich}, \citenamefont {Brosha}, \citenamefont {Huang},
  \citenamefont {Zhou}, \citenamefont {Zapf},\ and\ \citenamefont
  {Lee}}]{zhang23}%
  \BibitemOpen
  \bibfield  {author} {\bibinfo {author} {\bibfnamefont {S.}~\bibnamefont
  {Zhang}}, \bibinfo {author} {\bibfnamefont {S.}~\bibnamefont {Lee}}, \bibinfo
  {author} {\bibfnamefont {A.~J.}\ \bibnamefont {Woods}}, \bibinfo {author}
  {\bibfnamefont {W.~K.}\ \bibnamefont {Peria}}, \bibinfo {author}
  {\bibfnamefont {S.~M.}\ \bibnamefont {Thomas}}, \bibinfo {author}
  {\bibfnamefont {R.}~\bibnamefont {Movshovich}}, \bibinfo {author}
  {\bibfnamefont {E.}~\bibnamefont {Brosha}}, \bibinfo {author} {\bibfnamefont
  {Q.}~\bibnamefont {Huang}}, \bibinfo {author} {\bibfnamefont
  {H.}~\bibnamefont {Zhou}}, \bibinfo {author} {\bibfnamefont {V.~S.}\
  \bibnamefont {Zapf}},\ and\ \bibinfo {author} {\bibfnamefont
  {M.}~\bibnamefont {Lee}},\ }\bibfield  {title} {\bibinfo {title} {Electronic
  and magnetic phase diagrams of the Kitaev quantum spin liquid candidate
  ${\mathrm{Na}}_{2}{\mathrm{Co}}_{2}{\mathrm{TeO}}_{6}$},\ }\href
  {https://doi.org/10.1103/PhysRevB.108.064421} {\bibfield  {journal} {\bibinfo
   {journal} {Phys. Rev. B}\ }\textbf {\bibinfo {volume} {108}},\ \bibinfo
  {pages} {064421} (\bibinfo {year} {2023}{\natexlab{b}})}\BibitemShut
  {NoStop}%
\bibitem [{\citenamefont {Hong}\ \emph {et~al.}(2023)\citenamefont {Hong},
  \citenamefont {Gillig}, \citenamefont {Yao}, \citenamefont {Janssen},
  \citenamefont {Kocsis}, \citenamefont {Gass}, \citenamefont {Li},
  \citenamefont {Wolter}, \citenamefont {Büchner},\ and\ \citenamefont
  {Hess}}]{hong23}%
  \BibitemOpen
  \bibfield  {author} {\bibinfo {author} {\bibfnamefont {X.}~\bibnamefont
  {Hong}}, \bibinfo {author} {\bibfnamefont {M.}~\bibnamefont {Gillig}},
  \bibinfo {author} {\bibfnamefont {W.}~\bibnamefont {Yao}}, \bibinfo {author}
  {\bibfnamefont {L.}~\bibnamefont {Janssen}}, \bibinfo {author} {\bibfnamefont
  {V.}~\bibnamefont {Kocsis}}, \bibinfo {author} {\bibfnamefont
  {S.}~\bibnamefont {Gass}}, \bibinfo {author} {\bibfnamefont {Y.}~\bibnamefont
  {Li}}, \bibinfo {author} {\bibfnamefont {A.~U.~B.}\ \bibnamefont {Wolter}},
  \bibinfo {author} {\bibfnamefont {B.}~\bibnamefont {Büchner}},\ and\
  \bibinfo {author} {\bibfnamefont {C.}~\bibnamefont {Hess}},\ }\bibinfo
  {title} {Phonon thermal transport shaped by strong spin-phonon scattering in
  a Kitaev material Na$_2$Co$_2$TeO$_6$},\ \Eprint
  {https://arxiv.org/abs/2306.16963} {arXiv:2306.16963} \BibitemShut {NoStop}%
\bibitem [{\citenamefont {Regnault}\ \emph {et~al.}(1977)\citenamefont
  {Regnault}, \citenamefont {Burlet},\ and\ \citenamefont
  {Rossat-Mignod}}]{regnault77}%
  \BibitemOpen
  \bibfield  {author} {\bibinfo {author} {\bibfnamefont {L.}~\bibnamefont
  {Regnault}}, \bibinfo {author} {\bibfnamefont {P.}~\bibnamefont {Burlet}},\
  and\ \bibinfo {author} {\bibfnamefont {J.}~\bibnamefont {Rossat-Mignod}},\
  }\bibfield  {title} {\bibinfo {title} {Magnetic ordering in a planar X-Y
  model: BaCo$_2$(AsO$_4$)$_2$},\ }\href
  {https://doi.org/https://doi.org/10.1016/0378-4363(77)90635-0} {\bibfield
  {journal} {\bibinfo  {journal} {Physica B+C}\ }\textbf {\bibinfo {volume}
  {86-88}},\ \bibinfo {pages} {660} (\bibinfo {year} {1977})}\BibitemShut
  {NoStop}%
\bibitem [{\citenamefont {Regnault}\ \emph {et~al.}(2018)\citenamefont
  {Regnault}, \citenamefont {Boullier},\ and\ \citenamefont
  {Lorenzo}}]{regnault18}%
  \BibitemOpen
  \bibfield  {author} {\bibinfo {author} {\bibfnamefont {L.-P.}\ \bibnamefont
  {Regnault}}, \bibinfo {author} {\bibfnamefont {C.}~\bibnamefont {Boullier}},\
  and\ \bibinfo {author} {\bibfnamefont {J.}~\bibnamefont {Lorenzo}},\
  }\bibfield  {title} {\bibinfo {title} {Polarized-neutron investigation of
  magnetic ordering and spin dynamics in BaCo$_2$(AsO$_4$)$_2$ frustrated
  honeycomb-lattice magnet},\ }\href
  {https://doi.org/https://doi.org/10.1016/j.heliyon.2018.e00507} {\bibfield
  {journal} {\bibinfo  {journal} {Heliyon}\ }\textbf {\bibinfo {volume} {4}},\
  \bibinfo {pages} {e00507} (\bibinfo {year} {2018})}\BibitemShut {NoStop}%
\bibitem [{\citenamefont {Chaloupka}\ \emph {et~al.}(2010)\citenamefont
  {Chaloupka}, \citenamefont {Jackeli},\ and\ \citenamefont
  {Khaliullin}}]{chaloupka10}%
  \BibitemOpen
  \bibfield  {author} {\bibinfo {author} {\bibfnamefont {J.}~\bibnamefont
  {Chaloupka}}, \bibinfo {author} {\bibfnamefont {G.}~\bibnamefont {Jackeli}},\
  and\ \bibinfo {author} {\bibfnamefont {G.}~\bibnamefont {Khaliullin}},\
  }\bibfield  {title} {\bibinfo {title} {Kitaev-Heisenberg Model on a Honeycomb
  Lattice: Possible Exotic Phases in Iridium Oxides
  ${A}_{2}{\mathrm{IrO}}_{3}$},\ }\href
  {https://doi.org/10.1103/PhysRevLett.105.027204} {\bibfield  {journal}
  {\bibinfo  {journal} {Phys. Rev. Lett.}\ }\textbf {\bibinfo {volume} {105}},\
  \bibinfo {pages} {027204} (\bibinfo {year} {2010})}\BibitemShut {NoStop}%
\bibitem [{\citenamefont {Chaloupka}\ \emph {et~al.}(2013)\citenamefont
  {Chaloupka}, \citenamefont {Jackeli},\ and\ \citenamefont
  {Khaliullin}}]{chaloupka13}%
  \BibitemOpen
  \bibfield  {author} {\bibinfo {author} {\bibfnamefont {J.}~\bibnamefont
  {Chaloupka}}, \bibinfo {author} {\bibfnamefont {G.}~\bibnamefont {Jackeli}},\
  and\ \bibinfo {author} {\bibfnamefont {G.}~\bibnamefont {Khaliullin}},\
  }\bibfield  {title} {\bibinfo {title} {Zigzag Magnetic Order in the Iridium
  Oxide ${\mathrm{Na}}_{2}{\mathrm{IrO}}_{3}$},\ }\href
  {https://doi.org/10.1103/PhysRevLett.110.097204} {\bibfield  {journal}
  {\bibinfo  {journal} {Phys. Rev. Lett.}\ }\textbf {\bibinfo {volume} {110}},\
  \bibinfo {pages} {097204} (\bibinfo {year} {2013})}\BibitemShut {NoStop}%
\bibitem [{\citenamefont {Rau}\ \emph {et~al.}(2014)\citenamefont {Rau},
  \citenamefont {Lee},\ and\ \citenamefont {Kee}}]{rau14a}%
  \BibitemOpen
  \bibfield  {author} {\bibinfo {author} {\bibfnamefont {J.~G.}\ \bibnamefont
  {Rau}}, \bibinfo {author} {\bibfnamefont {E.~K.-H.}\ \bibnamefont {Lee}},\
  and\ \bibinfo {author} {\bibfnamefont {H.-Y.}\ \bibnamefont {Kee}},\
  }\bibfield  {title} {\bibinfo {title} {Generic Spin Model for the Honeycomb
  Iridates beyond the Kitaev Limit},\ }\href
  {https://doi.org/10.1103/PhysRevLett.112.077204} {\bibfield  {journal}
  {\bibinfo  {journal} {Phys. Rev. Lett.}\ }\textbf {\bibinfo {volume} {112}},\
  \bibinfo {pages} {077204} (\bibinfo {year} {2014})}\BibitemShut {NoStop}%
\bibitem [{\citenamefont {Rau}\ and\ \citenamefont {Kee}(2014)}]{rau14b}%
  \BibitemOpen
  \bibfield  {author} {\bibinfo {author} {\bibfnamefont {J.~G.}\ \bibnamefont
  {Rau}}\ and\ \bibinfo {author} {\bibfnamefont {H.-Y.}\ \bibnamefont {Kee}},\
  }\bibinfo {title} {Trigonal distortion in the honeycomb iridates: Proximity
  of zigzag and spiral phases in Na$_2$IrO$_3$},\ \Eprint
  {https://arxiv.org/abs/1408.4811} {arXiv:1408.4811} \BibitemShut {NoStop}%
\bibitem [{\citenamefont {Janssen}\ \emph {et~al.}(2016)\citenamefont
  {Janssen}, \citenamefont {Andrade},\ and\ \citenamefont {Vojta}}]{janssen16}%
  \BibitemOpen
  \bibfield  {author} {\bibinfo {author} {\bibfnamefont {L.}~\bibnamefont
  {Janssen}}, \bibinfo {author} {\bibfnamefont {E.~C.}\ \bibnamefont
  {Andrade}},\ and\ \bibinfo {author} {\bibfnamefont {M.}~\bibnamefont
  {Vojta}},\ }\bibfield  {title} {\bibinfo {title} {Honeycomb-Lattice
  Heisenberg-Kitaev Model in a Magnetic Field: Spin Canting, Metamagnetism, and
  Vortex Crystals},\ }\href {https://doi.org/10.1103/PhysRevLett.117.277202}
  {\bibfield  {journal} {\bibinfo  {journal} {Phys. Rev. Lett.}\ }\textbf
  {\bibinfo {volume} {117}},\ \bibinfo {pages} {277202} (\bibinfo {year}
  {2016})}\BibitemShut {NoStop}%
\bibitem [{\citenamefont {Chaloupka}\ and\ \citenamefont
  {Khaliullin}(2015)}]{chaloupka15}%
  \BibitemOpen
  \bibfield  {author} {\bibinfo {author} {\bibfnamefont {J.}~\bibnamefont
  {Chaloupka}}\ and\ \bibinfo {author} {\bibfnamefont {G.}~\bibnamefont
  {Khaliullin}},\ }\bibfield  {title} {\bibinfo {title} {Hidden symmetries of
  the extended Kitaev-Heisenberg model: Implications for the honeycomb-lattice
  iridates ${A}_{2}{\mathrm{IrO}}_{3}$},\ }\href
  {https://doi.org/10.1103/PhysRevB.92.024413} {\bibfield  {journal} {\bibinfo
  {journal} {Phys. Rev. B}\ }\textbf {\bibinfo {volume} {92}},\ \bibinfo
  {pages} {024413} (\bibinfo {year} {2015})}\BibitemShut {NoStop}%
\bibitem [{\citenamefont {Mermin}\ and\ \citenamefont
  {Wagner}(1966)}]{mermin66}%
  \BibitemOpen
  \bibfield  {author} {\bibinfo {author} {\bibfnamefont {N.~D.}\ \bibnamefont
  {Mermin}}\ and\ \bibinfo {author} {\bibfnamefont {H.}~\bibnamefont
  {Wagner}},\ }\bibfield  {title} {\bibinfo {title} {Absence of Ferromagnetism
  or Antiferromagnetism in One- or Two-Dimensional Isotropic Heisenberg
  Models},\ }\href {https://doi.org/10.1103/PhysRevLett.17.1133} {\bibfield
  {journal} {\bibinfo  {journal} {Phys. Rev. Lett.}\ }\textbf {\bibinfo
  {volume} {17}},\ \bibinfo {pages} {1133} (\bibinfo {year}
  {1966})}\BibitemShut {NoStop}%
\bibitem [{\citenamefont {Yang}\ \emph {et~al.}(2012)\citenamefont {Yang},
  \citenamefont {Albuquerque}, \citenamefont {Capponi}, \citenamefont
  {L{\"a}uchli},\ and\ \citenamefont {Schmidt}}]{yang12}%
  \BibitemOpen
  \bibfield  {author} {\bibinfo {author} {\bibfnamefont {H.-Y.}\ \bibnamefont
  {Yang}}, \bibinfo {author} {\bibfnamefont {A.~F.}\ \bibnamefont
  {Albuquerque}}, \bibinfo {author} {\bibfnamefont {S.}~\bibnamefont
  {Capponi}}, \bibinfo {author} {\bibfnamefont {A.~M.}\ \bibnamefont
  {L{\"a}uchli}},\ and\ \bibinfo {author} {\bibfnamefont {K.~P.}\ \bibnamefont
  {Schmidt}},\ }\bibfield  {title} {\bibinfo {title} {Effective spin couplings
  in the Mott insulator of the honeycomb lattice Hubbard model},\ }\href
  {https://doi.org/10.1088/1367-2630/14/11/115027} {\bibfield  {journal}
  {\bibinfo  {journal} {New J. Phys.}\ }\textbf {\bibinfo {volume} {14}},\
  \bibinfo {pages} {115027} (\bibinfo {year} {2012})}\BibitemShut {NoStop}%
\bibitem [{\citenamefont {Fernandes}\ \emph {et~al.}(2019)\citenamefont
  {Fernandes}, \citenamefont {Orth},\ and\ \citenamefont
  {Schmalian}}]{fernandes19}%
  \BibitemOpen
  \bibfield  {author} {\bibinfo {author} {\bibfnamefont {R.~M.}\ \bibnamefont
  {Fernandes}}, \bibinfo {author} {\bibfnamefont {P.~P.}\ \bibnamefont
  {Orth}},\ and\ \bibinfo {author} {\bibfnamefont {J.}~\bibnamefont
  {Schmalian}},\ }\bibfield  {title} {\bibinfo {title} {Intertwined Vestigial
  Order in Quantum Materials: Nematicity and Beyond},\ }\href
  {https://doi.org/10.1146/annurev-conmatphys-031218-013200} {\bibfield
  {journal} {\bibinfo  {journal} {Annu. Rev. Condens. Matter Phys.}\ }\textbf
  {\bibinfo {volume} {10}},\ \bibinfo {pages} {133} (\bibinfo {year}
  {2019})}\BibitemShut {NoStop}%
\bibitem [{\citenamefont {Kvashnin}\ \emph {et~al.}(2020)\citenamefont
  {Kvashnin}, \citenamefont {Bergman}, \citenamefont {Lichtenstein},\ and\
  \citenamefont {Katsnelson}}]{kvashnin20}%
  \BibitemOpen
  \bibfield  {author} {\bibinfo {author} {\bibfnamefont {Y.~O.}\ \bibnamefont
  {Kvashnin}}, \bibinfo {author} {\bibfnamefont {A.}~\bibnamefont {Bergman}},
  \bibinfo {author} {\bibfnamefont {A.~I.}\ \bibnamefont {Lichtenstein}},\ and\
  \bibinfo {author} {\bibfnamefont {M.~I.}\ \bibnamefont {Katsnelson}},\
  }\bibfield  {title} {\bibinfo {title} {Relativistic exchange interactions in
  $\mathrm{Cr}{X}_{3}$ ($X=\mathrm{Cl}$, Br, I) monolayers},\ }\href
  {https://doi.org/10.1103/PhysRevB.102.115162} {\bibfield  {journal} {\bibinfo
   {journal} {Phys. Rev. B}\ }\textbf {\bibinfo {volume} {102}},\ \bibinfo
  {pages} {115162} (\bibinfo {year} {2020})}\BibitemShut {NoStop}%
\bibitem [{\citenamefont {Fedorova}\ \emph {et~al.}(2015)\citenamefont
  {Fedorova}, \citenamefont {Ederer}, \citenamefont {Spaldin},\ and\
  \citenamefont {Scaramucci}}]{fedorova15}%
  \BibitemOpen
  \bibfield  {author} {\bibinfo {author} {\bibfnamefont {N.~S.}\ \bibnamefont
  {Fedorova}}, \bibinfo {author} {\bibfnamefont {C.}~\bibnamefont {Ederer}},
  \bibinfo {author} {\bibfnamefont {N.~A.}\ \bibnamefont {Spaldin}},\ and\
  \bibinfo {author} {\bibfnamefont {A.}~\bibnamefont {Scaramucci}},\ }\bibfield
   {title} {\bibinfo {title} {Biquadratic and ring exchange interactions in
  orthorhombic perovskite manganites},\ }\href
  {https://doi.org/10.1103/PhysRevB.91.165122} {\bibfield  {journal} {\bibinfo
  {journal} {Phys. Rev. B}\ }\textbf {\bibinfo {volume} {91}},\ \bibinfo
  {pages} {165122} (\bibinfo {year} {2015})}\BibitemShut {NoStop}%
\bibitem [{\citenamefont {Dalla~Piazza}\ \emph {et~al.}(2012)\citenamefont
  {Dalla~Piazza}, \citenamefont {Mourigal}, \citenamefont {Guarise},
  \citenamefont {Berger}, \citenamefont {Schmitt}, \citenamefont {Zhou},
  \citenamefont {Grioni},\ and\ \citenamefont {R\o{}nnow}}]{dallapiazza12}%
  \BibitemOpen
  \bibfield  {author} {\bibinfo {author} {\bibfnamefont {B.}~\bibnamefont
  {Dalla~Piazza}}, \bibinfo {author} {\bibfnamefont {M.}~\bibnamefont
  {Mourigal}}, \bibinfo {author} {\bibfnamefont {M.}~\bibnamefont {Guarise}},
  \bibinfo {author} {\bibfnamefont {H.}~\bibnamefont {Berger}}, \bibinfo
  {author} {\bibfnamefont {T.}~\bibnamefont {Schmitt}}, \bibinfo {author}
  {\bibfnamefont {K.~J.}\ \bibnamefont {Zhou}}, \bibinfo {author}
  {\bibfnamefont {M.}~\bibnamefont {Grioni}},\ and\ \bibinfo {author}
  {\bibfnamefont {H.~M.}\ \bibnamefont {R\o{}nnow}},\ }\bibfield  {title}
  {\bibinfo {title} {Unified one-band Hubbard model for magnetic and electronic
  spectra of the parent compounds of cuprate superconductors},\ }\href
  {https://doi.org/10.1103/PhysRevB.85.100508} {\bibfield  {journal} {\bibinfo
  {journal} {Phys. Rev. B}\ }\textbf {\bibinfo {volume} {85}},\ \bibinfo
  {pages} {100508} (\bibinfo {year} {2012})}\BibitemShut {NoStop}%
\bibitem [{\citenamefont {Larsen}\ \emph {et~al.}(2019)\citenamefont {Larsen},
  \citenamefont {R\o{}mer}, \citenamefont {Janas}, \citenamefont {Treue},
  \citenamefont {M\o{}nsted}, \citenamefont {Shaik}, \citenamefont
  {R\o{}nnow},\ and\ \citenamefont {Lefmann}}]{larsen19}%
  \BibitemOpen
  \bibfield  {author} {\bibinfo {author} {\bibfnamefont {C.~B.}\ \bibnamefont
  {Larsen}}, \bibinfo {author} {\bibfnamefont {A.~T.}\ \bibnamefont
  {R\o{}mer}}, \bibinfo {author} {\bibfnamefont {S.}~\bibnamefont {Janas}},
  \bibinfo {author} {\bibfnamefont {F.}~\bibnamefont {Treue}}, \bibinfo
  {author} {\bibfnamefont {B.}~\bibnamefont {M\o{}nsted}}, \bibinfo {author}
  {\bibfnamefont {N.~E.}\ \bibnamefont {Shaik}}, \bibinfo {author}
  {\bibfnamefont {H.~M.}\ \bibnamefont {R\o{}nnow}},\ and\ \bibinfo {author}
  {\bibfnamefont {K.}~\bibnamefont {Lefmann}},\ }\bibfield  {title} {\bibinfo
  {title} {Exact diagonalization study of the Hubbard-parametrized four-spin
  ring exchange model on a square lattice},\ }\href
  {https://doi.org/10.1103/PhysRevB.99.054432} {\bibfield  {journal} {\bibinfo
  {journal} {Phys. Rev. B}\ }\textbf {\bibinfo {volume} {99}},\ \bibinfo
  {pages} {054432} (\bibinfo {year} {2019})}\BibitemShut {NoStop}%
\bibitem [{\citenamefont {Fernandes}\ \emph {et~al.}(2012)\citenamefont
  {Fernandes}, \citenamefont {Chubukov}, \citenamefont {Knolle}, \citenamefont
  {Eremin},\ and\ \citenamefont {Schmalian}}]{fernandes12}%
  \BibitemOpen
  \bibfield  {author} {\bibinfo {author} {\bibfnamefont {R.~M.}\ \bibnamefont
  {Fernandes}}, \bibinfo {author} {\bibfnamefont {A.~V.}\ \bibnamefont
  {Chubukov}}, \bibinfo {author} {\bibfnamefont {J.}~\bibnamefont {Knolle}},
  \bibinfo {author} {\bibfnamefont {I.}~\bibnamefont {Eremin}},\ and\ \bibinfo
  {author} {\bibfnamefont {J.}~\bibnamefont {Schmalian}},\ }\bibfield  {title}
  {\bibinfo {title} {Preemptive nematic order, pseudogap, and orbital order in
  the iron pnictides},\ }\href {https://doi.org/10.1103/PhysRevB.85.024534}
  {\bibfield  {journal} {\bibinfo  {journal} {Phys. Rev. B}\ }\textbf {\bibinfo
  {volume} {85}},\ \bibinfo {pages} {024534} (\bibinfo {year}
  {2012})}\BibitemShut {NoStop}%
\bibitem [{\citenamefont {Shannon}\ \emph {et~al.}(2010)\citenamefont
  {Shannon}, \citenamefont {Penc},\ and\ \citenamefont {Motome}}]{shannon10}%
  \BibitemOpen
  \bibfield  {author} {\bibinfo {author} {\bibfnamefont {N.}~\bibnamefont
  {Shannon}}, \bibinfo {author} {\bibfnamefont {K.}~\bibnamefont {Penc}},\ and\
  \bibinfo {author} {\bibfnamefont {Y.}~\bibnamefont {Motome}},\ }\bibfield
  {title} {\bibinfo {title} {Nematic, vector-multipole, and plateau-liquid
  states in the classical $O(3)$ pyrochlore antiferromagnet with biquadratic
  interactions in applied magnetic field},\ }\href
  {https://doi.org/10.1103/PhysRevB.81.184409} {\bibfield  {journal} {\bibinfo
  {journal} {Phys. Rev. B}\ }\textbf {\bibinfo {volume} {81}},\ \bibinfo
  {pages} {184409} (\bibinfo {year} {2010})}\BibitemShut {NoStop}%
\bibitem [{\citenamefont {Janssen}\ and\ \citenamefont
  {Herbut}(2015)}]{janssen15}%
  \BibitemOpen
  \bibfield  {author} {\bibinfo {author} {\bibfnamefont {L.}~\bibnamefont
  {Janssen}}\ and\ \bibinfo {author} {\bibfnamefont {I.~F.}\ \bibnamefont
  {Herbut}},\ }\bibfield  {title} {\bibinfo {title} {Nematic quantum
  criticality in three-dimensional Fermi system with quadratic band touching},\
  }\href {https://doi.org/10.1103/PhysRevB.92.045117} {\bibfield  {journal}
  {\bibinfo  {journal} {Phys. Rev. B}\ }\textbf {\bibinfo {volume} {92}},\
  \bibinfo {pages} {045117} (\bibinfo {year} {2015})}\BibitemShut {NoStop}%
\bibitem [{\citenamefont {Chern}\ \emph {et~al.}(2012)\citenamefont {Chern},
  \citenamefont {Fernandes}, \citenamefont {Nandkishore},\ and\ \citenamefont
  {Chubukov}}]{chern12}%
  \BibitemOpen
  \bibfield  {author} {\bibinfo {author} {\bibfnamefont {G.-W.}\ \bibnamefont
  {Chern}}, \bibinfo {author} {\bibfnamefont {R.~M.}\ \bibnamefont
  {Fernandes}}, \bibinfo {author} {\bibfnamefont {R.}~\bibnamefont
  {Nandkishore}},\ and\ \bibinfo {author} {\bibfnamefont {A.~V.}\ \bibnamefont
  {Chubukov}},\ }\bibfield  {title} {\bibinfo {title} {Broken translational
  symmetry in an emergent paramagnetic phase of graphene},\ }\href
  {https://doi.org/10.1103/PhysRevB.86.115443} {\bibfield  {journal} {\bibinfo
  {journal} {Phys. Rev. B}\ }\textbf {\bibinfo {volume} {86}},\ \bibinfo
  {pages} {115443} (\bibinfo {year} {2012})}\BibitemShut {NoStop}%
\bibitem [{\citenamefont {Price}\ and\ \citenamefont
  {Perkins}(2012)}]{price12}%
  \BibitemOpen
  \bibfield  {author} {\bibinfo {author} {\bibfnamefont {C.~C.}\ \bibnamefont
  {Price}}\ and\ \bibinfo {author} {\bibfnamefont {N.~B.}\ \bibnamefont
  {Perkins}},\ }\bibfield  {title} {\bibinfo {title} {Critical Properties of
  the Kitaev-Heisenberg Model},\ }\href
  {https://doi.org/10.1103/PhysRevLett.109.187201} {\bibfield  {journal}
  {\bibinfo  {journal} {Phys. Rev. Lett.}\ }\textbf {\bibinfo {volume} {109}},\
  \bibinfo {pages} {187201} (\bibinfo {year} {2012})}\BibitemShut {NoStop}%
\bibitem [{\citenamefont {Price}\ and\ \citenamefont
  {Perkins}(2013)}]{price13}%
  \BibitemOpen
  \bibfield  {author} {\bibinfo {author} {\bibfnamefont {C.}~\bibnamefont
  {Price}}\ and\ \bibinfo {author} {\bibfnamefont {N.~B.}\ \bibnamefont
  {Perkins}},\ }\bibfield  {title} {\bibinfo {title} {Finite-temperature phase
  diagram of the classical Kitaev-Heisenberg model},\ }\href
  {https://doi.org/10.1103/PhysRevB.88.024410} {\bibfield  {journal} {\bibinfo
  {journal} {Phys. Rev. B}\ }\textbf {\bibinfo {volume} {88}},\ \bibinfo
  {pages} {024410} (\bibinfo {year} {2013})}\BibitemShut {NoStop}%
\bibitem [{\citenamefont {Chern}\ \emph {et~al.}(2017)\citenamefont {Chern},
  \citenamefont {Sizyuk}, \citenamefont {Price},\ and\ \citenamefont
  {Perkins}}]{chern17}%
  \BibitemOpen
  \bibfield  {author} {\bibinfo {author} {\bibfnamefont {G.-W.}\ \bibnamefont
  {Chern}}, \bibinfo {author} {\bibfnamefont {Y.}~\bibnamefont {Sizyuk}},
  \bibinfo {author} {\bibfnamefont {C.}~\bibnamefont {Price}},\ and\ \bibinfo
  {author} {\bibfnamefont {N.~B.}\ \bibnamefont {Perkins}},\ }\bibfield
  {title} {\bibinfo {title} {Kitaev-Heisenberg model in a magnetic field:
  Order-by-disorder and commensurate-incommensurate transitions},\ }\href
  {https://doi.org/10.1103/PhysRevB.95.144427} {\bibfield  {journal} {\bibinfo
  {journal} {Phys. Rev. B}\ }\textbf {\bibinfo {volume} {95}},\ \bibinfo
  {pages} {144427} (\bibinfo {year} {2017})}\BibitemShut {NoStop}%
\bibitem [{\citenamefont {Andrade}\ \emph {et~al.}(2020)\citenamefont
  {Andrade}, \citenamefont {Janssen},\ and\ \citenamefont {Vojta}}]{andrade20}%
  \BibitemOpen
  \bibfield  {author} {\bibinfo {author} {\bibfnamefont {E.~C.}\ \bibnamefont
  {Andrade}}, \bibinfo {author} {\bibfnamefont {L.}~\bibnamefont {Janssen}},\
  and\ \bibinfo {author} {\bibfnamefont {M.}~\bibnamefont {Vojta}},\ }\bibfield
   {title} {\bibinfo {title} {Susceptibility anisotropy and its disorder
  evolution in models for Kitaev materials},\ }\href
  {https://doi.org/10.1103/PhysRevB.102.115160} {\bibfield  {journal} {\bibinfo
   {journal} {Phys. Rev. B}\ }\textbf {\bibinfo {volume} {102}},\ \bibinfo
  {pages} {115160} (\bibinfo {year} {2020})}\BibitemShut {NoStop}%
\bibitem [{nhr()}]{nhr-alliance}%
  \BibitemOpen
  \bibinfo {note} {Computations were performed on the TAURUS and BARNARD
  clusters of NHR@TUD,
  \href{https://www.nhr-verein.de}{https://www.nhr-verein.de}}\BibitemShut
  {NoStop}%
\bibitem [{\citenamefont {Nienhuis}\ and\ \citenamefont
  {Nauenberg}(1975)}]{nienhuis75}%
  \BibitemOpen
  \bibfield  {author} {\bibinfo {author} {\bibfnamefont {B.}~\bibnamefont
  {Nienhuis}}\ and\ \bibinfo {author} {\bibfnamefont {M.}~\bibnamefont
  {Nauenberg}},\ }\bibfield  {title} {\bibinfo {title} {First-Order Phase
  Transitions in Renormalization-Group Theory},\ }\href
  {https://doi.org/10.1103/PhysRevLett.35.477} {\bibfield  {journal} {\bibinfo
  {journal} {Phys. Rev. Lett.}\ }\textbf {\bibinfo {volume} {35}},\ \bibinfo
  {pages} {477} (\bibinfo {year} {1975})}\BibitemShut {NoStop}%
\bibitem [{\citenamefont {Fisher}\ and\ \citenamefont
  {Berker}(1982)}]{fisher82}%
  \BibitemOpen
  \bibfield  {author} {\bibinfo {author} {\bibfnamefont {M.~E.}\ \bibnamefont
  {Fisher}}\ and\ \bibinfo {author} {\bibfnamefont {A.~N.}\ \bibnamefont
  {Berker}},\ }\bibfield  {title} {\bibinfo {title} {Scaling for first-order
  phase transitions in thermodynamic and finite systems},\ }\href
  {https://doi.org/10.1103/PhysRevB.26.2507} {\bibfield  {journal} {\bibinfo
  {journal} {Phys. Rev. B}\ }\textbf {\bibinfo {volume} {26}},\ \bibinfo
  {pages} {2507} (\bibinfo {year} {1982})}\BibitemShut {NoStop}%
\bibitem [{\citenamefont {Binder}(1987)}]{binder87}%
  \BibitemOpen
  \bibfield  {author} {\bibinfo {author} {\bibfnamefont {K.}~\bibnamefont
  {Binder}},\ }\bibfield  {title} {\bibinfo {title} {Theory of first-order
  phase transitions},\ }\href {https://doi.org/10.1088/0034-4885/50/7/001}
  {\bibfield  {journal} {\bibinfo  {journal} {Rep. Prog. Phys.}\ }\textbf
  {\bibinfo {volume} {50}},\ \bibinfo {pages} {783} (\bibinfo {year}
  {1987})}\BibitemShut {NoStop}%
\bibitem [{\citenamefont {Domany}\ and\ \citenamefont
  {Riedel}(1979)}]{domany79}%
  \BibitemOpen
  \bibfield  {author} {\bibinfo {author} {\bibfnamefont {E.}~\bibnamefont
  {Domany}}\ and\ \bibinfo {author} {\bibfnamefont {E.~K.}\ \bibnamefont
  {Riedel}},\ }\bibfield  {title} {\bibinfo {title} {Two-dimensional
  anisotropic $N$-vector models},\ }\href
  {https://doi.org/10.1103/PhysRevB.19.5817} {\bibfield  {journal} {\bibinfo
  {journal} {Phys. Rev. B}\ }\textbf {\bibinfo {volume} {19}},\ \bibinfo
  {pages} {5817} (\bibinfo {year} {1979})}\BibitemShut {NoStop}%
\bibitem [{\citenamefont {Nagai}(1985)}]{nagai85}%
  \BibitemOpen
  \bibfield  {author} {\bibinfo {author} {\bibfnamefont {K.}~\bibnamefont
  {Nagai}},\ }\bibfield  {title} {\bibinfo {title} {Phase diagrams of the
  corner cubic Heisenberg model and its site-diluted version on a triangular
  lattice: Renormalization-group treatment},\ }\href
  {https://doi.org/10.1103/PhysRevB.31.1570} {\bibfield  {journal} {\bibinfo
  {journal} {Phys. Rev. B}\ }\textbf {\bibinfo {volume} {31}},\ \bibinfo
  {pages} {1570} (\bibinfo {year} {1985})}\BibitemShut {NoStop}%
\bibitem [{\citenamefont {Br\'ezin}\ and\ \citenamefont
  {Zinn-Justin}(1976)}]{brezin76}%
  \BibitemOpen
  \bibfield  {author} {\bibinfo {author} {\bibfnamefont {E.}~\bibnamefont
  {Br\'ezin}}\ and\ \bibinfo {author} {\bibfnamefont {J.}~\bibnamefont
  {Zinn-Justin}},\ }\bibfield  {title} {\bibinfo {title} {Spontaneous breakdown
  of continuous symmetries near two dimensions},\ }\href
  {https://doi.org/10.1103/PhysRevB.14.3110} {\bibfield  {journal} {\bibinfo
  {journal} {Phys. Rev. B}\ }\textbf {\bibinfo {volume} {14}},\ \bibinfo
  {pages} {3110} (\bibinfo {year} {1976})}\BibitemShut {NoStop}%
\bibitem [{\citenamefont {Kim}(1994)}]{kim94}%
  \BibitemOpen
  \bibfield  {author} {\bibinfo {author} {\bibfnamefont {J.-K.}\ \bibnamefont
  {Kim}},\ }\bibfield  {title} {\bibinfo {title} {Asymptotic scaling of the
  mass gap in the two-dimensional O(3) nonlinear \ensuremath{\sigma} model: A
  numerical study},\ }\href {https://doi.org/10.1103/PhysRevD.50.4663}
  {\bibfield  {journal} {\bibinfo  {journal} {Phys. Rev. D}\ }\textbf {\bibinfo
  {volume} {50}},\ \bibinfo {pages} {4663} (\bibinfo {year}
  {1994})}\BibitemShut {NoStop}%
\bibitem [{\citenamefont {All\'es}\ \emph {et~al.}(1999)\citenamefont
  {All\'es}, \citenamefont {Cella}, \citenamefont {Dilaver},\ and\
  \citenamefont {G\"und\"u\ifmmode~\mbox{\c{c}}\else \c{c}\fi{}}}]{alles99}%
  \BibitemOpen
  \bibfield  {author} {\bibinfo {author} {\bibfnamefont {B.}~\bibnamefont
  {All\'es}}, \bibinfo {author} {\bibfnamefont {G.}~\bibnamefont {Cella}},
  \bibinfo {author} {\bibfnamefont {M.}~\bibnamefont {Dilaver}},\ and\ \bibinfo
  {author} {\bibfnamefont {Y.}~\bibnamefont
  {G\"und\"u\ifmmode~\mbox{\c{c}}\else \c{c}\fi{}}},\ }\bibfield  {title}
  {\bibinfo {title} {Testing fixed points in the 2D $O(3)$ nonlinear
  $\ensuremath{\sigma}$ model},\ }\href
  {https://doi.org/10.1103/PhysRevD.59.067703} {\bibfield  {journal} {\bibinfo
  {journal} {Phys. Rev. D}\ }\textbf {\bibinfo {volume} {59}},\ \bibinfo
  {pages} {067703} (\bibinfo {year} {1999})}\BibitemShut {NoStop}%
\bibitem [{\citenamefont {Binder}\ and\ \citenamefont
  {Landau}(1976)}]{binder76}%
  \BibitemOpen
  \bibfield  {author} {\bibinfo {author} {\bibfnamefont {K.}~\bibnamefont
  {Binder}}\ and\ \bibinfo {author} {\bibfnamefont {D.~P.}\ \bibnamefont
  {Landau}},\ }\bibfield  {title} {\bibinfo {title} {Critical properties of the
  two-dimensional anisotropic Heisenberg model},\ }\href
  {https://doi.org/10.1103/PhysRevB.13.1140} {\bibfield  {journal} {\bibinfo
  {journal} {Phys. Rev. B}\ }\textbf {\bibinfo {volume} {13}},\ \bibinfo
  {pages} {1140} (\bibinfo {year} {1976})}\BibitemShut {NoStop}%
\bibitem [{\citenamefont {Rebbi}\ and\ \citenamefont
  {Swendsen}(1980)}]{rebbi80}%
  \BibitemOpen
  \bibfield  {author} {\bibinfo {author} {\bibfnamefont {C.}~\bibnamefont
  {Rebbi}}\ and\ \bibinfo {author} {\bibfnamefont {R.~H.}\ \bibnamefont
  {Swendsen}},\ }\bibfield  {title} {\bibinfo {title} {Monte Carlo
  renormalization-group studies of $q$-state Potts models in two dimensions},\
  }\href {https://doi.org/10.1103/PhysRevB.21.4094} {\bibfield  {journal}
  {\bibinfo  {journal} {Phys. Rev. B}\ }\textbf {\bibinfo {volume} {21}},\
  \bibinfo {pages} {4094} (\bibinfo {year} {1980})}\BibitemShut {NoStop}%
\bibitem [{\citenamefont {Cardy}\ \emph {et~al.}(1980)\citenamefont {Cardy},
  \citenamefont {Nauenberg},\ and\ \citenamefont {Scalapino}}]{cardy80}%
  \BibitemOpen
  \bibfield  {author} {\bibinfo {author} {\bibfnamefont {J.~L.}\ \bibnamefont
  {Cardy}}, \bibinfo {author} {\bibfnamefont {M.}~\bibnamefont {Nauenberg}},\
  and\ \bibinfo {author} {\bibfnamefont {D.~J.}\ \bibnamefont {Scalapino}},\
  }\bibfield  {title} {\bibinfo {title} {Scaling theory of the Potts-model
  multicritical point},\ }\href {https://doi.org/10.1103/PhysRevB.22.2560}
  {\bibfield  {journal} {\bibinfo  {journal} {Phys. Rev. B}\ }\textbf {\bibinfo
  {volume} {22}},\ \bibinfo {pages} {2560} (\bibinfo {year}
  {1980})}\BibitemShut {NoStop}%
\bibitem [{\citenamefont {Salas}\ and\ \citenamefont {Sokal}(1997)}]{salas97}%
  \BibitemOpen
  \bibfield  {author} {\bibinfo {author} {\bibfnamefont {J.}~\bibnamefont
  {Salas}}\ and\ \bibinfo {author} {\bibfnamefont {A.~D.}\ \bibnamefont
  {Sokal}},\ }\bibfield  {title} {\bibinfo {title} {Logarithmic Corrections and
  Finite-Size Scaling in the Two-Dimensional 4-State Potts Model},\ }\href
  {https://doi.org/10.1023/B:JOSS.0000015164.98296.85} {\bibfield  {journal}
  {\bibinfo  {journal} {J. Stat. Phys.}\ }\textbf {\bibinfo {volume} {88}},\
  \bibinfo {pages} {567} (\bibinfo {year} {1997})}\BibitemShut {NoStop}%
\bibitem [{\citenamefont {Lefran\ifmmode~\mbox{\c{c}}\else \c{c}\fi{}ois}\
  \emph {et~al.}(2016)\citenamefont {Lefran\ifmmode~\mbox{\c{c}}\else
  \c{c}\fi{}ois}, \citenamefont {Songvilay}, \citenamefont {Robert},
  \citenamefont {Nataf}, \citenamefont {Jordan}, \citenamefont {Chaix},
  \citenamefont {Colin}, \citenamefont {Lejay}, \citenamefont {Hadj-Azzem},
  \citenamefont {Ballou},\ and\ \citenamefont {Simonet}}]{lefrancois16}%
  \BibitemOpen
  \bibfield  {author} {\bibinfo {author} {\bibfnamefont {E.}~\bibnamefont
  {Lefran\ifmmode~\mbox{\c{c}}\else \c{c}\fi{}ois}}, \bibinfo {author}
  {\bibfnamefont {M.}~\bibnamefont {Songvilay}}, \bibinfo {author}
  {\bibfnamefont {J.}~\bibnamefont {Robert}}, \bibinfo {author} {\bibfnamefont
  {G.}~\bibnamefont {Nataf}}, \bibinfo {author} {\bibfnamefont
  {E.}~\bibnamefont {Jordan}}, \bibinfo {author} {\bibfnamefont
  {L.}~\bibnamefont {Chaix}}, \bibinfo {author} {\bibfnamefont {C.~V.}\
  \bibnamefont {Colin}}, \bibinfo {author} {\bibfnamefont {P.}~\bibnamefont
  {Lejay}}, \bibinfo {author} {\bibfnamefont {A.}~\bibnamefont {Hadj-Azzem}},
  \bibinfo {author} {\bibfnamefont {R.}~\bibnamefont {Ballou}},\ and\ \bibinfo
  {author} {\bibfnamefont {V.}~\bibnamefont {Simonet}},\ }\bibfield  {title}
  {\bibinfo {title} {Magnetic properties of the honeycomb oxide
  ${\mathrm{Na}}_{2}{\mathrm{Co}}_{2}{\mathrm{TeO}}_{6}$},\ }\href
  {https://doi.org/10.1103/PhysRevB.94.214416} {\bibfield  {journal} {\bibinfo
  {journal} {Phys. Rev. B}\ }\textbf {\bibinfo {volume} {94}},\ \bibinfo
  {pages} {214416} (\bibinfo {year} {2016})}\BibitemShut {NoStop}%
\bibitem [{\citenamefont {Bera}\ \emph {et~al.}(2017)\citenamefont {Bera},
  \citenamefont {Yusuf}, \citenamefont {Kumar},\ and\ \citenamefont
  {Ritter}}]{bera17}%
  \BibitemOpen
  \bibfield  {author} {\bibinfo {author} {\bibfnamefont {A.~K.}\ \bibnamefont
  {Bera}}, \bibinfo {author} {\bibfnamefont {S.~M.}\ \bibnamefont {Yusuf}},
  \bibinfo {author} {\bibfnamefont {A.}~\bibnamefont {Kumar}},\ and\ \bibinfo
  {author} {\bibfnamefont {C.}~\bibnamefont {Ritter}},\ }\bibfield  {title}
  {\bibinfo {title} {Zigzag antiferromagnetic ground state with anisotropic
  correlation lengths in the quasi-two-dimensional honeycomb lattice compound
  $\mathrm{N}{\mathrm{a}}_{2}\mathrm{C}{\mathrm{o}}_{2}\mathrm{Te}{\mathrm{O}}_{6}$},\
  }\href {https://doi.org/10.1103/PhysRevB.95.094424} {\bibfield  {journal}
  {\bibinfo  {journal} {Phys. Rev. B}\ }\textbf {\bibinfo {volume} {95}},\
  \bibinfo {pages} {094424} (\bibinfo {year} {2017})}\BibitemShut {NoStop}%
\bibitem [{\citenamefont {Swendsen}(1979)}]{swendsen79}%
  \BibitemOpen
  \bibfield  {author} {\bibinfo {author} {\bibfnamefont {R.~H.}\ \bibnamefont
  {Swendsen}},\ }\bibfield  {title} {\bibinfo {title} {Monte Carlo
  Renormalization Group},\ }\href {https://doi.org/10.1103/PhysRevLett.42.859}
  {\bibfield  {journal} {\bibinfo  {journal} {Phys. Rev. Lett.}\ }\textbf
  {\bibinfo {volume} {42}},\ \bibinfo {pages} {859} (\bibinfo {year}
  {1979})}\BibitemShut {NoStop}%
\bibitem [{\citenamefont {Wellegehausen}\ \emph {et~al.}(2014)\citenamefont
  {Wellegehausen}, \citenamefont {K{\"o}rner},\ and\ \citenamefont
  {Wipf}}]{wellegehausen14}%
  \BibitemOpen
  \bibfield  {author} {\bibinfo {author} {\bibfnamefont {B.~H.}\ \bibnamefont
  {Wellegehausen}}, \bibinfo {author} {\bibfnamefont {D.}~\bibnamefont
  {K{\"o}rner}},\ and\ \bibinfo {author} {\bibfnamefont {A.}~\bibnamefont
  {Wipf}},\ }\bibfield  {title} {\bibinfo {title} {Asymptotic safety on the
  lattice: The nonlinear O($N$) sigma model},\ }\href
  {https://doi.org/https://doi.org/10.1016/j.aop.2014.06.024} {\bibfield
  {journal} {\bibinfo  {journal} {Ann. Phys.}\ }\textbf {\bibinfo {volume}
  {349}},\ \bibinfo {pages} {374} (\bibinfo {year} {2014})}\BibitemShut
  {NoStop}%
\bibitem [{\citenamefont {{Berezinski{\u{i}}}}(1971)}]{berezinskii71}%
  \BibitemOpen
  \bibfield  {author} {\bibinfo {author} {\bibfnamefont {V.~L.}\ \bibnamefont
  {{Berezinski{\u{i}}}}},\ }\bibfield  {title} {\bibinfo {title} {{Destruction
  of long-range order in one-dimensional and two-dimensional systems having a
  continuous symmetry group I. Classical systems}},\ }\href
  {http://jetp.ras.ru/cgi-bin/e/index/e/32/3/p493?a=list} {\bibfield  {journal}
  {\bibinfo  {journal} {Sov. Phys. JETP}\ }\textbf {\bibinfo {volume} {32}},\
  \bibinfo {pages} {493} (\bibinfo {year} {1971})}\BibitemShut {NoStop}%
\bibitem [{\citenamefont {Kosterlitz}\ and\ \citenamefont
  {Thouless}(1973)}]{kosterlitz73}%
  \BibitemOpen
  \bibfield  {author} {\bibinfo {author} {\bibfnamefont {J.~M.}\ \bibnamefont
  {Kosterlitz}}\ and\ \bibinfo {author} {\bibfnamefont {D.~J.}\ \bibnamefont
  {Thouless}},\ }\bibfield  {title} {\bibinfo {title} {Ordering, metastability
  and phase transitions in two-dimensional systems},\ }\href
  {https://doi.org/10.1088/0022-3719/6/7/010} {\bibfield  {journal} {\bibinfo
  {journal} {J. Phys. C Solid State Phys.}\ }\textbf {\bibinfo {volume} {6}},\
  \bibinfo {pages} {1181} (\bibinfo {year} {1973})}\BibitemShut {NoStop}%
\bibitem [{\citenamefont {Shannon}\ \emph {et~al.}(2006)\citenamefont
  {Shannon}, \citenamefont {Momoi},\ and\ \citenamefont
  {Sindzingre}}]{shannon06}%
  \BibitemOpen
  \bibfield  {author} {\bibinfo {author} {\bibfnamefont {N.}~\bibnamefont
  {Shannon}}, \bibinfo {author} {\bibfnamefont {T.}~\bibnamefont {Momoi}},\
  and\ \bibinfo {author} {\bibfnamefont {P.}~\bibnamefont {Sindzingre}},\
  }\bibfield  {title} {\bibinfo {title} {Nematic Order in Square Lattice
  Frustrated Ferromagnets},\ }\href
  {https://doi.org/10.1103/PhysRevLett.96.027213} {\bibfield  {journal}
  {\bibinfo  {journal} {Phys. Rev. Lett.}\ }\textbf {\bibinfo {volume} {96}},\
  \bibinfo {pages} {027213} (\bibinfo {year} {2006})}\BibitemShut {NoStop}%
\bibitem [{\citenamefont {Zhao}\ \emph {et~al.}(2012)\citenamefont {Zhao},
  \citenamefont {Xu}, \citenamefont {Chen}, \citenamefont {Wei}, \citenamefont
  {Qin}, \citenamefont {Zhang},\ and\ \citenamefont {Xiang}}]{zhao12}%
  \BibitemOpen
  \bibfield  {author} {\bibinfo {author} {\bibfnamefont {H.~H.}\ \bibnamefont
  {Zhao}}, \bibinfo {author} {\bibfnamefont {C.}~\bibnamefont {Xu}}, \bibinfo
  {author} {\bibfnamefont {Q.~N.}\ \bibnamefont {Chen}}, \bibinfo {author}
  {\bibfnamefont {Z.~C.}\ \bibnamefont {Wei}}, \bibinfo {author} {\bibfnamefont
  {M.~P.}\ \bibnamefont {Qin}}, \bibinfo {author} {\bibfnamefont {G.~M.}\
  \bibnamefont {Zhang}},\ and\ \bibinfo {author} {\bibfnamefont
  {T.}~\bibnamefont {Xiang}},\ }\bibfield  {title} {\bibinfo {title} {Plaquette
  order and deconfined quantum critical point in the spin-1
  bilinear-biquadratic Heisenberg model on the honeycomb lattice},\ }\href
  {https://doi.org/10.1103/PhysRevB.85.134416} {\bibfield  {journal} {\bibinfo
  {journal} {Phys. Rev. B}\ }\textbf {\bibinfo {volume} {85}},\ \bibinfo
  {pages} {134416} (\bibinfo {year} {2012})}\BibitemShut {NoStop}%
\bibitem [{\citenamefont {Pohle}\ \emph {et~al.}(2023)\citenamefont {Pohle},
  \citenamefont {Shannon},\ and\ \citenamefont {Motome}}]{pohle23}%
  \BibitemOpen
  \bibfield  {author} {\bibinfo {author} {\bibfnamefont {R.}~\bibnamefont
  {Pohle}}, \bibinfo {author} {\bibfnamefont {N.}~\bibnamefont {Shannon}},\
  and\ \bibinfo {author} {\bibfnamefont {Y.}~\bibnamefont {Motome}},\
  }\bibfield  {title} {\bibinfo {title} {Spin nematics meet spin liquids:
  Exotic quantum phases in the spin-1 bilinear-biquadratic model with Kitaev
  interactions},\ }\href {https://doi.org/10.1103/PhysRevB.107.L140403}
  {\bibfield  {journal} {\bibinfo  {journal} {Phys. Rev. B}\ }\textbf {\bibinfo
  {volume} {107}},\ \bibinfo {pages} {L140403} (\bibinfo {year}
  {2023})}\BibitemShut {NoStop}%
\bibitem [{\citenamefont {Svistov}\ \emph {et~al.}(2006)\citenamefont
  {Svistov}, \citenamefont {Smirnov}, \citenamefont {Prozorova}, \citenamefont
  {Petrenko}, \citenamefont {Micheler}, \citenamefont {B\"uttgen},
  \citenamefont {Shapiro},\ and\ \citenamefont {Demianets}}]{svistov06}%
  \BibitemOpen
  \bibfield  {author} {\bibinfo {author} {\bibfnamefont {L.~E.}\ \bibnamefont
  {Svistov}}, \bibinfo {author} {\bibfnamefont {A.~I.}\ \bibnamefont
  {Smirnov}}, \bibinfo {author} {\bibfnamefont {L.~A.}\ \bibnamefont
  {Prozorova}}, \bibinfo {author} {\bibfnamefont {O.~A.}\ \bibnamefont
  {Petrenko}}, \bibinfo {author} {\bibfnamefont {A.}~\bibnamefont {Micheler}},
  \bibinfo {author} {\bibfnamefont {N.}~\bibnamefont {B\"uttgen}}, \bibinfo
  {author} {\bibfnamefont {A.~Y.}\ \bibnamefont {Shapiro}},\ and\ \bibinfo
  {author} {\bibfnamefont {L.~N.}\ \bibnamefont {Demianets}},\ }\bibfield
  {title} {\bibinfo {title} {Magnetic phase diagram, critical behavior, and
  two-dimensional to three-dimensional crossover in the triangular lattice
  antiferromagnet
  $\mathrm{Rb}\mathrm{Fe}{(\mathrm{Mo}{\mathrm{O}}_{4})}_{2}$},\ }\href
  {https://doi.org/10.1103/PhysRevB.74.024412} {\bibfield  {journal} {\bibinfo
  {journal} {Phys. Rev. B}\ }\textbf {\bibinfo {volume} {74}},\ \bibinfo
  {pages} {024412} (\bibinfo {year} {2006})}\BibitemShut {NoStop}%
\bibitem [{\citenamefont {Vojta}(2018)}]{vojta18}%
  \BibitemOpen
  \bibfield  {author} {\bibinfo {author} {\bibfnamefont {M.}~\bibnamefont
  {Vojta}},\ }\bibfield  {title} {\bibinfo {title} {Frustration and quantum
  criticality},\ }\href {https://doi.org/10.1088/1361-6633/aab6be} {\bibfield
  {journal} {\bibinfo  {journal} {Rep. Progr. Phys.}\ }\textbf {\bibinfo
  {volume} {81}},\ \bibinfo {pages} {064501} (\bibinfo {year}
  {2018})}\BibitemShut {NoStop}%
\bibitem [{\citenamefont {Sun}\ \emph {et~al.}(2023)\citenamefont {Sun},
  \citenamefont {Ye}, \citenamefont {Huang}, \citenamefont {Zhou},
  \citenamefont {Huang}, \citenamefont {Li}, \citenamefont {Ye}, \citenamefont
  {Nnokwe}, \citenamefont {Deng}, \citenamefont {Mandrus}, \citenamefont
  {Meng}, \citenamefont {Sun}, \citenamefont {Du}, \citenamefont {He},\ and\
  \citenamefont {Zhao}}]{sun23}%
  \BibitemOpen
  \bibfield  {author} {\bibinfo {author} {\bibfnamefont {Z.}~\bibnamefont
  {Sun}}, \bibinfo {author} {\bibfnamefont {G.}~\bibnamefont {Ye}}, \bibinfo
  {author} {\bibfnamefont {M.}~\bibnamefont {Huang}}, \bibinfo {author}
  {\bibfnamefont {C.}~\bibnamefont {Zhou}}, \bibinfo {author} {\bibfnamefont
  {N.}~\bibnamefont {Huang}}, \bibinfo {author} {\bibfnamefont
  {Q.}~\bibnamefont {Li}}, \bibinfo {author} {\bibfnamefont {Z.}~\bibnamefont
  {Ye}}, \bibinfo {author} {\bibfnamefont {C.}~\bibnamefont {Nnokwe}}, \bibinfo
  {author} {\bibfnamefont {H.}~\bibnamefont {Deng}}, \bibinfo {author}
  {\bibfnamefont {D.}~\bibnamefont {Mandrus}}, \bibinfo {author} {\bibfnamefont
  {Z.~Y.}\ \bibnamefont {Meng}}, \bibinfo {author} {\bibfnamefont
  {K.}~\bibnamefont {Sun}}, \bibinfo {author} {\bibfnamefont {C.}~\bibnamefont
  {Du}}, \bibinfo {author} {\bibfnamefont {R.}~\bibnamefont {He}},\ and\
  \bibinfo {author} {\bibfnamefont {L.}~\bibnamefont {Zhao}},\ }\bibinfo
  {title} {Dimensionality crossover to 2D vestigial nematicity from 3D zigzag
  antiferromagnetism in an XY-type honeycomb van der Waals magnet},\ \Eprint
  {https://arxiv.org/abs/2311.03493} {arXiv:2311.03493} \BibitemShut {NoStop}%
\bibitem [{\citenamefont {Fisher}(1974)}]{fisher74}%
  \BibitemOpen
  \bibfield  {author} {\bibinfo {author} {\bibfnamefont {M.~E.}\ \bibnamefont
  {Fisher}},\ }\bibfield  {title} {\bibinfo {title} {The renormalization group
  in the theory of critical behavior},\ }\href
  {https://doi.org/10.1103/RevModPhys.46.597} {\bibfield  {journal} {\bibinfo
  {journal} {Rev. Mod. Phys.}\ }\textbf {\bibinfo {volume} {46}},\ \bibinfo
  {pages} {597} (\bibinfo {year} {1974})}\BibitemShut {NoStop}%
\bibitem [{\citenamefont {Kosterlitz}(1974)}]{kosterlitz74}%
  \BibitemOpen
  \bibfield  {author} {\bibinfo {author} {\bibfnamefont {J.~M.}\ \bibnamefont
  {Kosterlitz}},\ }\bibfield  {title} {\bibinfo {title} {The critical
  properties of the two-dimensional xy model},\ }\href
  {https://doi.org/10.1088/0022-3719/7/6/005} {\bibfield  {journal} {\bibinfo
  {journal} {J. Phys. C Solid State Phys.}\ }\textbf {\bibinfo {volume} {7}},\
  \bibinfo {pages} {1046} (\bibinfo {year} {1974})}\BibitemShut {NoStop}%
\bibitem [{\citenamefont {Calabrese}\ and\ \citenamefont
  {Celi}(2002)}]{calabrese02}%
  \BibitemOpen
  \bibfield  {author} {\bibinfo {author} {\bibfnamefont {P.}~\bibnamefont
  {Calabrese}}\ and\ \bibinfo {author} {\bibfnamefont {A.}~\bibnamefont
  {Celi}},\ }\bibfield  {title} {\bibinfo {title} {Critical behavior of the
  two-dimensional $N$-component Landau-Ginzburg Hamiltonian with cubic
  anisotropy},\ }\href {https://doi.org/10.1103/PhysRevB.66.184410} {\bibfield
  {journal} {\bibinfo  {journal} {Phys. Rev. B}\ }\textbf {\bibinfo {volume}
  {66}},\ \bibinfo {pages} {184410} (\bibinfo {year} {2002})}\BibitemShut
  {NoStop}%
\bibitem [{\citenamefont {Binder}\ and\ \citenamefont
  {Deutsch}(1992)}]{binder92}%
  \BibitemOpen
  \bibfield  {author} {\bibinfo {author} {\bibfnamefont {K.}~\bibnamefont
  {Binder}}\ and\ \bibinfo {author} {\bibfnamefont {H.-P.}\ \bibnamefont
  {Deutsch}},\ }\bibfield  {title} {\bibinfo {title} {Crossover Phenomena and
  Finite-Size Scaling Analysis of Numerical Simulations},\ }\href
  {https://doi.org/10.1209/0295-5075/18/8/001} {\bibfield  {journal} {\bibinfo
  {journal} {Europhys. Lett.}\ }\textbf {\bibinfo {volume} {18}},\ \bibinfo
  {pages} {667} (\bibinfo {year} {1992})}\BibitemShut {NoStop}%
\bibitem [{\citenamefont {Tomita}(2014)}]{tomita14}%
  \BibitemOpen
  \bibfield  {author} {\bibinfo {author} {\bibfnamefont {Y.}~\bibnamefont
  {Tomita}},\ }\bibfield  {title} {\bibinfo {title} {Finite-size scaling
  analysis of pseudocritical region in two-dimensional continuous-spin
  systems},\ }\href {https://doi.org/10.1103/PhysRevE.90.032109} {\bibfield
  {journal} {\bibinfo  {journal} {Phys. Rev. E}\ }\textbf {\bibinfo {volume}
  {90}},\ \bibinfo {pages} {032109} (\bibinfo {year} {2014})}\BibitemShut
  {NoStop}%
\bibitem [{\citenamefont {Burgelman}\ \emph {et~al.}(2023)\citenamefont
  {Burgelman}, \citenamefont {Devos}, \citenamefont {Vanhecke}, \citenamefont
  {Verstraete},\ and\ \citenamefont {Vanderstraeten}}]{burgelman23}%
  \BibitemOpen
  \bibfield  {author} {\bibinfo {author} {\bibfnamefont {L.}~\bibnamefont
  {Burgelman}}, \bibinfo {author} {\bibfnamefont {L.}~\bibnamefont {Devos}},
  \bibinfo {author} {\bibfnamefont {B.}~\bibnamefont {Vanhecke}}, \bibinfo
  {author} {\bibfnamefont {F.}~\bibnamefont {Verstraete}},\ and\ \bibinfo
  {author} {\bibfnamefont {L.}~\bibnamefont {Vanderstraeten}},\ }\bibfield
  {title} {\bibinfo {title} {Contrasting pseudocriticality in the classical
  two-dimensional Heisenberg and ${\mathrm{RP}}^{2}$ models: Zero-temperature
  phase transition versus finite-temperature crossover},\ }\href
  {https://doi.org/10.1103/PhysRevE.107.014117} {\bibfield  {journal} {\bibinfo
   {journal} {Phys. Rev. E}\ }\textbf {\bibinfo {volume} {107}},\ \bibinfo
  {pages} {014117} (\bibinfo {year} {2023})}\BibitemShut {NoStop}%
\bibitem [{\citenamefont {Harada}(2011)}]{harada11}%
  \BibitemOpen
  \bibfield  {author} {\bibinfo {author} {\bibfnamefont {K.}~\bibnamefont
  {Harada}},\ }\bibfield  {title} {\bibinfo {title} {Bayesian inference in the
  scaling analysis of critical phenomena},\ }\href
  {https://doi.org/10.1103/PhysRevE.84.056704} {\bibfield  {journal} {\bibinfo
  {journal} {Phys. Rev. E}\ }\textbf {\bibinfo {volume} {84}},\ \bibinfo
  {pages} {056704} (\bibinfo {year} {2011})}\BibitemShut {NoStop}%
\end{thebibliography}%

\end{document}